\documentclass[12pt,a4paper]{article}

\usepackage{geometry}
\usepackage{setspace}

\usepackage[fontsize=12pt]{fontsize}

\usepackage{times}
\usepackage{graphicx}
\graphicspath{{Figures/}}
\usepackage{bm}
\usepackage{amssymb,amsmath,amsthm}
\usepackage{booktabs}
\usepackage{todonotes}
\usepackage{bbm}
\usepackage{caption}
\usepackage{multirow}
\usepackage{natbib}
\usepackage{authblk}
\usepackage{hyperref}

\renewcommand{\P}{\mathbb{P}}
\newcommand{\E}{\mathbb{E}}
\newcommand{\R}{\mathbb{R}}

\newcommand{\N}{\mathbb{N}}

\newcommand{\eqdis}{\stackrel{\text{\tiny d}}{=}}

\newcommand{\as}{\mbox{a.s.}}

\newcommand{\Ind}[1]{I_{\{#1\}}}

\newcommand{\cov}{\operatorname{cov}}

 \newcommand{\rd}{\mathrm{d}}

\renewcommand{\leq}{\leqslant}
\renewcommand{\geq}{\geqslant}
\renewcommand{\le}{\leqslant}
\renewcommand{\ge}{\geqslant}
\newcommand{\indicator}[1]{\ensuremath{I_{\{#1\}}}}

\newcommand{\asequal}{\ensuremath{\stackrel{\footnotesize{\as}}{=}}}

\newcommand{\gtrans}{g}
\newcommand{\htrans}{h}
\newcommand{\vtrans}{\ensuremath{T_\vee}}
\newcommand{\legendre}{P}

\newcommand{\orthobase}{B}
\newcommand{\orthosystem}{\mathcal{B}}
\newcommand{\pbase}{F}
\newcommand{\dbase}{f}
\newcommand{\udp}{T}

\newcommand{\slegendre}{L}
\newcommand{\blegendre}{\Lambda}

\newcommand{\ocos}{\Omega}
\newcommand{\psionAm}{\psi\vert_{A_m}}
\newcommand{\udponAl}{\udp\vert_{A_\ell}}

\newtheorem{proposition}{Proposition}
\newtheorem{theorem}{Theorem}

\newtheorem{assump}{Assumption}

\newtheorem{assumpP}{Assumption}
\newtheorem{assumpC}{Assumption}

\renewcommand\theassumpP{P\arabic{assumpP}}

\newtheorem{lemma}{Lemma}
\newtheorem{corollary}{Corollary}
\newtheorem{definition}{Definition}
\newtheorem{remark}{Remark}
\newtheorem{example}{Example}

\begin{document}

\newcommand{\myThanks}{\thanks{Address correspondence to
     Alexander J.~McNeil, The School for Business and Society,
     University of York, Heslington, York YO10 5DD, UK, 
     \texttt{alexander.mcneil@york.ac.uk}.}}

\title{Measures and Models of Non-Monotonic Dependence}

\author{Alexander J. McNeil\myThanks}
\affil{The School for Business and Society, University of York, UK}

\author{Johanna G. Ne\v{s}lehov\'a}
\affil{Department of Mathematics and Statistics, McGill University,
  Montr\'eal, Canada}

\author{Andrew D. Smith}
\affil{University College Dublin, Ireland}

\maketitle

%
%
%
%
%
%
%
%
%
%



\setstretch{1}
\begin{abstract}A margin-free measure of bivariate association generalizing Spearman's
 rho to the case of non-monotonic dependence is defined in terms of two square integrable
 functions on the unit interval. Properties
 of generalized Spearman correlation are investigated when the functions are piecewise
 continuous and strictly monotonic, with particular focus on the
 special cases where the functions are
 drawn from orthonormal bases of Legendre polynomials and
 cosine functions. For continuous random variables, generalized Spearman
 correlation is treated as a copula-based measure and shown to
 depend on a pair of uniform-distribution-preserving (udp) transformations
 determined by the underlying functions. Bounds for generalized
 Spearman correlation are derived and a novel technique referred to
 as stochastic inversion of udp transformations is used to
 construct singular copulas that attain the bounds and parametric
 copulas with densities that
 interpolate between the bounds and model different degrees of
 non-monotonic dependence. Sample analogues of generalized Spearman
 correlation are proposed and their asymptotic
 properties are investigated. Potential applications of the theory are demonstrated
 including: exploratory analyses of the dependence structures of
 datasets and their symmetries; elicitation of functions
 maximizing generalized Spearman correlation; and construction of tractable probability densities to model a
 wide variety of non-monotonic dependencies.
 \end{abstract}
 
 \noindent \textbf{Keywords:} copulas, Correlation bounds, Cosine basis, Legendre polynomial basis, Non-monotonic dependence, Spearman's rho, Uniform-distribution-preserving transformations

\maketitle


\section{Introduction}

\begin{figure}[t!]
  \centering
    \includegraphics[width=15cm,height=6.5cm]{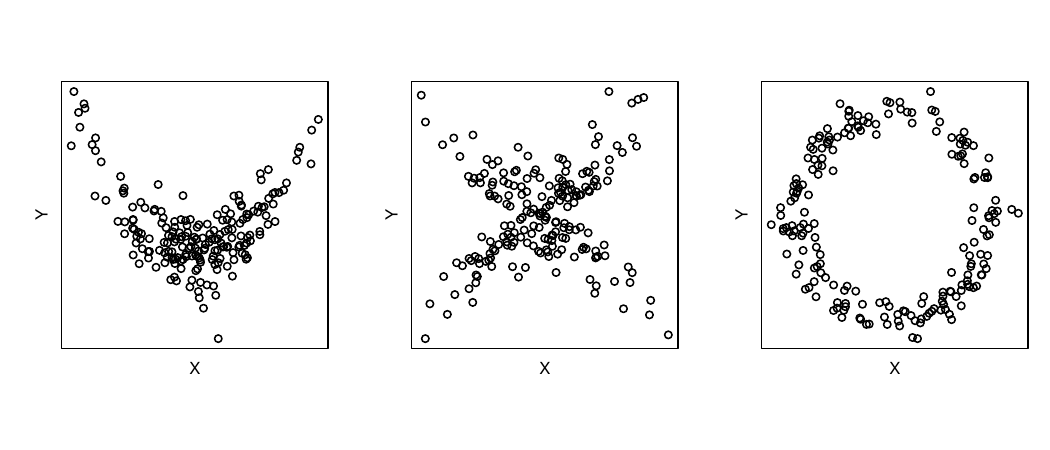}\vspace{-1cm}
  \caption{\label{fig:motivate}Models of the form $h(Y) = g(X) +
    \epsilon$. Left: $X \sim N(0,1)$ and $Y = X^2+ \epsilon$ where $\epsilon \sim
N(0,1)$. Middle: $X \sim N(0,1)$ and $Y= \pm \sqrt{X^2 +\epsilon}$
where $\epsilon = |Z|$ and $Z \sim N(0,1.5)$; hence
$Y^2 = X^2 + \epsilon$. Right: $\Theta \sim U(0,2\pi), X =
\cos(\Theta)Z, Y = \sin(\Theta)Z$ where $Z \sim N(1,0.1)$; hence $Y^2
= 1 - X^2 + \epsilon$ where $\epsilon = Z^2-1$. 200 points are randomly
simulated from each model.}
\end{figure}

In many bivariate datasets an important
dependency between two variables $X$ and $Y$ may be missed 
if we apply a standard measure of association or correlation. 
This occurs, in particular, when the dependency manifests itself in
a non-monotonic relationship between $X$ and $Y$.
The three pictures
in Figure~\ref{fig:motivate} show stylized non-monotonic dependencies. All three
pictures may be subsumed in the general model $\htrans(Y) = \gtrans(X) +
\epsilon$ where $\epsilon$ is random noise and $\gtrans$ and $\htrans$ are
piecewise strictly monotonic functions; details are found in
the caption. In all three cases the linear correlation of $X$ and $Y$ is
modest but the dependence between $X$ and $Y$ is clearly strong.

Non-monotonic dependencies of the kind illustrated in
Figure~\ref{fig:motivate} can arise in many
contexts. Economics and finance abound in u-shaped or inverted
u-shaped relationships. The Laffer curve~\citep{bib:laffer-04} posits an inverse u-shaped
relationship between tax
revenue (Y) and tax rates (X).
Time series of stock prices
show a u-shaped relationship between volatility on one day (Y) and returns the previous day (X), a feature that is exploited by 
stochastic volatility models like ARCH and
GARCH~\citep{bib:engle-82}. They also show a cross-shaped
relationship between returns on successive days, reminiscent of the
middle picture in Figure~\ref{fig:motivate}, a phenomenon described in papers
by~\citet{bib:loaiza-maya-et-al-18}
and~\citet{bib:bladt-mcneil-21}. 
Out-of-phase periodic
phenomena can give
rise to circular patterns\footnote{They can also give rise to 
  more complex so-called Lissajous curves.} of the kind suggested by the right picture in Figure~\ref{fig:motivate}; see, for example,~\citet{bib:sekiguchi-et-al-06} on the relationship between
Schumann resonance and global surface temperature.
Although Figure~\ref{fig:motivate} illustrates situations where there is an
association between $\gtrans(X)$ and $\htrans(Y)$ for 
piecewise strictly monotonic $g$ and $h$ with at most a single turning
point, multiple turning points are also possible. An example from ecology is the bimodal, m-shaped relationship between marine species richness 
and latitude documented by~\citet{bib:chaudhary-saeedi-costello-16}. Many bivariate datasets giving the
appearance of randomness may mask more
complicated non-monotonic dependencies, an issue addressed
by~\citet{bib:zhang-19}.


This article has two main aims: first, to investigate the
properties of a particular margin-free measure of functional correlation between
$g(X)$ and $h(Y)$ for piecewise continuous and strictly monotonic $g$ and $h$; second, to show how this
measure can be used to inform the construction of flexible parametric copula
models for data showing complex non-monotonic behaviour.

The concept of a functional correlation can be traced back to
the classical Hirschfeld-Gebelein-R\'{e}nyi (HGR)
maximal correlation~\citep{bib:gebelein-41} given by $\sup
\rho(g(X),h(Y))$ where $\rho$ denotes Pearson correlation
and the supremum is taken over measurable $g$ and $h$ such
that $g(X)$ and $h(Y)$ have finite variance. The determination of $g$
and $h$ typically uses techniques from the field of
nonlinear canonical correlation analysis, such as the
ACE algorithm~\citep{bib:breiman-friedman-85,bib:buja-90}, or neural
networks~\citep{bib:lai-fyfe-00}. Literature on the properties of
functional covariances and correlations and their relation to density expansions
include the work of Cuadras (see~\citet{bib:cuadras-15} and references
therein) which builds on~\citeauthor{lancaster:1958}'s (\citeyear{lancaster:1958})
work on the Pearson contingency coefficient and canonical correlations.
Measures related to maximal correlation are explored in~\citet{bib:holzmann-klar-24}.

We consider margin-free functional correlations for
continuous random variables $X$ and $Y$ given by
$\rho_{\{\gtrans,\htrans\}}(X,Y) = \rho(\gtrans(F_X(X)),
  \htrans(F_Y(Y)))$
where $F_X$ and $F_Y$ denote the distributions functions (dfs) of $X$
and $Y$
and $g$ and $h$ are piecewise continuous and strictly monotonic,
square integrable functions on $[0,1]$. The measure
$\rho_{\{\gtrans,\htrans\}}$ generalizes 
Spearman's rho, which is the case $\gtrans(x) = \htrans(x) =
x$, and we refer to it
as generalized Spearman correlation. Clearly $\rho_{\{\gtrans,\htrans\}}\left(X,Y \right)$ can also
be expressed in the form $
\rho(\tilde \gtrans(X),  \tilde \htrans(Y))$ for functions $\tilde \gtrans$ and $\tilde \htrans$ that are
piecewise strictly monotonic on the support of $X$ and $Y$. However, the
advantage of the margin-free definition is that $\rho_{\{\gtrans,\htrans\}}(X,Y)$
depends only on $\gtrans$, $\htrans$ and the unique copula $C$ of $(X,Y)^\top$,  which is the distribution
 function of $(U,V)^\top$ where $U = F_X(X)$ and $V =
 F_Y(Y)$~\citep{sklar:1959}. Thus the proposed generalized Spearman correlation is a copula-based
 measure.


Of particular interest is the case where the functions
$g$ and $h$ are drawn from an orthonormal basis of functions on the
unit interval. The best-known such basis consists of the shifted Legendre
polynomials although we also consider an alternative basis
related to cosine functions. When $\gtrans$ and $\htrans$ are functions taken
from the same basis system,
we refer to the resulting correlations as basis correlations, or as Legendre or
cosine correlations according to the name of the basis.
Since every square integrable function has a mean-square
convergent expansion in an orthonormal basis, this admits the idea 
that 
generalized Spearman correlations can be approximated in terms of basis correlations.
Legendre polynomials have previously been used by~\citet{bib:kallenberg-ledwina-99}
and~\citet{bib:kallenberg-08} to measure and model
dependencies. Empirical measures based on Legendre polynomials are used in the actuarial literature to calibrate copulas,
particularly under the name
arachnitude~\citep{bib:androschuck-et-al-17}\footnote{The concept is
 based on
  the work of Richard Shaw, Andrew Smith and Grigory Spivak as
  presented at various actuarial conferences; see, for example,
  https://www.actuaries.org.uk/system/files/documents/pdf/b3.pdf.}.

While the maximal correlation literature concerns the maximization of
the functional correlation $\rho(g(X),h(Y))$ with respect to $g$ and
$h$ for fixed $(X,Y)^\top$, we
investigate
Spearman correlations $\rho_{\{g,h\}}(X,Y)$ 
for fixed $g$ and $h$ and find their sharp upper and lower bounds with
respect to the Fr\'{e}chet class defined by $F_X$ and $F_Y$. 
Our main
tools are
uniform-distribution-preserving (udp) transformations and we show that
piecewise strictly monotonic $\gtrans$ and $\htrans$ can be associated with udp
functions $T_\gtrans$ and $T_\htrans$ and the generalized Spearman
correlation of the copula $C$ of $(X,Y)^\top$ can be understood as a
measure of monotonic dependence for an alternative copula which is
the distribution of $(T_\gtrans \circ F_X(X), T_\htrans \circ F_Y(Y))^\top$. We
develop a technique that we call stochastic inversion of a
udp transformation which allows us to create tractable models that attain the
bounds for generalized Spearman correlation and interpolate
between them.  Our results generalize
those of~\cite{bib:mcneil-20} and~\citet{bib:bladt-mcneil-21} for the special udp functions known as v-transforms. 
The constructions differ from those of~\cite{quessy:2024} and~\cite{hofert/pang:2025} which also rely on udp
transformations.

We propose a rank-based estimator of generalized Spearman
correlation and derive the asymptotic properties of
matrices of sample basis correlations. Estimates of these matrices
open the door to a variety of possible applications and we outline
three possible directions.
First, we suggest that sample basis correlation matrices are 
a powerful tool for exploratory analyses. They reveal the relative
strengths of
different kinds of non-monotonic dependency as well as the presence or
absence of certain forms of symmetry in the underlying copula, such as
exchangeability, radial symmetry or joint symmetry. This idea points
the way to new formal tests of symmetry. Second, they can be used to find piecewise-monotonic functions
$\gtrans$ and $\htrans$ that approximate maximal generalized Spearman
correlation and thus reveal the strongest form of non-monotonic
dependency in a given dataset. This gives important information for model building, which is our third application.
We use stochastic inversion to construct tractable parametric bivariate
copula models for datasets showing non-monotonic
dependencies. The models have
explicit densities and may be easily fitted to data using
standard likelihood techniques. Their flexible forms greatly extend
the range of phenomena that can be modelled with bivariate copulas or
vine copulas. Special cases incorporating so-called
v-transforms have previously been used in applications
to time series~\citep{bib:bladt-mcneil-21,bib:dias-han-mcneil-25}.

The paper is organized as follows. In Section~\ref{sec:2} we define generalized
Spearman correlation, investigate its properties 
and explain why the special cases we refer to as basis
correlations are of particular interest. In Section~\ref{sec:udp} we
address the
question of finding sharp upper and lower bounds for generalized Spearman
correlation and introduce the aforementioned udp transformations. 
The powerful technique of stochastic inversion of udp transformations is
presented in Section~\ref{sec:bounds} along with the
  novel model construction technique it facilitates.
Estimation of generalized Spearman correlations and an outlook
on applications are found in Section~\ref{sec:persp-appl}. The Supplementary Material contains proofs,
technical material, and additional examples.

\section{Generalized Spearman correlation} \label{sec:2}
  
  \subsection{Definitions, assumptions, and basic properties}\label{sec:nonmoncor}

Throughout, we denote by $\mathcal{L}^2([0,1])$ the space of square integrable functions on $[0,1]$ with respect to the unit weight function, that is, the space of functions $\psi : [0,1] \to \mathbb{R}$ with $\int_0^1 \psi^2(u) \rd u < \infty$. In what follows, we will consider transformations in $\mathcal{L}^2([0,1])$ with certain properties, introduced in the following definition. 

\begin{definition}\label{def:psiproperties}
A function $\psi \in \mathcal{L}^2([0,1])$ is said to be:
\begin{enumerate}
\item[(i)] \emph{standardized} if 
\[
\int_0^1 \psi(u) \rd u = 0, \quad \int_0^1 \psi^2(u) \rd u = 1;
\]
\item[(ii)] \emph{piecewise continuous and strictly monotonic} if
  there exists a finite partition $0 = u_0 < u_1 < \dots < u_M = 1$ so that for all $m \in \{1,\dots, M\}$, the restriction of $\psi$ on $(u_{m-1}, u_m)$ is continuous and strictly monotonic;
\item[(iii)] \emph{regular} if it is continuous on $[0,1]$,
  continuously differentiable on $(0,1)$, and if its derivative is
  bounded on $(0,1)$ and satisfies $\psi^\prime(u) = 0$ iff $u$ is a
  turning point.\footnote{The exclusion of saddle points for $\psi$
    simplifies later proofs but could be relaxed. However, none of our
    examples require the consideration of saddle points.}
\end{enumerate}
\end{definition}

\noindent The generalized Spearman correlation previewed in the Introduction is defined next.
\begin{definition}[Generalized Spearman correlation]\label{def:nmcor} Let $X$ and $Y$ be continuous random variables with
distribution functions $F_X$ and $F_Y$.
The generalized Spearman correlation of $X$ and $Y$ with respect to $\gtrans, \htrans \in \mathcal{L}^2([0,1])$ is given in terms of Pearson's correlation operator $\rho$ by
\begin{equation}
  \label{eq:64}
\rho_{\{\gtrans,\htrans\}}\left(X,Y \right) = \rho\left(\gtrans(F_X(X)),
  \htrans(F_Y(Y))\right). 
\end{equation}
\end{definition}
In view of the fact that generalized Spearman correlation is based on Pearson's correlation, we can standardize $\gtrans$ and $\htrans$ without loss of generality. 
\begin{assump}\label{assA1.new}
The transformations $\gtrans$ and $\htrans$ are standardized in the
sense of Definition~\ref{def:psiproperties}(i).
\end{assump}
Working under 
Assumption~\ref{assA1.new} allows for the convenient expression
\begin{equation}\label{eq:expectation}
\rho_{\{\gtrans,\htrans\}}\left(X,Y \right) = \E\left(\gtrans(F_X(X)) \htrans(F_Y(Y))\right). 
\end{equation}
In principle, the transformations $\gtrans$ and $\htrans$ in Definition~\ref{def:nmcor} can be any square integrable functions on the unit interval. In practical applications however, it makes most sense to generalize measures of monotone dependence by focusing on piecewise monotonic and continuous transformations. We articulate this requirement as another assumption. 

\begin{assump}\label{assA2.new}
The transformations $\gtrans$ and $\htrans$ are piecewise continuous
and strictly monotonic in the sense of
Definition~\ref{def:psiproperties}(ii) with partitions that may differ.
\end{assump}

Some results later on will require the following greater degree of smoothness of $\gtrans$ and $\htrans$.

\begin{assump}\label{assA3.new}
The transformations $\gtrans$ and $\htrans$ are regular in the sense of Definition~\ref{def:psiproperties}(iii).
\end{assump}

\begin{example}\label{ex:gandh} The functions given, for all $u \in [0,1]$, by
$u \mapsto u$ and $u \mapsto |2u-1|^2$
satisfy Assumptions~\ref{assA2.new} and \ref{assA3.new} and can be readily standardized to obtain $g(u) = \sqrt{3}(2u-1)$ and $h(u) = (\sqrt{5}/2)(3|2u-1|^2 - 1)$ fulfilling Assumption~\ref{assA1.new}. Utilizing $\gtrans$ and $\htrans$ in \eqref{eq:64} leads to three known measures of dependence.

First, $\rho_{\{\gtrans,\gtrans\}}(X,Y)$ is the population version of Spearman's rho. Second,  $\rho_{\{\htrans,\htrans\}}(X,Y)$ is the population version of the
 sample ``arachnitude''  based on the work of Richard Shaw, Andrew Smith and Grigory Spivak \citep{bib:androschuck-et-al-17}. It measures the
 tendency of  values $(F_X(X),F_Y(Y))^\top$
 to accumulate on the two arms of a cross formed by the diagonals of
 the unit square, and hence we refer to it as a measure of
 cruciformity.
It can also be thought of as measuring the propensity for extreme values
of $X$ to be associated with extreme values of $Y$. 

Finally, $\rho_{\{\gtrans, \htrans\}}(X,Y)$ and $\rho_{\{\htrans, \gtrans\}}(X,Y)$ are the population versions of the so-called
 sample rank convexity also considered by Richard Shaw, Andrew Smith and Grigory Spivak. They
  measure the strength of u-shaped
 dependencies of $X$ on $Y$ and $Y$ on $X$ respectively. We describe them as
 measures of angularity since they  measure the
 tendency of  values $(F_X(X),F_Y(Y))^\top$
 to accumulate on v-shaped sets, for example $\{(u,v) : v = |2u-1|\}$
 in the case of $\rho_{\{\htrans, \gtrans\}}(X,Y)$.
\end{example}

\begin{example}\label{ex:asymmetrich}
To introduce asymmetry into the measurement of the strength of v-shaped and cross-shaped relationships between $F_X(X)$ and $F_Y(Y)$, we consider the generalized function
\begin{equation}\label{eq:adymmetric-u-shaped}
\htrans(u) =
\begin{cases}
   \left(\frac{\delta - u}{\delta} \right)^q, & u \leq \delta,\\
   \left(\frac{u-\delta}{1-\delta}\right)^p,& u > \delta,
\end{cases}\qquad 0 < \delta < 1,\, p >1,\, q>1,
\end{equation}
which also satisfies Assumptions~\ref{assA2.new} and \ref{assA3.new}
and reduces to the function $\htrans$ in Example~\ref{ex:gandh} when
$\delta=0.5$ and $p=q=2$. For $\delta \neq 0.5$ or $p \neq q$ the
function is asymmetric. 
\end{example}
For a random vector
$(X,Y)^\top$ with continuous margins,
$\rho_{\{\gtrans,\htrans\}}(X,Y)$ depends only on the unique underlying copula $C$.  Under
Assumption~\ref{assA1.new}, this leads to the integral
\[
\rho_{\{\gtrans,\htrans\}}\left(X,Y \right) = \int_0^1 \int_0^1 \gtrans(u) \htrans(v) \rd C(u,v) = \int_0^1 \int_0^1 \gtrans(u) \htrans(v) c(u,v) \rd u \rd v,
\]
where the second equality holds if $C$ has a density $c$. Although the integral involving the copula density seems convenient, it is rarely explicit and typically requires numerical integration. 

We derive an alternative expression that can be
applied to any copula $C$, regardless of whether it has a density or
not. This formula also provides 
a faster and more stable method of computing $\rho_{\{\gtrans,\htrans\}}(X,Y)$; the computational advantages derive from the
boundedness of the copula function $C$ and are particularly important
when strong tail dependencies are present.

\begin{proposition}\label{prop:HardyKrause}
Under Assumptions~\ref{assA1.new} and \ref{assA3.new}, the generalized Spearman correlation equals
\begin{equation}\label{eq:13general}
\rho_{\{\gtrans,\htrans\}}(X,Y) = \int_0^1 \int_0^1 C(u,v) \gtrans^\prime(u) \htrans^\prime(v) \rd u \rd v - \gtrans(1) \htrans(1),
\end{equation}
where $\gtrans^\prime$ and $\htrans^\prime$ denote the derivatives of $\gtrans$ and $\htrans$, respectively.
\end{proposition}

\subsection{Generalized Spearman correlation for orthonormal systems}\label{sec:basecor}

A broad class of generalized Spearman
correlation measures ensues when
$\gtrans$ and $\htrans$ are chosen from a complete orthonormal system
$\orthosystem = \{\orthobase_j(u), j \in \N_0\}$  in
$\mathcal{L}^2([0,1])$. Systems that include a constant function are particularly convenient, which leads to the following terminology.

\begin{definition}[Correlation basis]\label{def:correlation-basis}
  A correlation basis is a complete orthonormal system
  $\orthosystem = \{\orthobase_j(u), j \in \N_0\}$ of functions in $\mathcal{L}^2([0,1])$ with respect to a
  unit weight function such that $\orthobase_0(u) \equiv 1$
      for all $u \in [0,1]$.
  \end{definition}
  
The inclusion of $\orthobase_0(u) =1$ rules out
some complete orthonormal systems indexed by $\N$ but have the
advantage that, for any pair of integers $j,k \in \N$,
Assumption~\ref{assA1.new} holds with $\gtrans=\orthobase_j$ and
$\htrans=\orthobase_k$ so that the resulting generalized Spearman correlation is
the expectation~\eqref{eq:expectation}. This follows easily from the
orthogonality relations
\begin{equation}
\int_{0}^1 \orthobase_j(u) \orthobase_k(u) \mathrm{d} u = 
\delta_{jk}, \quad j,k \in \N_0,\label{eq:orthog}
\end{equation}
where $\delta_{jk}$ denotes the Kronecker delta ($\delta_{jk} =1$ if
$j=k$ and $\delta_{jk}=0$ if $j\neq k$). As well as a computational
advantage, we show in Section~\ref{sec:expansions} that this has an
interpretational advantage in that the densities of certain copulas
can be written as convergent expansions of products of the basis
functions with the corresponding generalized Spearman correlations appearing as the coefficients.
 We introduce the following notion to simplify terminology and notation.

\begin{definition}\label{def:basis-correlation}
Let $X$ and $Y$ be continuous random variables with
distribution functions $F_X$ and $F_Y$ and let $\orthosystem$ be a
correlation basis. Then,  for $j,k \in \N$, the generalized Spearman correlation $\rho_{\{\orthobase_j,\orthobase_k\}}(X,Y)$ is called the
 basis correlation of order $(j,k)$ 
 and denoted $  \rho^B_{jk}(X, Y)$.
\end{definition}

Before we elaborate on the special role that basis correlations play in the study of non-monotonic dependence in Section~\ref{sec:expansions}, we introduce three bases conforming to
Definition~\ref{def:correlation-basis}. 

\begin{figure}[t]
  \centering
   \includegraphics[width=\textwidth,height=3in]{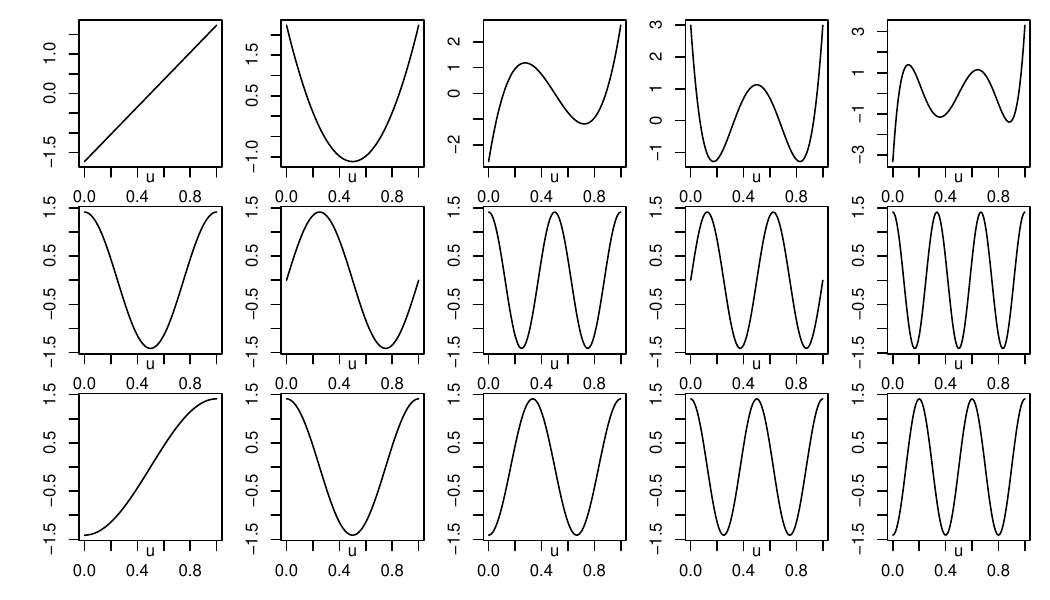}
   \caption{\label{fig:basis} Basis functions $\{\orthobase_j(u),
     j=1,\ldots,5\}$ for three correlation bases: Legendre (top row),
     Fourier (middle row); cosine (bottom row).}
\end{figure}

\begin{example}[Legendre polynomial basis]\label{ex:legendre}
 This basis is obtained by taking scaled versions of
  the shifted Legendre polynomials (top row of Figure~\ref{fig:basis}). If $\{ \legendre_j(x) ,
  j \in \N_0\}$ is the set of Legendre polynomials on $[-1,1]$ then
  the shifted Legendre polynomials are the functions
  $\slegendre_j(u) = \legendre_j(2u-1)$, 
  $j \in \N_0$ on $[0,1]$ and the basis functions are  $\blegendre_j(u) =  \sqrt{2j+1}\slegendre_j(u), j \in \N_0$.
\end{example}

  \begin{example}[Fourier basis]\label{ex:sine}
The basis functions (middle row of Figure~\ref{fig:basis}) are
    given by
  $\Psi_0(u) =1$ and $\Psi_j(u) =\sqrt{2}\cos((j+1) \pi u)$ when $j$ is odd, and $
      \Psi_j(u) =  \sqrt{2}\sin(j \pi u)$ when $j$ is even.
    \end{example}

  \begin{example}[Cosine basis]\label{ex:cosine}  An alternative basis on $[0,1]$ can be constructed using only cosine functions
    (bottom row of Figure~\ref{fig:basis}) with
  $\ocos_0(u) =1$ and $\ocos_j(u) = (-1)^{j} \sqrt{2}\cos\left(j \pi u\right)$,  $j \in \N$.
  \end{example}


As with the wider class of generalized Spearman correlations, it will be useful to
consider refined regularity conditions and this leads to the following notion.

\begin{definition}\label{def:regularbasis}
A correlation basis $\mathcal{B}$ is called \emph{regular} if for all
$j \in \N$, $\orthobase_j$ is piecewise continuous and strictly
monotonic and regular in the sense of parts (ii) and (iii) of
Definition~\ref{def:psiproperties}.
\end{definition}

The above Legendre, Fourier and cosine bases are easily seen to be
regular; an example of a correlation basis which is not regular is the
Haar system \citep[Section 5.3]{walnut:2013}. The advantage of a regular correlation basis is that for each $j,k \in \N$, the pair of transformations $\gtrans=\orthobase_j$ 
and $\htrans=\orthobase_k$ fulfills Assumptions~\ref{assA1.new}, \ref{assA2.new} and \ref{assA3.new}. In particular, this allows us to use Proposition~\ref{prop:HardyKrause} to calculate basis correlations.

\begin{example}[Legendre polynomial basis]\label{ex:LegendreFormula}
Since $\blegendre_j(1) = \sqrt{2j +1}$ for $j \in \N$,~\eqref{eq:13general} becomes
  \begin{equation}
    \label{eq:13}
    \rho_{jk}^{\blegendre}(X,Y) = \sqrt{2j+1}\sqrt{2k+1}\left(
\int_0^1 \int_0^1 \slegendre^\prime_j(u) \slegendre^\prime_k(v) C(u,v) \rd u
\rd v
 -1     \right),
\end{equation}
where $\slegendre^\prime_j$ is the derivative of the shifted
Legendre polynomial $\slegendre_j$. 
Note that $\blegendre_1(u) = \sqrt{3}(2u-1)$ and $\blegendre_2(u) =
\sqrt{5}(3 (2u-1)^2 -1)/2$ are precisely the normalized versions of
the functions $\gtrans$ and $\htrans$ in
Example~\ref{ex:gandh}. Consequently, $\rho_{11}^\Lambda$ is
Spearman's rho and $\rho_{22}^\Lambda$, $\rho_{12}^\Lambda$, and
$\rho_{21}^\Lambda$ are the measures of cruciformity and angularity,
respectively.
  \end{example} 

\begin{example}[Cosine basis]\label{ex:CosFormula}
In this case we have for each $j \in \N$ that $\ocos_j(1) = \sqrt{2}$ while $\ocos^\prime_j(u) = (-1)^{j+1} (j \pi \sqrt{2}) \sin(j \pi u)$. An application of \eqref{eq:13general} gives
\[
\rho_{jk}^{\ocos}(X,Y) = (-1)^{j+k} 2 jk \pi^2 \int_0^1 \int_0^1 \sin(j\pi u) \sin(k \pi v) C(u,v) \rd u \rd v - 2.
\]
\end{example}


We will sometimes impose additional assumptions that
  lead to 
appealing interpretations and properties of the resulting basis correlations.
This motivates the notion of a natural correlation basis, which applies to the Legendre
and the cosine bases, but not the Fourier basis.

\begin{definition}\label{def:naturalbasis}
A correlation basis $\mathcal{B}$ is called \emph{natural} if it is a
regular correlation basis and if the basis functions $\orthobase_j$, $j \in \N$, satisfy the following additional properties:
\begin{enumerate}
\item[(i)]  $\orthobase_j^\prime(u) > 0$ on an
interval $(1-\epsilon_j,1)$ for some $\epsilon_j > 0$;
\item[(ii)] $\orthobase_j$ has exactly $j-1$ turning points;
\item[(iii)] $\orthobase_j(1-u) = (-1)^j \orthobase_j(u)$
       for all $u \in [0,1]$.
\end{enumerate}
\end{definition}
%

Property (i) in Definition~\ref{def:naturalbasis} makes
our basis systems explicit by choosing
signs such that, for $j\in\N$, $\orthobase_j(u)$ is always an increasing function at
the right end of the interval $[0,1]$.
Without this assumption, there is some indeterminacy in a correlation basis $\{
\orthobase_j(u) , j \in \N_0\}$ satisfying Definition~\ref{def:correlation-basis}, since we can
replace any function $\orthobase_j(u)$ by
$-\orthobase_j(u)$ and preserve orthogonality.

Property (ii) in Definition~\ref{def:naturalbasis} implies that the the first non-constant basis function
$\orthobase_1$ has no turning point and the basis
correlation $\rho_{11}^B$ measures monotonic dependence; in
conjunction with property (i)  it measures positive dependence.
In contrast, the functions $\orthobase_j$ of the
Fourier basis have $j$ turning points and therefore lack a simple measure of
monotonic dependence. 

Property~(iii) in Definition~\ref{def:naturalbasis} is a symmetry property which says that the
shifted functions $\tilde B_j$ defined on $[-0.5,0.5]$ by $\tilde B_j(x) = \orthobase_j(x+0.5)$ are odd functions when $j$ is odd and even functions when $j$
is even. In Section~\ref{sec:basis-corr-spec}
we show that this is helpful for recognizing
certain symmetry properties of dependence structures.
The Fourier basis does not satisfy property~(iii).

\subsection{Expansions in terms of correlation bases and basis correlations}\label{sec:expansions}

We now explore two reasons why basis correlations play a special role in the study of non-monotonic dependence. Throughout this section, let $\mathcal{B}$ be some given correlation basis.

 The first reason stems from the fact that correlation bases can be used to expand square integrable functions
on $[0,1]^2$. This obviously applies to any square integrable copula density.

\begin{lemma}\label{lem:1}
Let $\mathcal{B}$ be an arbitrary correlation basis. Suppose that $(U,V)$ is a random vector distributed as a copula $C$ with density $c$ such that $\int_0^1 \int_0^1 c(u,v)^2 \rd
u \rd v < \infty$. Then the density $c$ has a 
mean-square
convergent representation
\begin{equation}\label{eq:27}
  c(u,v) = 1 + \sum_{j=1}^\infty
  \sum_{k=1}^\infty \rho_{jk}^B(U,V) \orthobase_j(u) \orthobase_k(v).
\end{equation}
\end{lemma}


The proof of Lemma~\ref{lem:1} in the Supplementary Material uses the representation of
  basis correlations as expectations, which partly influences our requirement that $\orthobase_0(u)
  \equiv 1$ in Definition~\ref{def:correlation-basis}.
In the expansion~\eqref{eq:27}, the basis correlations play the role of the coefficients
of the products of the orthonormal functions $\orthobase_j$
and $\orthobase_k$. In such cases we can interpret
the set of basis correlations
$(\rho_{jk}^B(U,V))_{j,k\in\N}$ as a
representation of the dependence structure implicit in the copula
density $c$. For copulas $C$ with square integrable densities the
expansion~\eqref{eq:27}
implies an expansion for
$C$ itself; see Remark~\ref{rem:Ssupp:expansionC} in the Supplementary Material.

  
The expansion~\eqref{eq:27} with Legendre basis functions has been used in the
  literature to approximate copula densities, for example
  in~\cite{bib:kallenberg-08}. However, such approximations
  may be very inaccurate in cases where $c$ is not square
  integrable. 
Unfortunately, copulas with square integrable densities cannot have
tail dependence~\citep{bib:beare-10} which rules out many important examples.
An alternative is to apply expansions to the log
  density of the copula, as in~\cite{bib:kallenberg-ledwina-99}, but
  this does not lead to readily interpretable coefficients.
 
  \begin{remark}\label{rem:lancaster}
Lemma~\ref{lem:1} can also be viewed as a special case of Theorem~1 in
\cite{lancaster:1958} and the quantity $\int_0^1 \int_0^1 c(u,v)^2 \rd
u \rd v -1$ is the Pearson contingency coefficient $\phi^2$ in the
general result. The expansion~\eqref{eq:27} differs from the diagonal
expansion in \citet[Section~3]{lancaster:1958}, which is used in correspondence analysis and the ACE
algorithm \citep{bib:buja-90}. This would take the form
$
  c(u,v) = 1 + \sum_{j=1}^\infty \rho_{j}^D(U,V) \eta_j(u) \xi_j(v)
$
for two correlation bases $\{\eta_j(u), j \in \N_0\}$ and
$\{\xi_j(u), j \in \N_0\}$ satisfying $\E(
\eta_j(U)\xi_k(V)) =0 $ for $k \neq j$ and $\E(
\eta_j(U)\xi_j(V)) =\rho_{j}^D(U,V)$ for $ j \in \N, $ where the
$\rho_j^D$ are
the canonical
correlations ordered by decreasing size. The canonical
correlation bases $\{\eta_j\}$ and $\{\xi_j\}$ depend explicitly on the copula $C$
and are related to its eigenvalues~\citep{bib:cuadras-15}, but are generally
difficult to derive. In contrast, we use a single correlation basis that is
tractable and chosen independently of $C$.
  \end{remark}

The second reason for our interest in basis correlations is that the
general transformations $\gtrans$ and $\htrans$ used to define the
generalized Spearman correlation  in Definition~\ref{def:nmcor}
can also be expanded in the functions of a correlation basis.  This is helpful because the practical utility of $\rho_{\{\gtrans,\htrans\}}$ largely depends on the particular form of $\gtrans$ and $\htrans$, and a suitable choice may not always be obvious from the context; we illustrate this in Section~\ref{sec:three-applications}.
\begin{lemma}\label{lem:2}
Suppose that $\mathcal{B}$ is a correlation basis, $\gtrans, \htrans \in \mathcal{L}^2([0,1])$ satisfy Assumption~\ref{assA1.new} and $X$ and $Y$ are random variables with continuous distributions $F_X$ and $F_Y$. Then 
\begin{equation}\label{eq:rhoexpansion}
\rho_{\{\gtrans,\htrans\}} (X,Y) = \sum_{j=1}^\infty \sum_{k=1}^\infty \alpha_j(\gtrans) \alpha_k(\htrans)  \rho^B_{jk}(X, Y),
\end{equation}
where for any $j \in \N_0$ and $\psi \in \mathcal{L}^2([0,1])$,
$\alpha_j(\psi)$ is the generalized Fourier coefficient given by $\int_0^1
\psi(u) \orthobase_j(u) \rd u$ and $\sum_{j=1}^\infty
  \alpha_j(\gtrans)^2 = \sum_{j=1}^\infty
  \alpha_j(\htrans)^2 = 1$.
\end{lemma}
The identity~\eqref{eq:rhoexpansion} suggests that, for large $N \in
\N$, we may find approximating functions $g_N(u) =\sum_{j=1}^N
\alpha_{1j} B_j(u)$ and $h_N(u) =\sum_{j=1}^N
  \alpha_{2j} B_j(u)$ in terms of unit weight vectors $\bm{\alpha}_1 =
  (\alpha_{11},\ldots,\alpha_{1N})^\top$ and $\bm{\alpha}_2 =
  (\alpha_{21},\ldots,\alpha_{2N})^\top$ such that
  \begin{equation}\label{eq:3}
  \rho_{\{\gtrans,\htrans\}} (X,Y) \approx
  \rho_{\{\gtrans_N,\htrans_N\}} (X,Y) = \bm{\alpha}_1^\top P^B_N
  \bm{\alpha}_2
  \end{equation}
where $P^B_N = P^B_N(X,Y) = (\rho^B_{jk}(X,Y))_{j,k \in \{1,\ldots,N\}}$
is an $N \times N$ basis correlation matrix.
We can use~\eqref{eq:3} to find functions $\gtrans_N$ and $\htrans_N$
 that maximize the generalized Spearman correlation
 $\rho_{\{g_N,h_N\}}(X,Y)$ and
 thus approximate the maximal generalized Spearman correlation
 $\sup_{g,h \in \mathcal{L}^2([0,1])}\rho_{\{g,h\}}(X,Y)$, an idea taken up in Section~\ref{sec:three-applications}.

\subsection{Properties of basis correlations in natural bases}\label{sec:basis-corr-spec}

We close Section~\ref{sec:2} by exploring some properties of basis
correlations in natural correlation bases for 
dependence structures that are simple or have particular forms
of symmetry.
\begin{proposition}\label{prop:legendre-perfect-dependence}
  Let $\orthosystem$ be a natural correlation basis.
If $X$ and $Y$ are comonotonic random variables then   $ \rho^B_{jk}(X,
Y) = \delta_{jk}$. If $X$ and $Y$ are countermonotonic then  $ \rho^B_{jk}(X,
Y) = (-1)^j \delta_{jk}$.
\end{proposition}
Following~\cite{bib:nelsen-99} and~\citet{bib:dias-han-mcneil-25} we
make the following definitions.
 \begin{definition}\label{def:symmetry}
 A random vector $(X,Y)^\top$ is said to be:
 \begin{enumerate}
     \item[(i)] h-symmetric (v-symmetric), if
     $(X,Y-a)^\top \eqdis (X,a-Y)^\top$ for some $a \in \R$
  (if $(X-b,Y)^\top \eqdis (b-X,Y)^\top$  for some $b\in\R$);
         \item[(ii)] radially symmetric, if $(X-b,Y-a)^\top \eqdis (b-X,a-Y)^\top$  for
           $a,b\in\R$;
 \item[(iii)] jointly symmetric, if $(X,Y)^\top$ is h-symmetric and v-symmetric.
   \end{enumerate}
 \end{definition}
\noindent If $(X,Y)^\top$ is both h-symmetric and v-symmetric then it is also
radially symmetric, but not vice versa. Thus joint symmetry implies
radial symmetry, but not vice versa.
 
\begin{figure}[htb]
  \centering
   \includegraphics[width=0.4\textwidth]{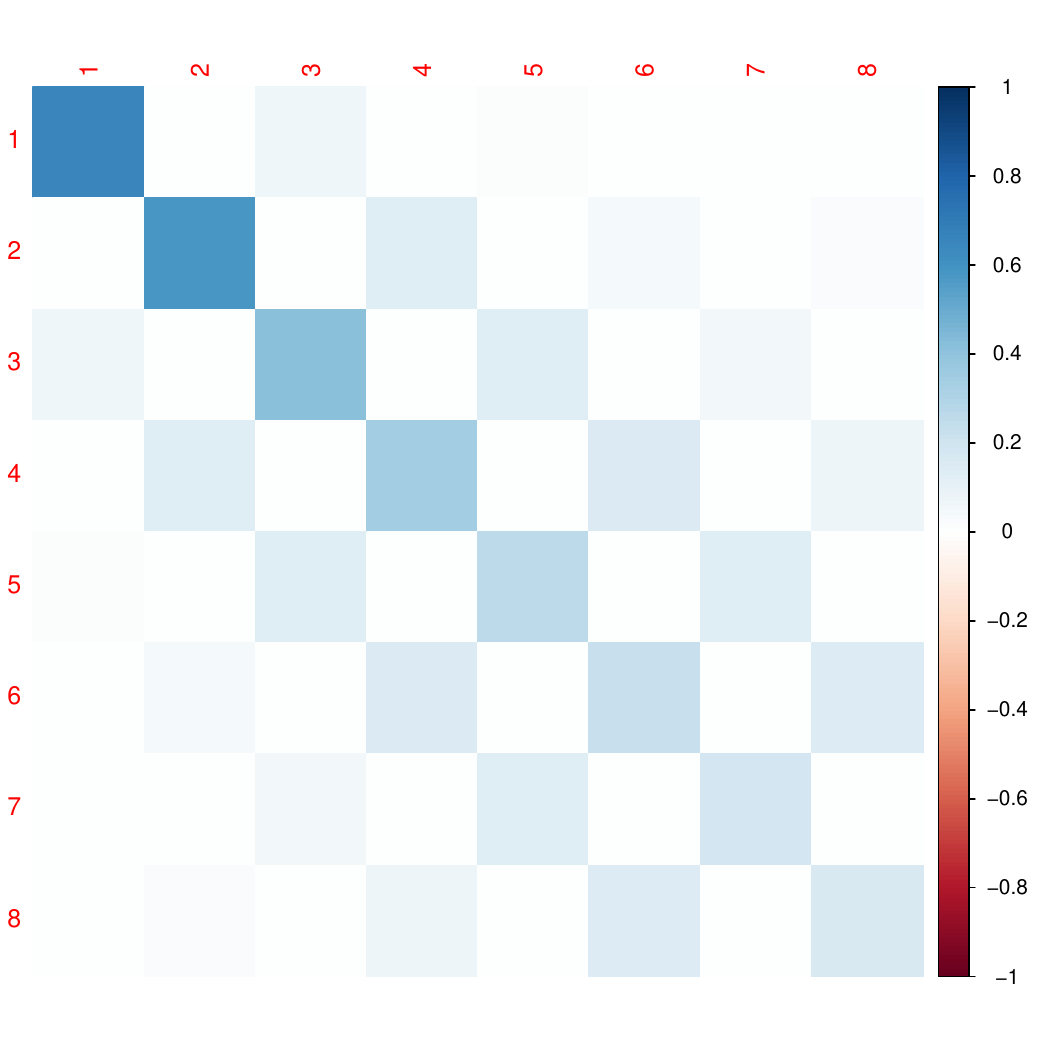}
  \includegraphics[width=0.4\textwidth]{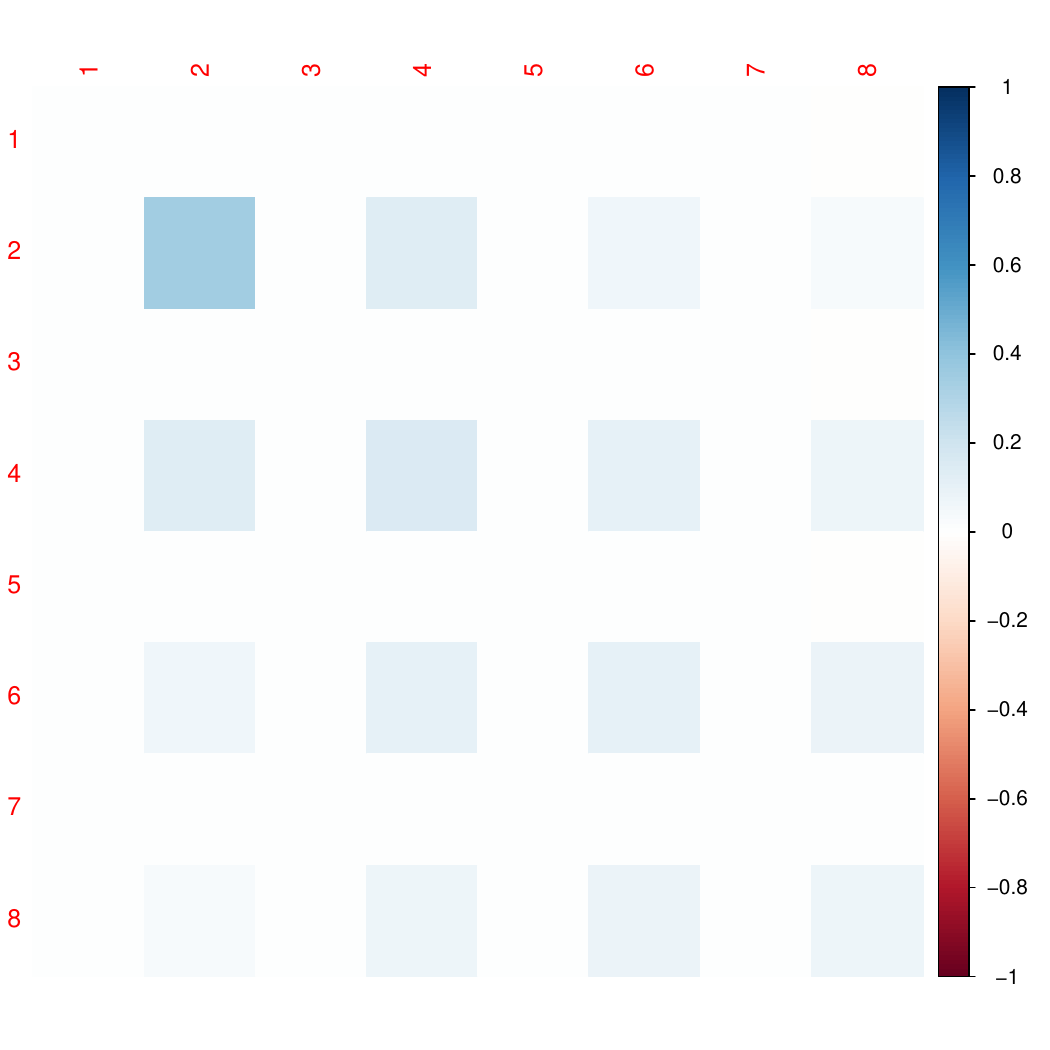}
   \caption{\label{fig:symmetry} Heat maps of the matrix of Legendre
   correlations $P^\blegendre_8$ for the radially symmetric t-copula
   $C^t_{0.7,2}$ (left) and jointly symmetric spherical t-copula
   $C^t_{0,2}$ (right).}
\end{figure}

 If $X$ and $Y$ are continuous then h-symmetry implies that the variables $U = F_X(X)$ and $V = F_Y(Y)$ satisfy
 $(U,V)^\top \eqdis (U,1-V)^\top$, while v-symmetry implies
 $(U,V)^\top \eqdis (1-U,V)^\top$ and radial symmetry implies
 $(U,V)^\top \eqdis (1-U,1-V)^\top$. This leads to the following
 result.
 \begin{proposition}\label{prop:symmetry}
   For continuous random variables $X$ and $Y$ and a natural correlation basis
   $\orthosystem$,
   \begin{enumerate}
   \item[(i)] h-symmetry (v-symmetry) implies $\rho^B_{jk}(X,Y) = 0$
     for $k$ odd ($j$ odd);
    \item[(ii)] radial symmetry implies $\rho^B_{jk}(X,Y) = 0$ for $j+k$
      odd;
    \item[(iii)] joint symmetry implies $\rho^B_{jk}(X,Y) = 0$ for $j$ odd or
      $k$ odd.
      \end{enumerate}
    \end{proposition}
Figure~\ref{fig:symmetry} illustrates this result in the case of the
Legendre polynomial basis using the t-copula
$C^t_{\rho,\nu}$, which is radially symmetric for all values of
$\rho$. Calculations are carried out using~\eqref{eq:13}. Plotting the Legendre correlation matrix $P^\blegendre_8$ when $\rho=0.7$ leads to the chessboard pattern in the left
picture. When $\rho=0$ the copula is jointly symmetric, leading to the
pattern in the right picture where odd rows and columns are all zero;
however, the copula has high cruciformity.
An exchangeable copula leads to a symmetric Legendre
correlation matrix, as is the case for both illustrations in Figure~\ref{fig:symmetry}.

\section{Bounds for generalized Spearman correlation}\label{sec:udp}

   Although $\rho_{\{\gtrans,\htrans\}}(X,Y) \in [-1,1]$, the fact that
  the functions $\gtrans$ and $\htrans$ may not be monotonic 
  means that it is not a priori clear that all values between $-1$ and $1$
  are attainable.
We already know that the extrema do not always
  correspond to simple dependence structures such as counter- or
  comonotonicity; for example,
  Proposition~\ref{prop:legendre-perfect-dependence} shows that the
  minimum value of a basis correlation $\rho^\orthobase_{jj}(X,Y)$ for $j$ even cannot
  correspond to countermonotonicity.

To provide a complete answer to
  the question of bounds in Sections~\ref{sec:corbounds} and
  \ref{sec:support-sets-copulas}, we first study the connection
  between piecewise continuous and strictly monotonic functions and uniform-distribution preserving
  (udp) transformations in
  Section~\ref{sec:unif-distr-pres-1}.

\subsection{Udp  transformations}\label{sec:unif-distr-pres-1}

\begin{definition}\label{def:udp}
A function $\udp : [0,1] \to [0,1]$ is called uniform distribution preserving (udp) if $T(U) \sim \mathcal{U}(0,1)$ for $U \sim \mathcal{U}(0,1)$.
\end{definition}

In the next result we see how udp functions emerge in the study
of generalized Spearman correlations under the key Assumption~\ref{assA2.new} of piecewise
continuity and strict monotonicity of $\gtrans$ and $\htrans$. To ease notation, we denote the distribution function of $\gtrans(U)$ and $\htrans(U)$ with $U \sim \mathcal{U}(0,1)$ by $F_\gtrans$ and $F_\htrans$, respectively. Furthermore, for a distribution function $F$, we introduce the generalized inverse given, for all $u \in (0,1)$, by $F^{-1}(u) = \inf\{ x \in \mathbb{R} : F(x) \ge u\}$. 

\begin{proposition}\label{prop:nonmonotonic-udp}
Under Assumption~\ref{assA2.new} the following holds. 
\begin{enumerate}
\item[(i)] The distribution functions $F_\gtrans$ and $F_\htrans$ are continuous and the functions 
$\udp_\gtrans= \pbase_{\gtrans} \circ \gtrans$ and $\udp_\htrans= \pbase_{\htrans} \circ \htrans$
are udp transformations.
\item[(ii)] If $\gtrans$ and $\htrans$ are continuous, then $F_\gtrans$ and $F_\htrans$  are strictly increasing on the range of $\gtrans$ and $\htrans$, respectively.
\item[(iii)] Suppose that  $X$ and $Y$ are continuous random variables  and set $U = F_X(X)$ and $V = F_Y(Y)$. Then 
$
\rho_{\{\gtrans,\htrans\}}(X,Y) = \rho(\gtrans(U), \htrans(V)) = \rho(F^{-1}_\gtrans (T_\gtrans(U)), F^{-1}_\htrans(T_\htrans(V))).
$
\end{enumerate}
\end{proposition}
Proposition~\ref{prop:nonmonotonic-udp} gives us two ways of looking
at generalized Spearman correlations. On the one hand, they are measures of
non-monotonic dependence defined on the copula of $(X,Y)^\top$,
that is the distribution of $(U,V)^\top = (F_X(X),F_Y(Y))^\top$. On
the other hand, if $\gtrans$ and $\htrans$ are continuous,
$F_{\gtrans}^{-1}$ and
$F_{\htrans}^{-1}$ are strictly increasing and
$\rho_{\{\gtrans,\htrans\}}(X,Y)$ is a measure of
monotonic dependence defined on an alternative copula, which is the
distribution of $(T_\gtrans(U), T_\htrans(V))^\top$. If $\gtrans$ and $\htrans$ satisfy both Assumptions~\ref{assA1.new} and \ref{assA2.new}, we also have that $  \rho_{\{\gtrans,\htrans\}}(X,Y)$ equals
  \begin{equation}\label{eq:53}
  \rho\left(F_{\gtrans}^{-1}\left(T_\gtrans(U)\right),  F_{\htrans}^{-1}\left(T_\htrans(V)\right) \right)
  = \E\left(F_{\gtrans}^{-1}\left(T_\gtrans(U)\right)  F_{\htrans}^{-1}\left(T_\htrans(V)\right) \right).
\end{equation}

If
$\gtrans$ is a continuous and
strictly increasing function then $T_\gtrans(u) = u$, the identity transformation. If $\htrans$ is continuous with
a single turning point at $0 < \delta < 1$, and is
strictly decreasing on $[0,\delta]$, strictly increasing on
$[\delta,1]$ and satisfies $\lim_{u \to 0}\htrans(u) = \lim_{u \to 1}
\htrans(u)$, we can show that
$T_\htrans$ is a v-transform~\citep{bib:mcneil-20} with fulcrum $\delta$, that is a function
of the form
 \begin{equation}
    \label{eq:v-transform}
   \udp_\htrans(u) =
    \begin{cases}
(1-u) - (1-\delta) \Psi\left( \frac{u}{\delta} \right) & u \leq
\delta, \\
u - \delta \Psi^{-1}\left( \frac{1-u}{1-\delta} \right) & u > \delta,
\end{cases}
\end{equation}
where $\Psi$ is a continuous and strictly increasing distribution function
on $[0,1]$ called the generator of $\udp$. If $\htrans$ is symmetric about its
turning point then $T_\htrans(u) = T_\vee(u) = |2u-1|$, a special
v-transform that we call the symmetric v-transform.
\begin{proposition}\label{prop:udp-vtransform}
Let $\htrans$ be a continuous function on $[0,1]$ with $\lim_{u \to 0}\htrans(u) = \lim_{u \to 1}
\htrans(u)$ and assume that $h$ strictly decreasing on $[0,\delta]$ and strictly increasing on $[\delta,1]$ for some $\delta \in (0,1)$.
Write $\htrans(u) = \htrans_\ell(u)$ for $u \in [0,\delta]$ and $\htrans(u) = \htrans_r(u)$ for $u \in [\delta,1]$. Then $T_\htrans = F_{\htrans}
\circ \htrans$ is a v-transform with fulcrum $\delta$ and generator $\Psi(x) =
(1-\delta)^{-1}(1- \htrans_r^{-1} \circ \htrans_\ell(\delta x))$. If
$\htrans$ is symmetric about $\delta = 0.5$,
$\Psi(x) = x$ and $\udp_\htrans = \udp_\vee$.
\end{proposition}
\begin{example}\label{ex:ex8}
For the asymmetric u-shaped transformation in Example~\ref{ex:asymmetrich} it is easy to
check that the generator of the v-transform is $\Psi(x) = 1 -
(1-x)^{q/p}$ and for the special case in Example~\ref{ex:gandh} we have symmetry
about $\delta= 0.5$ yielding the symmetric v-transform.
\end{example}
\begin{remark}
More examples of udp transformations can be found in \cite{quessy:2024} and \cite{hofert/pang:2025}. In the former paper, they are referred to as uniform-to-uniform transformations and we note that the so called bi-monotone mappings defined therein are in fact v-transforms. 
The so-called W-transform with base distribution $F_X$ in Definition~2.1 of \cite{hofert/pang:2025} is udp if $F_X$ is continuous. Under Assumption~\ref{assA2.new}, $T_\gtrans$ and $T_\htrans$ are easily seen to be W-transforms whose base distribution is uniform on $(0,1)$.
\end{remark}

For some of our later results concerning stochastic inverses of udp transformations, we will need the following regularity condition on the udp
transformations. 

\begin{definition}\label{def:udp-regular}
A uniform-distribution preserving transformation is called regular if there exists a finite partition $a_0 = 0 < a_1 < \dots < a_{L-1} < a_L = 1$ for $L \in \{1, 2, \dots\}$ such that $\udp$ is continuously differentiable on $A_\ell =(a_{\ell-1}, a_{\ell})$ for all $\ell \in \{1,\dots, L\}$. 
\end{definition}

If $\gtrans$ and $\htrans$ are continuous, $F_\gtrans$ and $F_\htrans$ are continuous and strictly increasing by Proposition~\ref{prop:nonmonotonic-udp}(i) and (ii), and hence the functions $\udp_\gtrans$ and $\udp_\htrans$ in Proposition~\ref{prop:nonmonotonic-udp} (i) inherit the property of
piecewise, strict monotonicity from $\gtrans$ and $\htrans$.  If we also impose Assumption~\ref{assA3.new}, then the udp transformations are guaranteed to be regular. We summarize these insights as follows.

\begin{proposition}\label{prop:udp-orthosystem-regular}
Under Assumptions~\ref{assA2.new} and~\ref{assA3.new} the udp transformations
$T_\gtrans$ and $T_\htrans$ in Proposition~\ref{prop:nonmonotonic-udp} (i) are regular.
\end{proposition}

We now investigate the udp transformations associated with
natural correlation bases. First, we simplify notation as follows. For a correlation basis $\mathcal{B}$, we denote the distribution function of $\orthobase_j(U)$ for $U \sim \mathcal{U}(0,1)$ by $F^{\orthobase}_j$ and write  $\udp_j^\orthobase
= F^{\orthobase}_j \circ \orthobase_j$ for the associated udp transformation in Proposition~\ref{prop:nonmonotonic-udp} (i).
Deriving the forms of $T_j^\orthobase$ for $j > 2$ requires calculation
of $F^{\orthobase}_j$ in general, but the following observations are straightforward.

\begin{lemma}\label{lemma:tv}
  For a natural correlation
basis 
$\udp_1^\orthobase(u) = u$ and
$\udp_2^\orthobase(u) = \udp_\vee(u)$ for $u \in [0,1]$.
\end{lemma}

\begin{figure}[htb]
    \centering
     \includegraphics[width=\textwidth]{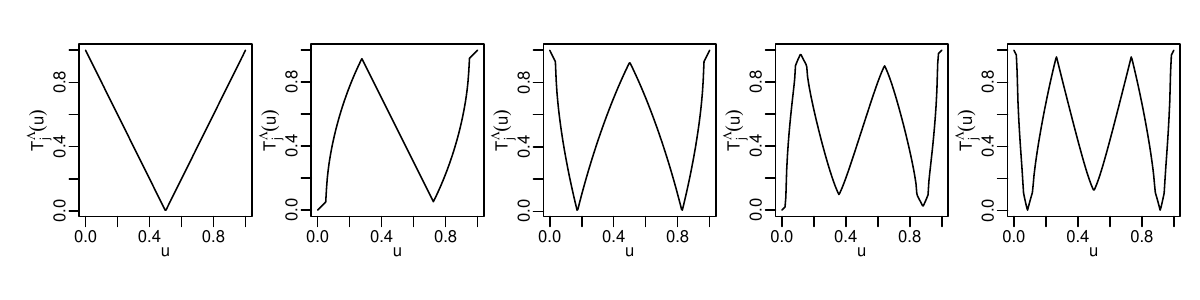}\\\vspace{-1cm}
     \includegraphics[width=\textwidth]{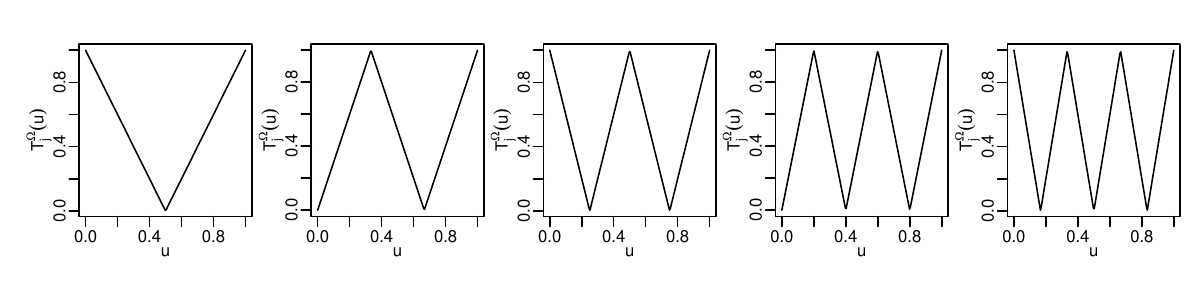}  
   \caption{\label{fig:Tplots}  Plots of $\udp^\orthobase_j$ corresponding to the
     Legendre polynomial basis (top row) and cosine basis (bottom row) for $j=2,\ldots,6$.}
 \end{figure}

\begin{example}[Legendre polynomial basis]\label{ex:legendre-udp}
The Legendre polynomial basis leads to the udp functions
$\udp^{\blegendre}_j$ shown in the
top row of Figure~\ref{fig:Tplots} for $j=2,\ldots,6$. The distribution functions
$\pbase^{\blegendre}_j$ are calculated in
Example~\ref{ex:Slegendre-pbase} in Section~\ref{sec:Spiec-strictly-monot}.
Notice in passing that $\udp^{\blegendre}_j$ is indeed piecewise
monotonic with $j-1$ turning points which are also the turning points
of $\blegendre_j$, and piecewise continuously differentiable, as
proved in Proposition~\ref{prop:udp-orthosystem-regular}. The
partitions on which these piecewise properties hold differ; the latter
is finer. Easy calculations show that $(F_2^\blegendre)^{-1}(u) =
\sqrt{5}(3u^2 -1)/2$ and, together with~\eqref{eq:53}  and Lemma~\ref{lemma:tv}, this yields the appealing
formulas $  \rho_{12}^\blegendre(U,V)
= \rho(U,\udp_\vee(V)^2)$ and $\rho_{22}^\blegendre(U,V) =\rho(\udp_\vee(U)^2,\udp_\vee(V)^2)$.
\end{example}


\begin{example}[Cosine basis]\label{ex:cosine-udp}
 In Example~\ref{ex:Scosine-pbase} it is shown that, for any $j
  \in\N$,  the distribution functions
  $\pbase^\ocos_j$ are all equal to $\pbase^\ocos_1(x) =
  1 - \pi^{-1} \arccos(x/\sqrt{2})$. This leads to the explicit
  formula
  $ \udp^{\ocos}_j(u) = 1- \pi^{-1}\arccos((-1)^j \cos(j\pi u))$ which
  is the piecewise-linear (zigzag) function shown
  in the lower row of Figure~\ref{fig:Tplots}. Moreover, by noting
  that $F^{\ocos}_1(u)=\ocos_1^{-1}(u) $, we obtain the
 useful identity
 $ \ocos_1 \circ \udp^{\ocos}_j = \ocos_1\circ \pbase_j^\ocos\circ
 \ocos_j = \ocos_1\circ \pbase_1^\ocos\circ \ocos_j = \ocos_j$
which implies that the base correlations in the cosine system satisfy $\rho_{jk}^\ocos(U,V)  = \rho(\ocos_j(U),\ocos_k(V)) = 
          \rho_{11}^\ocos(\udp^{\ocos}_j(U), \udp^{\ocos}_k(V))$
        so that they can be calculated from the first order
        measure $\rho_{11}^\ocos$ and the udp functions $ \udp^{\ocos}_j$.
      \end{example}

      \subsection{Bounds for generalized Spearman correlation}\label{sec:corbounds}
      We are now in a position to calculate bounds
      for generalized Spearman correlation.

\begin{theorem}\label{prop:nonmonotonic-bounds}
If $\gtrans$ and $\htrans$ are transformations satisfying
Assumptions~\ref{assA1.new} and~\ref{assA2.new}
and $X$ and $Y$ are continuous random variables, then the upper bound $\overline{\rho}_{\{\gtrans,\htrans\}}$
for $\rho_{\{\gtrans,\htrans\}}(X,Y)$ is attained when the random variables $U =
                    F_X(X)$ and $V = F_Y(Y)$ satisfy
                    $\udp_\gtrans(U) \asequal   \udp_\htrans(V)$ and
the lower bound  $\underline{\rho}_{\{\gtrans,\htrans\}}$ is attained
when $
  \udp_\gtrans(U) \asequal   1 - \udp_\htrans(V)$. The bounds are
  given by
\begin{equation}
 \overline{ \rho}_{\{\gtrans,\htrans\}} = \int_0^1 F_{\gtrans}^{-1}(u)
                      F_{\htrans}^{-1}(u) \rd u\quad\text{and}\quad   \underline{\rho}_{\{\gtrans,\htrans\}}  = \int_0^1 F_{\gtrans}^{-1}(u)
                       F_{\htrans}^{-1}(1-u) \rd u .
                       \label{eq:24a}
                    \end{equation}
Moreover, $ \overline{ \rho}_{\{\gtrans,\htrans\}} =1$ if $F_{\gtrans} =
F_{\htrans}$ and $ \underline{ \rho}_{\{\gtrans,\htrans\}} =-1$ if $F_{\htrans}^{-1}(1-u) = -
F_{\gtrans}^{-1}(u)$ for all $u \in [0,1]$.
\end{theorem}

We apply these ideas to determine bounds for basis correlations in
natural correlation bases, such as the Legendre and cosine bases,
using the notation
$\overline{\rho}^{B}_{jk} = \overline{\rho}_{\{B_j,B_k\}}$ and $ \underline{\rho}^{B}_{jk}=\underline{\rho}_{\{B_j,B_k\}}$. The
following corollary to Proposition~\ref{prop:nonmonotonic-bounds}
follows easily from the observation that in a natural
 correlation basis, for $j$ odd, $F^{\orthobase}_j$ is
symmetric around 0 and 
$(\pbase^\orthobase_j)^{-1}(1-u)= -(\pbase^\orthobase_j)^{-1}(u)$.

\begin{corollary}\label{cor:max-min-relation}
If the functions $\gtrans$ and $\htrans$ in Proposition~\ref{prop:nonmonotonic-bounds} are functions $B_j$ and $B_k$
drawn from a natural correlation basis, then we attain the lower bound
of -1 when $k=j$ and $j$ is odd. Moreover, we always have $ \underline{\rho}^{B}_{jk} =  -\overline{\rho}^{B}_{jk}$ when $j$ is odd or $k$
is odd.
\end{corollary}

\begin{example}[Cosine basis]
 This case is simple since
 $\pbase^{\ocos}_j = \pbase^{\ocos}_1$, as shown in
 Example~\ref{ex:Scosine-pbase}, and $F^{\ocos}_1$ is symmetric around
 0.
 The bounds are $ \overline{\rho}^{\ocos}_{jk} =1$ and $ \underline{\rho}^{\ocos}_{jk} =-1$ for any pair $(j,k)$.
\end{example}

\begin{example}[Legendre polynomial basis]
 We obtain the values
  given in
Table~\ref{tab:maxLminL} by evaluating the integrals~\eqref{eq:24a} after calculating
  $\pbase^{\blegendre}_j$ using a method described in
  Example~\ref{ex:Slegendre-pbase} in the Supplementary Material.
Some exact values can be obtained using the insights of
Example~\ref{ex:legendre-udp}: setting  $U
\sim \mathcal{U}(0,1)$, we obtain $\underline{\rho}_{22}^\blegendre
=\rho(U^2, (1-U)^2) = -7/8$  and
$\overline{\rho}_{12}^\blegendre = \rho(U, U^2) = \sqrt{15/16}$.
\end{example}
\begin{table}[ht]
\centering
{\small
\begin{tabular}{rrrrrrrrrrrr}
  \toprule
  \multicolumn{6}{c}{matrix of maximum values
  $(\overline{\rho}^{\blegendre}_{jk})$} & \multicolumn{6}{c}{matrix of minimum values
  $(\underline{\rho}^{\blegendre}_{jk})$} \\
  \cmidrule(r){1-6} \cmidrule(l){7-12}
1.000 & 0.968 & 0.984 & 0.976 & 0.979 & 0.977 &  -1.000 & -0.968 & -0.984 & -0.976 & -0.979 & -0.977\\ 
 & 1.000 & 0.952 & 0.978 & 0.948 & 0.963 &    & -0.875 & -0.952 & -0.913 & -0.948 & -0.929\\ 
&  & 1.000 & 0.980 & 0.994 & 0.986 &   &  & -1.000 & -0.980 & -0.994 & -0.986\\ 
&  &  & 1.000 & 0.974 & 0.992 &   &  &  & -0.932 & -0.974 & -0.945\\ 
 &  &  &  & 1.000 & 0.984 &   &  &  &  & -1.000 & -0.984\\ 
  &  &  &  &  & 1.000 &    &  &  &  &  & -0.951\\ 
   \bottomrule
\end{tabular}
}
\caption{\label{tab:maxLminL}Upper triangles of symmetric $6\times6$ matrices of
  maximum Legendre correlations
(columns 1--6) and  minimum Legendre correlations
(columns 7--12) to 3 decimal places.}
\end{table}

\subsection{Support sets for copulas attaining
  bounds}\label{sec:support-sets-copulas}

Under Assumptions~\ref{assA1.new} and~\ref{assA2.new}, Theorem~\ref{prop:nonmonotonic-bounds} shows that the sets on which the
copulas attaining $\overline{\rho}_{\{g,h\}}$ and $\underline{\rho}_{\{g,h\}}$ 
are concentrated are, respectively, $ \mathsf{U}_{\{g,h\}} = \{(u,v) : \udp_\htrans(v) =
\udp_\gtrans(u)\}$ and $\mathsf{L}_{\{g,h\}} = \{(u,v) : \udp_\htrans(v) = 1 -
\udp_\gtrans(u)\}$. If $g$ and $h$ are continuous,
Proposition~\ref{prop:nonmonotonic-udp} (ii) lets us rewrite these
sets as $\mathsf{U}_{\{g,h\}} = \{(u,v) : \htrans(v) - \pbase_{\htrans}^{-1} \circ \pbase_{\gtrans} \left( \gtrans(u)
  \right) = 0\}$ and $\mathsf{L}_{\{g,h\}} =  \{(u,v) : \htrans(v) -
  \pbase_{\htrans}^{-1} \left( 1 - \pbase_{\gtrans} \left(\gtrans(u)
    \right)\right) = 0\}$. The sets can thus be determined by fixing
  values of $u$ and applying root finding techniques to 
  equations of the form $h(v) - c(u) =0$ for constants $c(u)$.
  
For $\gtrans =
  \htrans$ we have the further simplification $\mathsf{U}_{\{g,g\}} = \{(u,v) :\gtrans(v) - \gtrans(u) =
  0\}$, which can be determined without evaluating
  $\pbase_\gtrans$. This allows us to easily calculate the sets
  corresponding to maximal basis correlations
  $\overline{\rho}_{jj}^\orthobase$ for $j \in \N$.
 When $\gtrans$
  has the symmetry property $\gtrans(1-u) = -\gtrans(u)$ for all $u
  \in [0,1]$, we obtain  $\mathsf{L}_{\{g,h\}} =  \{(u,v) : \htrans(v) -
  \pbase_{\htrans}^{-1} \circ \pbase_{\gtrans} \left( \gtrans(1-u)
  \right) = 0\}$ and, in the case $\gtrans =
  \htrans$, this
  simplifies to $\mathsf{L}_{\{g,g\}} =  \{(u,v) : \gtrans(v) + \gtrans(u) =
  0\}$. This allows us to easily calculate the sets for 
  minimal basis correlations
  $\underline{\rho}_{jj}^\orthobase$ in natural
  correlation bases  for $j$ odd.

\begin{figure}[htb]
      \centering
       \includegraphics[width=0.4\textwidth]{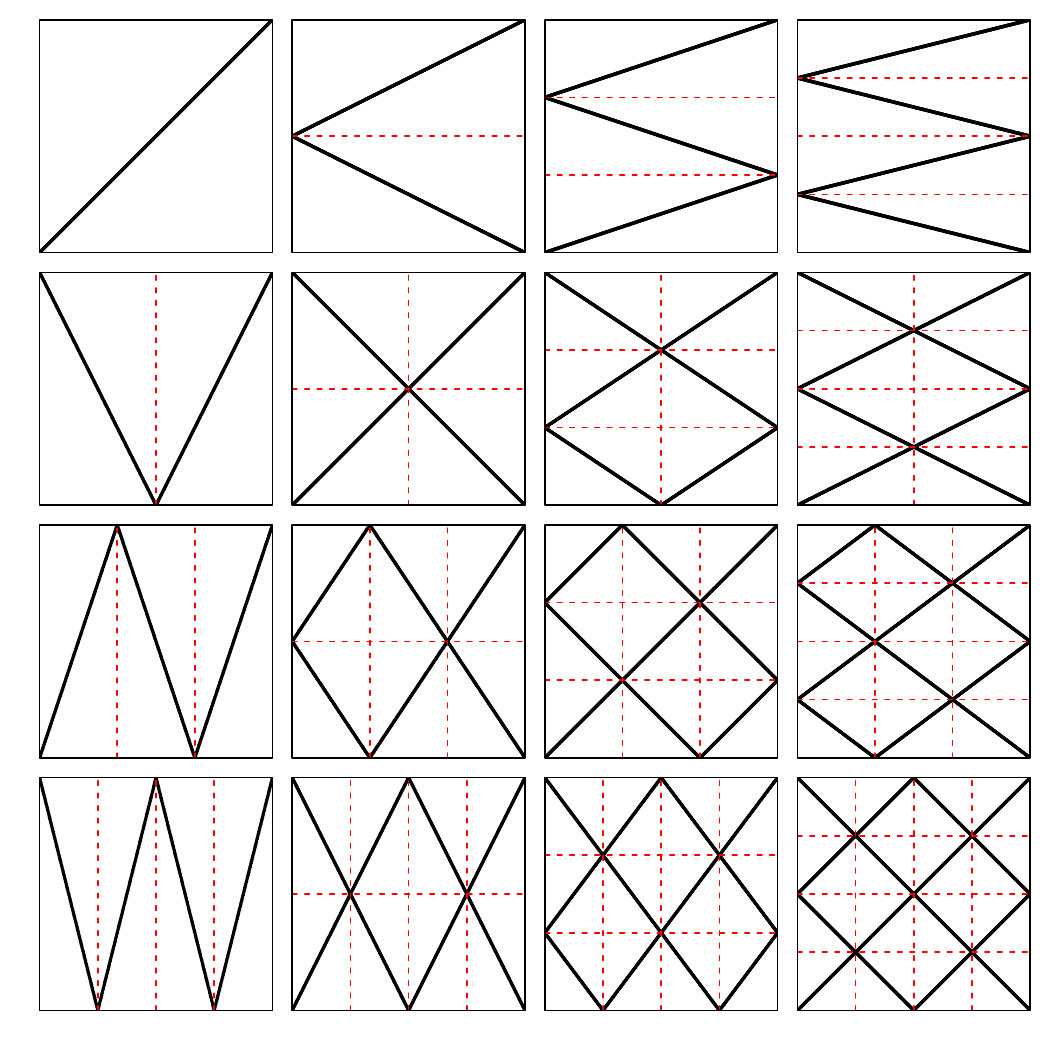}
  \includegraphics[width=0.4\textwidth]{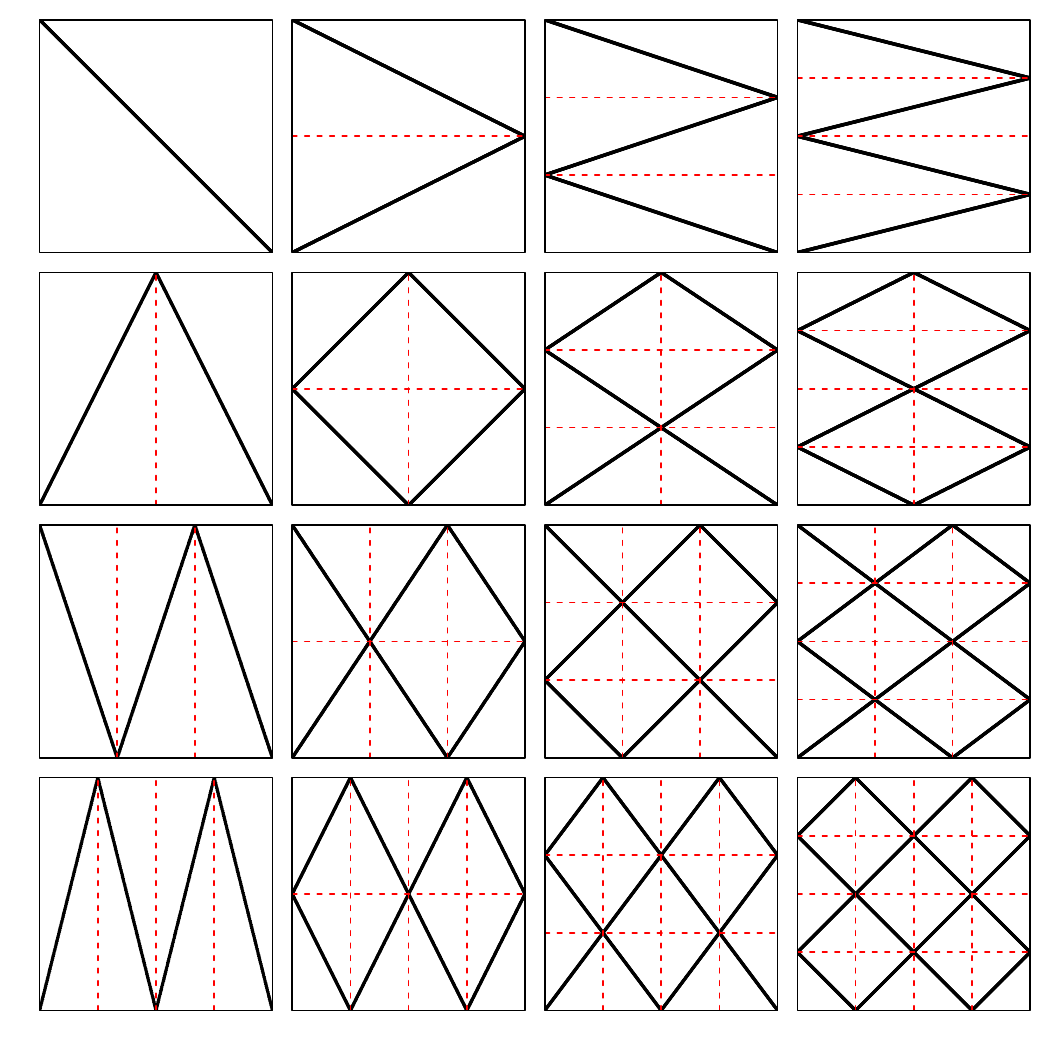}
   \caption{\label{fig:cosine-maxmin} Matrices of plots showing
     $\mathsf{U}^\ocos_{jk}$ (left) and  $\mathsf{L}^\ocos_{jk}$
     (right) for $j=1,\ldots,4$ and $k=1,\ldots,4$. Dashed red grid lines show turning points of
   $\ocos_j$ (vertical) and $\ocos_k$ (horizontal).}
\end{figure}

\begin{example}[Cosine basis]
Since $\pbase^{\ocos}_j(u)$ =
$\pbase^{\ocos}_1(u)$ for any $j \in \N$ and $1 -
  \pbase^{\ocos}_1(u) = \pbase^{\ocos}_1(-u)$, the sets may be written as
  $\mathsf{U}^\ocos_{jk} := \{(u,v) : \ocos_k(v) -\ocos_j(u) = 0\}$
  and $\mathsf{L}^\ocos_{jk} := \{(u,v) : \ocos_k(v) + \ocos_j(u) = 0\}$.
  For $j=1,\ldots,4$ and $k = 1,\ldots,4$ we obtain the matrices of
  illustrations in the left and right panels of
  Figure~\ref{fig:cosine-maxmin} consisting of straight line segments.
\end{example}

\begin{figure}[htb]
     \centering
      \includegraphics[width=0.4\textwidth]{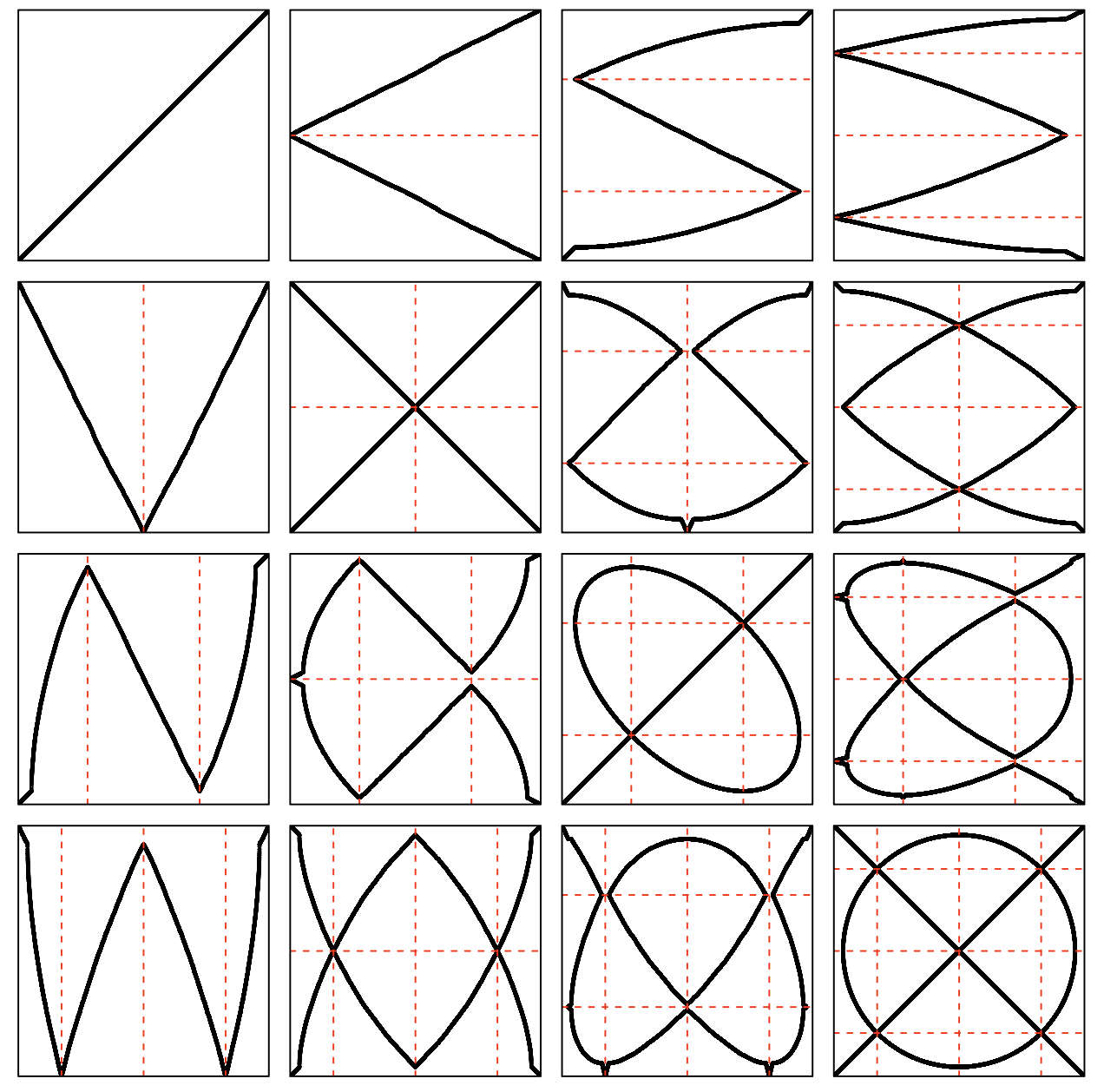} 
  \includegraphics[width=0.4\textwidth]{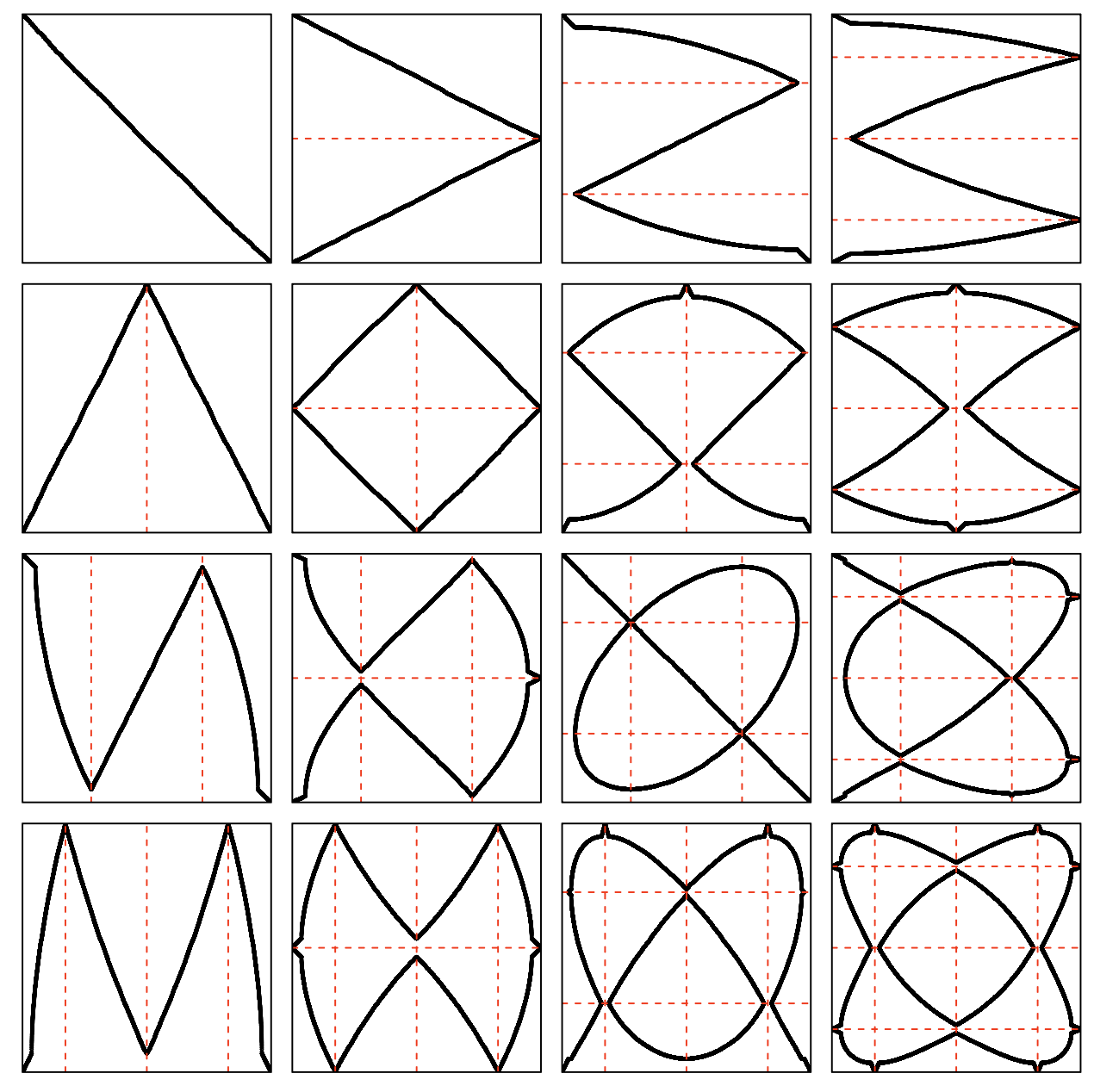}
   \caption{\label{fig:allmaximaminima} Matrices of plots showing
     $\mathsf{U}^\blegendre_{jk}$ (left) and
     $\mathsf{L}^\blegendre_{jk}$ (right) for $j=1,\ldots,4$ and
     $k=1,\ldots,4$. Dashed red grid lines show turning points of
   $\blegendre_j$ (vertical) and $\blegendre_k$ (horizontal).}
 \end{figure}

\begin{example}[Legendre polynomial basis]
We use polynomial root finding to
determine $\mathsf{U}^\blegendre_{jk}$ and $\mathsf{L}^\blegendre_{jk}$
 and we obtain the matrices of
  illustrations in the left and right panels of
  Figure~\ref{fig:allmaximaminima}.
  For $j > 2$ or $k > 2$, the sets contain curved segments

  It is
  interesting to note that
  $\mathsf{U}^\blegendre_{33}$ and $\mathsf{U}^\blegendre_{44}$
  contain elliptical and circular parts, respectively. To see this, observe that
   $\mathsf{U}^\blegendre_{33} =\{(u,v) :\blegendre_3(v) -
  \blegendre_3(u) = 0\} = \{(u,v): (v-u)(20v^2 + 20uv
    + 20 u^2 -30v -30u +12) =0\}$.
The first term corresponds to the leading diagonal while the
equation $20v^2 + 20uv
    + 20 u^2 -30v -30u +12 =0$ may be shown to be that of an ellipse centred at $(0.5,0.5)$ and
    rotated through $\pi/4$ degrees  with
    semi-major axis $\sqrt{3/10}$ and semi-minor axis
    $\sqrt{1/10}$. 
    Similarly,
$\mathsf{U}^\blegendre_{44} =\{(u,v) :
(v-u)(v+u-1)(70v^2 -70v
    + 70 u^2 -70u +20) =0\}
$.
  The first two terms correspond to the two diagonals of the unit
  square while the equation $70v^2 -70v
    + 70 u^2 -70u +20 = 0$ may be rewritten as $(v - 0.5)^2 + (u
    - 0.5)^2 = 3/14$,  which is the equation of a circle centred at
    $(0.5,0.5)$ with radius $\sqrt{3/14}$.
\end{example}



 The following observations, prompted by
 Figures~\ref{fig:cosine-maxmin} and~\ref{fig:allmaximaminima}, 
 apply to the support sets $\mathsf{U}^B_{jk}$ and
$\mathsf{L}^B_{jk}$ corresponding to any natural
basis. First note that $\mathsf{U}^B_{j1}= \{(u,v) : v = T_j^B(u)\}$ is simply an image of the
function $T_j^B$. Note also that, for any pair $(j,k)$ with $j,k \in \{1,2\}$, every natural
basis yields the
same sets $\mathsf{U}^B_{jk}$and
$\mathsf{L}^B_{jk}$ as a consequence of Lemma~\ref{lemma:tv}. The
forms of these sets
underscore why $\rho_{12}^\orthobase$ and
$\rho_{21}^\orthobase$ are measures of angularity and
$\rho_{22}^\orthobase$ is a measure of cruciformity.


The line $v=u$ is contained in
$\mathsf{U}^\orthobase_{jj}$ for all $j\in\N$ (see the diagonals of the left panels of both figures). 
This is expected since comonotonic uniform variables satisfying $V
\asequal U$ also satisfy $T^\orthobase_j(U) \asequal
T^\orthobase_j(V)$  for all $j\in\N$. The line
$v = 1 - u$ is in $\mathsf{U}^\orthobase_{jj}$ for $j$ even and in
$\mathsf{L}^\orthobase_{jj}$ for $j$
odd. This is also expected since countermonotonic uniform variables satisfying $V
\asequal 1- U$ also satisfy $T^\orthobase_j(U) \asequal
T^\orthobase_j(V)$ when $j$ is even and  $T^\orthobase_j(U) \asequal
1 - T^\orthobase_j(V)$ when $j$ is odd. 

$\mathsf{U}^B_{kj}$ is obtained by reflection of
$\mathsf{U}^B_{jk}$ in the line $v=u$; the same
statement holds for $\mathsf{L}^B_{kj}$ and $\mathsf{L}^B_{jk}$. This is due to the fact that if
 $(U,V)^\top$ maximizes (minimizes) $\rho_{jk}^B$
 then $(V,U)^\top$ maximizes (minimizes) $\rho_{kj}^B$.
For $j$ odd, $\mathsf{U}^B_{jk}$ is obtained by reflection of
$\mathsf{L}^B_{jk}$ in the line $u=0.5$.
 This follows from the observation that
  $\rho_{jk}^\orthobase (1-U,V) = -\rho^\orthobase_{jk}(U,V)$ and 
Corollary~\ref{cor:max-min-relation}. Similarly, for $k$ odd, $\mathsf{U}^B_{jk}$ is obtained by reflection of
$\mathsf{L}^B_{jk}$ in the line $v=0.5$.

A final observation, which is proved formally in
Section~\ref{sec:Ssupplement-support-sets} of the Supplementary material,
is that, both $\mathsf{U}^B_{jk}$ and
$\mathsf{L}^B_{jk}$ consist of segments that are monotonic functions
within the boxes of a grid formed by horizontal and vertical grid lines at the turning points of
$\orthobase_j$ and $\orthobase_k$. The only points in $(0,1)^2$
where lines and curves cross lie on this grid.

\section{Stochastic inversion 
 and copula construction}\label{sec:bounds}

The task of constructing
a copula to maximize or minimize generalized Spearman correlation may not have a unique solution. Consider, for example, the problem
of constructing a copula to attain the upper bound
$\overline{\rho}_{44}^\blegendre$. The discussion of Section~\ref{sec:support-sets-copulas} explains why both the
comonotonicity copula $M(u,v) = \min(u,v)$ and the countermonotonicity
$W(u,v) = \max(u+v-1,0)$ are possible solutions. Any copula that is
concentrated on any subset of $\mathsf{U}^\blegendre_{44} $ provides a solution and we
explain the general construction in Section~\ref{sec:mult-stoch-invers}.
The key concept for constructing
bound-attaining copulas is the stochastic
inverse of a regular udp
transformation described in
Section~\ref{sec:stoch-invers-udp}. However, the applicability of the
stochastic inversion technique is not
restricted to the solution of the bound problem and in
Section~\ref{sec:copul-dens-under} we show that it can be used to
create a rich variety of new bivariate copulas with differing degrees of
non-monotonic dependence. 

\subsection{Stochastic inverses of udp transformations}\label{sec:stoch-invers-udp}
Since a udp transformation is not necessarily injective,
the notion of an inverse that we propose utilizes a stochastic component, i.e.~a randomizer.

  \begin{definition}\label{def:si}
Let $\udp$ be a regular udp transformation, $Z \sim \mathcal{U}(0,1)$, $A$ be the
union of open sets on which $\udp$ is continuously differentiable (see
Definition~\ref{def:udp-regular}), and $0=b_{0} < \dots  < b_{K}=1$ be the so-called $T$-partition of $[0,1]$ consisting of  ordered distinct elements of the set
  $\{T_{+}(a_{\ell-1}), T_{-}(a_\ell), \ell = 1,\dots, L\}$ where
  $T_{+}$ and $T_{-}$ denote the one-sided limits of $T$ and $K \in
\{1,2,\dots\}$. Let also $B = \cup_{k=1}^K(b_{k-1},b_{k})$ and for any $x \in B$, set $\udp^{-1}(\{x\}) = \{u : \udp(u) = x, u \in A\} = \{r_1(x),
\ldots, r_{n(x)}(x)\}$ be the pre-image of the point $x$ intersected with $A$. 
For $x\in B$, assign the inverse value 
$
\udp^\leftarrow(x,Z) = G_x^{-1}(Z)
$,
where $G_x^{-1}$ is the generalized inverse of the distribution function $G_x$ of the discrete random variable with support $\{r_1(x),
\ldots, r_{n(x)}(x)\}$ and probabilities $1/|\udp^\prime(r_i(x))|$, $i
\in \{1,\dots, n(x)\}$.  Otherwise, for $x \not \in B$, set
$\udp^\leftarrow(x,Z)$ to be an arbitrary value from $[0,1]$, say $0$ without loss of generality.
\end{definition}
Propositions~\ref{prop:ST-partition} and \ref{prop:Supd-properties} in
  Section~\ref{sec:Sudp-theory} of the Supplementary Material guarantees that
Definition~\ref{def:si} is meaningful. The essential idea is that, on
all but a null set of points, we construct the inverse by finding the
roots $\{r_1(x),
\ldots, r_{n(x)}(x)\}$ of the equation $\udp(u) = x$ and selecting a root by multinomial sampling according to a set of probabilities 
$\{|\udp^\prime(r_1(x))|^{-1},\ldots,
|\udp^\prime(r_{n(x)}(x))|^{-1}\}$, which sum to one by Proposition~\ref{prop:Supd-properties} (iv).
Under this construction we have for almost all $x$ that $ \udp ( \udp^\leftarrow(x,Z)) = x$,
almost surely. However, $\udp^\leftarrow(T(u),Z)$ is not necessarily equal to $u$
a.s., only with a certain probability. The main result of this section
shows that $\udp^\leftarrow$ is uniformity preserving
and generalizes a result for the stochastic inverse of
  v-transforms in~\citet{bib:mcneil-20}.

\begin{theorem}\label{theorem:stochinverse}
If $\udp$ is a regular udp transformation and $V\sim \mathcal{U}(0,1)$, $Z \sim \mathcal{U}(0,1)$ are independent,
 $\udp^\leftarrow(V,Z) \sim \mathcal{U}(0,1)$.
\end{theorem}

For applying stochastic inversion to the udp functions
  $\udp_\gtrans$ associated with transformations satisfying
  Assumptions~\ref{assA2.new} and~\ref{assA3.new}, we need to
  be able to calculate derivatives $\udp_\gtrans^\prime$. This is
  possible for transformations from both the cosine and Legendre correlation bases.
\begin{example}[Cosine basis]\label{ex:16}
Example~\ref{ex:cosine-udp} implies that, where $\udp^{\ocos}_j$ is differentiable, $
   (\udp_j^{\ocos})^\prime(u) = (-1)^{j+1} j \sin(j \pi u)/|\sin(j \pi u)|$
 so that $|(\udp^{\ocos}_j)^\prime(u)| = j$. We
  construct the stochastic inverse $(\udp^{\ocos}_j)^\leftarrow(x, Z)$ by choosing
  each of the $j$ values satisfying $\udp^{\ocos}_j(u) = x$ with 
  probability $1/j$.
\end{example}
\begin{example}[Legendre polynomial basis]
This case is more complicated since,
for $j > 2$, 
the udp functions $\udp^{\blegendre}_j$ are not piecewise linear and the
probabilities attached to the members of the sets of inverse values
$(\udp^{\blegendre}_j)^{-1}(\{x\})$ are different. At points where $\udp^{\blegendre}_j$ is differentiable we
can calculate the derivative $ (\udp^{\blegendre}_j)^\prime(u) =
\dbase^\blegendre_j(\blegendre_j(u))\blegendre_j^\prime(u)$ using the chain rule
and calculations for the density $\dbase^\blegendre_j$ of $\pbase^\blegendre_j$ discussed in
Example~\ref{ex:Slegendre-pbase} in the Supplementary Material.
\end{example}

\subsection{Multivariate stochastic inversion}\label{sec:mult-stoch-invers}
A corollary to Theorem~\ref{theorem:stochinverse} explains what happens when possibly dependent uniform random variables $V_1,\dots, V_d$
are transformed simultaneously by stochastic inverses of regular udp
transformations. While each such transformation is uniformity
preserving, additional dependence can be introduced through the
randomizers $Z_1,\dots, Z_d$. 

\begin{corollary}\label{cor:stochinversemult}
Suppose that $V_1,\dots, V_d, Z_1,\dots, Z_d$ are possibly dependent random variables, each uniformly distributed on $[0,1]$. Further assume that $\udp_{1},\dots, \udp_{d}$ are regular udp transformations. If for each $j \in \{1,\dots, d\}$, $V_j$ and $Z_j$ are independent,   $\udp_{1}^\leftarrow(V_1,Z_1), \dots, \udp_{d}^\leftarrow(V_d,Z_d)$ are possibly dependent random variables, each uniformly distributed on $[0,1]$ and the random vector $(\udp_{1}(\udp_{1}^\leftarrow(V_1,Z_1)), \dots, \udp_{d}(\udp_{d}^\leftarrow(V_d,Z_d)))$ is equal to $(V_1,\dots, V_d)$ almost surely.
\end{corollary}

\noindent This result implies that the distribution
  of $(\udp_{1}^\leftarrow(V_1,Z_1), \dots,
  \udp_{d}^\leftarrow(V_d,Z_d))^\top$ is not fully determined by the
  distribution of $(V_1,\ldots,V_d)^\top$; the joint conditional distribution of the
  randomizers $Z_1,\ldots,Z_d$ given $(V_1,\ldots,V_d)^\top$ also plays an important role. 
The case where the randomizers $Z_1,\ldots,Z_d$ are mutually
  independent uniform variables that are also independent of
  $(V_1,\ldots,V_d)^\top$ will be referred to as independent
  multivariate stochastic inversion. However, there are many other
  possibilities and we develop intuition in the next example.
  
\begin{example}\label{ex:mult-stoch-invers-Gauss}
  We consider the case where $d=2$, the random vector
  $(V_1,V_2)^\top$ is distributed according to the Gaussian copula $C^{\text{Ga}}_\rho$
  with correlation parameter $\rho = 0.85$ and where $\udp_{1} =
  \udp_{2} = \udp_\vee$, the symmetric v-transform. We generate points
  from three different distributions for $(\udp_\vee^\leftarrow(V_1,
  Z_1),\udp_\vee^\leftarrow(V_2, Z_2))^\top$. The first panel is the
  case of independent stochastic inversion. For the second and third
  panels, let $Z_1$ be independent of $(V_1,V_2)^\top$. In the second panel we set
  $Z_2 = Z_1$ so that the randomizers are comonotonic. In the
  third panel we set $Z_2=Z_1$ if $\max(V_1,V_2) > 0.6$ and $Z_2 =
  1-Z_1$ otherwise; this gives a model where both $Z_1$ and $Z_2$ are
  independent of $(V_1,V_2)^\top$ but their copula is not.
 \begin{figure}[htb]
    \centering
     \includegraphics[width=15cm,height=5cm]{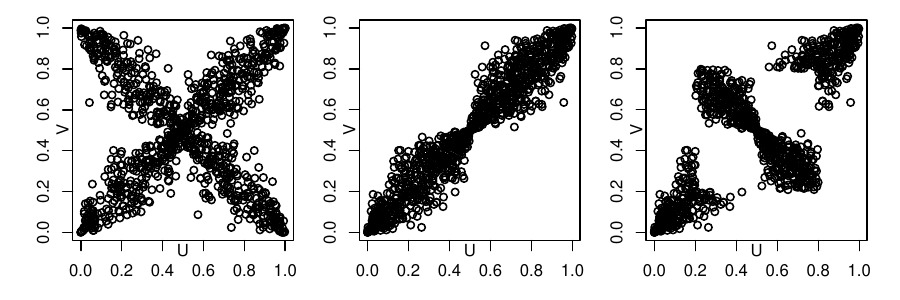}
   \caption{\label{fig:udp_biv} Three different distributions for $(\udp_\vee^\leftarrow(V_1, Z_1),\udp_\vee^\leftarrow(V_2, Z_2))^\top$ when $(V_1,V_2)^\top$ is distributed
  according to a Gaussian copula. See Example~\ref{ex:mult-stoch-invers-Gauss} for details.
   }
 \end{figure}
\end{example}

Corollary~\ref{cor:stochinversemult} can be applied to the comonotonicity and
countermonotonicity copulas to construct copulas that maximize and
minimize generalized Spearman correlations.

\begin{corollary}\label{sec:mult-stoch-invers-2}
  Let $\gtrans$ and $\htrans$ be transformations
  satisfying Assumptions~\ref{assA2.new} and~\ref{assA3.new} and let $U^*$, $Z_1$ and $Z_2$ be uniform variables such that $Z_1
  \perp U^*$ and $Z_2 \perp U^*$. Let $\udp_\gtrans = F_\gtrans \circ
  \gtrans$ and $\udp_\htrans = F_\htrans \circ
  \htrans$. Set $U = \udp_\gtrans^\leftarrow(U^*, Z_1)$, $V =
  \udp_\htrans^\leftarrow(U^*, Z_2)$ and $\tilde V =
  \udp_\htrans^\leftarrow(1-U^*, Z_2)$. Then the distributions of $(U,V)^\top$ and $(U,\tilde V)^\top$ attain
  $\overline{\rho}_{\{\gtrans,\htrans\}}$ 
  and
 $\underline{\rho}_{\{\gtrans,\htrans\}}$ respectively.
 \end{corollary}
 We can construct different
  solutions to the problem of attaining the bounds, by applying Corollary~\ref{sec:mult-stoch-invers-2} and choosing
  different distributions for $(U^*,Z_1,Z_2)^\top$ such that
 $Z_1
 \perp U^*$ and $Z_2 \perp U^*$. In Section~\ref{sec:Srho44} of the Supplementary Material we
 construct two new copulas which maximize
  $\rho_{44}^\blegendre$ but do not correspond to the independent stochastic inversion case. Both copulas correspond to a situation where the copula of $Z_1$
and $Z_2$ depends on $U^*$.

We focus on the case of independent bivariate stochastic
  inversion which always gives rise to copulas distributed over the entire
  support sets $\mathsf{U}_{\{g,h\}}$ and
    $\mathsf{L}_{\{g,h\}}$. Moreover, if $\gtrans$ or $\htrans$ satisfy Assumptions~\ref{assA2.new} and~\ref{assA3.new} and are
      symmetric so that $g(1-u)=g(u)$ and $h(1-u) = h(u)$, the following result implies the copulas
      obtained conform to symmetry concepts
  introduced in Definition~\ref{def:symmetry}.

 \begin{proposition}\label{prop:mult-stoch-invers-symmetry}
Let $U^*$ and $V^*$ be standard uniform with
arbitrary dependence structure
and let $\udp_{1}$ and $\udp_{2}$ be two regular udp functions. If
the uniform variables $U =\udp_{1}^\leftarrow(U^*, Z_1)$ and $V =
\udp_{2}^\leftarrow(V^*, Z_2)$
are constructed by independent stochastic inversion 
then the
distribution of 
$(U,V)^\top$ is (i) v-symmetric if $\udp_{1}$ is symmetric,
(ii) h-symmetric if $\udp_{2}$ is symmetric and (iii) jointly symmetric if
both $\udp_{1}$ and $\udp_{2}$ are symmetric.
\end{proposition}

For symmetric $\gtrans$ and $\htrans$
satisfying Assumptions~\ref{assA2.new} and~\ref{assA3.new}, the udp transformations $T_1 = F_\gtrans \circ \gtrans$ and $T_2
=F_\htrans \circ \htrans$ are regular and symmetric. Hence
the
distribution of $(U,V)^\top$ resulting from independent bivariate stochastic
inversion is jointly symmetric; if only one of $\gtrans$ or
$\htrans$ is symmetric, we have v-symmetry or h-symmetry,
respectively.

For example, suppose that $\gtrans =B_j$ and $\htrans = B_k$ for functions $B_j$ and $B_k$
from a natural correlation basis. If $j$ is even ($k$ is even), the distribution of
$(U,V)^\top$  is v-symmetric (h-symmetric); if both are even we obtain joint
symmetry. These symmetries are also reflected in the symmetries of the
support sets in Figures~\ref{fig:cosine-maxmin} and~\ref{fig:allmaximaminima}.
Moreover, Proposition~\ref{prop:symmetry} implies that if either $j$
or $k$ is even we must have $\rho_{11}^B(U,V) = 0$ so that
the measure of first-order monotonic dependence is zero for the copula
obtained by independent stochastic inversion.

\subsection{Using stochastic inversion to construct new copula
  families}\label{sec:copul-dens-under}

Stochastic inversion is extremely useful for
  building copula models to describe phenomena exhibiting
  non-monotonic dependencies. An example is found in~\citet{bib:dias-han-mcneil-25} where 
  copulas constructed using bivariate stochastic inversion of v-transforms are
  used to describe the serial dependencies in financial time series,
  which are characterized by high cruciformity.

For the purposes of estimating such copulas, it is
 desirable to have closed form expressions for the densities.
In the case of multivariate stochastic inversion of each component we
can get an explicit representation of the density $c$ of  $(\udp_{1}^\leftarrow(V_1,Z_1), \dots,  \udp_{d}^\leftarrow(V_d,Z_d))^\top$, as shown in \cite{bib:mcneil-neslehova-26}. When $(V_1,\dots, V_d)$ are distributed according to a copula $C^*$ with density $c^*$, the randomizers $Z_1,\dots, Z_d$ are i.i.d.\ uniform variables independent of $V_1,\dots, V_d)$, and the udp transformations are regular, we obtain the expression $c(u_1,\ldots,u_d) = c^*(\udp_{1}(u_1), \ldots, \udp_{d}(u_d))$ for almost all $u_1,\dots, u_d \in (0,1)$. 
This particularly simple relationship generalizes
  a result for v-transforms in~\citet{bib:mcneil-20}\footnote{Although the copula density takes a simple
form, the expression for the copula is generally less tractable. The cosine basis is relatively
convenient, due to the
piecewise linearity of the implied udp transformations.}.

A natural approach to creating useful new parametric copula families, and the
 one used in~\citet{bib:dias-han-mcneil-25}, is to build models that can
 interpolate between the maximum and minimum generalized Spearman
 correlations, or between the maximum value and zero if it is a priori
 clear that the sign of the generalized Spearman correlation measure should be
 positive.

Suppose that
  $(U^*,V^*)^\top$ is distributed according to a copula $C^*_\theta$
  in a parametric
 family
  $\{C^*_\theta : \theta \in [\theta_L, \theta_U]\}$ such that
  $\lim_{\theta \to \theta_U}C^*_\theta(u,v) = M(u,v)$
  (comonotonicity) and $\lim_{\theta
    \to \theta_L}C^*_\theta(u,v) = W(u,v)$ (countermonotonicity) or $\lim_{\theta
    \to \theta_L}C^*_\theta(u,v) = uv$ (independence). We will refer to this
  family as the base copula family. Suitable choices might be the Gaussian or
  Frank copula families for interpolating between comonotonicity and
  countermonotonicity, and the Clayton, Gumbel and Joe families for
  interpolating between comonotonicity and independence.

Construction of a
  bivariate stochastic inverse $(U,V)^\top =
  (\udp_\gtrans^\leftarrow(U^*,Z_1), \udp_\htrans^\leftarrow(V^*,Z_2))^\top$
  leads to the creation of a copula $C_\theta$ which may be viewed
  as a member of a parametric family  $\{C_\theta : \theta
  \in [\theta_L, \theta_U]\}$ which we refer to as the
  interpolating family for $\rho_{\{\gtrans,\htrans\}}$. Writing $\rho_{\{\gtrans,\htrans\}}(C_\theta)$
  for the basis correlation of a random vector
  distributed according to $C_\theta$, the interpolating family has the property
  that $\lim_{\theta \to \theta_U}\rho_{\{\gtrans,\htrans\}}(C_\theta) =
  \overline{\rho}_{\{\gtrans,\htrans\}}$ and $\lim_{\theta \to
    \theta_L}\rho_{\{\gtrans,\htrans\}}(C_\theta) =
 \rho_L$ where
 $ \rho_L = \underline{\rho}_{\{\gtrans,\htrans\}}$ if
 $C^*_{\theta_L}(u,v) =W(u,v)$ or $\rho_L =
 0$ if $C^*_{\theta_L}(u,v) = uv$.
 By the multivariate density result in~\citet{bib:mcneil-neslehova-26} we
  know that, when $C^*_\theta$ has a density $c^*_\theta$ and we carry
  out independent stochastic inversion, the density $c_\theta$ of
  $C_\theta$ is given by $c_\theta(u,v) =
  c_\theta^*(\udp_\gtrans(u),\udp_\htrans(v))$.
In the final example we create an interpolating
  family for cosine basis correlations, which turn out to have a particularly straightforward
  interpretation. 
  
  \begin{example}
Let $(U^*, V^*)^\top \sim C^*_\theta$ for a parametric
base family  $\{C^*_\theta : \theta \in [\theta_L, \theta_U]\}$. Let $(U,V)^\top =
  ((\udp_j^\ocos)^{\leftarrow}(U^*,Z_1),
  (\udp_k^\ocos)^{\leftarrow}(V^*,Z_2))^\top \sim C_\theta$ and recall that
  $(\udp_j^\ocos(U),\udp_k^\ocos(V))^\top = (U^*,V^*)^\top \sim C^*_\theta$. From
  Example~\ref{ex:cosine-udp} we have
$
 \rho_{jk}^\ocos(C_\theta) =   \rho_{jk}^\ocos(U,V) =
    \rho_{11}^\ocos(\udp_j^\ocos(U),\udp_k^\ocos(V)) =
    \rho_{11}(U^*,V^*) =\rho_{11}^\ocos(C_\theta^*).
$
That is, the $(j,k)$ correlation measure for the interpolating
  family $C_\theta$ for $\rho_{jk}^\orthobase$ is identical to the measure of monotonic
  dependence for the base copula family $C_\theta^*$.
\end{example}


  \begin{remark}
The
copulas obtained from a random vector $(U_1,\dots,
U_d)^\top$ via the stochastic inversion construction $(\udp_{1}^\leftarrow(U_1,Z_1)$, $\dots$,
$\udp_{d}^\leftarrow(U_d,Z_d))^\top$ differ from those studied by
~\cite{quessy:2024} and \cite{hofert/pang:2025} which use the ``forward
construction'' $(\udp_1(U_1),\dots,
\udp_d(U_d))^\top$. \cite{quessy:2024} shows that the forward
construction with particular choices of $T_1,\ldots,T_d$ leads to
various known copula families, such
as the chi-squared copulas
and their non-central
generalizations 
, v-copulas 
or squared copulas
; see \cite{quessy:2024} for references to these models.
However, our construction leads to models that have a tractable
density and are well suited
to datasets of the kind shown in Figure~\ref{fig:motivate} as we
explain in Section~\ref{sec:persp-appl}.
  \end{remark}

\section{An outlook on applications}\label{sec:persp-appl}

Although the focus of this article is on
concepts and theory, we close the paper with an outlook on
applications. In Section~\ref{sec:three-applications} we describe three
applications and illustrate these using the motivating datasets of
  Figure~\ref{fig:motivate}. The starting point for the applications
  is a matrix of sample basis correlations and we provide the necessary
  estimation theory in Section~\ref{sec:estim-gener-spearm}.

  \subsection{Estimating generalized Spearman correlations}\label{sec:estim-gener-spearm}


Since generalized Spearman correlation  $\rho_{\{\gtrans,\htrans\}}(X,Y)$ is a margin-free measure of dependence calculated from the copula of $(X,Y)^\top$, it is natural to search for sample analogues that depend on
ranks. 
 Let $\{(X_i, Y_i)^\top, i = 1,\ldots,n\}$ denote a paired data sample
 from
some underlying bivariate distribution with continuous margins and a copula $C$, and let 
the ranks of $X_1,\ldots,X_n$ be denoted $R_1
\leq \cdots  \leq R_n$ and the ranks of $Y_1,\ldots,Y_n$ be
denoted $S_1
\leq \cdots \leq S_n$.


There are a number of different possibilities for estimating generalized Spearman
correlations $\rho_{\{\gtrans,\htrans\}}$ and we consider a choice
that subsumes some estimators of Legendre correlations proposed in the literature. 
Because $\rho_{\{\gtrans,\htrans\}}(X,Y) =
\rho\left(\gtrans(F_X(X)),\htrans(F_Y(Y))\right)$ we could simply
replace the unknown marginal distributions by their empirical counterparts $F_n^{(X)}$ and $F_n^{(Y)}$ 
and approximate $\rho$ by the usual sample Pearson correlation
coefficient $\text{corr}\{\cdot\}$  of a set of paired
values. However, since $\{F_n^{(X)}(X_i), i = 1,\dots, n\}=\{R_i/n, i
= 1,\dots, n\}$ are not symmetrically distributed on $[0,1]$, we
divide the ranks by $n+1$ to obtain the estimator
\begin{equation} 
     \hat{\rho}_{\{\gtrans,\htrans\}} =  \text{corr} \left\{  \left( \gtrans\left(\frac{R_i
   }{n+1}\right), \htrans\left(\frac{S_i
                                }{n+1}\right) 
                                 \right)^\top, i = 1,\ldots,n
                                  \right\}.  \label{eq:B(3)} 
 \end{equation}                               
Upon setting $\gtrans=\orthobase_j$,
$\htrans=\orthobase_k$ for $j, k \in \mathbb{N}$, for functions $\orthobase_j$ and $\orthobase_k$ taken from a
correlation basis $\orthosystem$, the estimator~\eqref{eq:B(3)} yields the empirical basis correlation $ \hat{\rho}_{jk}^B$.

\begin{example}
In the case of Legendre basis correlations, we retrieve some existing rank-based estimators. First, recall from Example~\ref{ex:LegendreFormula} that $\rho^\blegendre_{11}$ is Spearman's rho; it is easy to verify
that  $\hat{\rho}_{11}$ in ~\eqref{eq:B(3)} reduces to the classical Spearman's rank correlation coefficient.
The estimates $\hat{\rho}_{jk}$ for $j\in \{1,2\}$ and $k=2$ are identical to the rank convexity and arachnitude first proposed by~\cite{bib:shaw-smith-spivak-11}. Their estimate of rank convexity is the sample
correlation of $\{ (R_i,(2S_i - n - 1)^2)^\top, i =
1,\ldots,n\}$, and we can verify that
\begin{equation}
  \label{eq:31}
   \hat{\rho}_{12} = \frac{\sqrt{12}\sqrt{45}}{2n(n^2-1)\sqrt{n^2-4}} \sum_{i=1}^n
 \left( R_i - \frac{n+1}{2}\right)\left(\left(2S_i - n - 1\right)^2 -
  \frac{n^2-1}{3} \right).
\end{equation} 
Similarly, $\hat{\rho}_{22}$ is equivalent to the sample correlation coefficient of $\{ ((2R_i - n - 1)^2,(2S_i - n - 1)^2)^\top, i =
1,\ldots,n\}$ which is~\citeauthor{bib:shaw-smith-spivak-11}'s
\citeyearpar{bib:shaw-smith-spivak-11} estimate of arachnitude given by
\begin{equation*}
  \label{eq:30}
  \hat{\rho}_{22} = \frac{45}{4n(n^2-1)(n^2-4)} \sum_{i=1}^n
 \left( \left(2R_i - n - 1\right)^2 - \frac{n^2-1}{3}\right)\left(\left(2S_i - n - 1\right)^2 -
  \frac{n^2-1}{3} \right).
\end{equation*}
Finally, after some algebra, $\hat{\rho}_{jk}$ can be seen to be asymptotically equivalent to the estimator $(1/n) \sum_{i=1}^n \blegendre_j\{(R_i -0.5)/n\} \blegendre_k\{(S_i -0.5)/n\}$ of~\cite{bib:kallenberg-ledwina-99}.
\end{example}

The estimator $\hat{\rho}_{\{\gtrans,\htrans\}}$ in \eqref{eq:B(3)} is asymptotically normal. Because we will rely on the empirical copula process, we need to make the following assumption from \cite{Segers:2012}, which is commonly made in rank-based inference for copula models, although it can be weakened at the cost of more complex arguments \citep{Buecher/Segers/Volgushev:2014,Genest/Neslehova/Remillard:2017}.

\begin{assumpC}\label{assumption:copula}
For each $j \in \{1,2\}$, the first order partial derivative of $C$ with respect to $u_j$, denoted $\dot C_j = (\partial/\partial u_j) C$, exists and is continuous on the set $\{(u_1,u_2) \in [0,1]^2 : 0 < u_j < 1\}$.
\end{assumpC} 


\begin{theorem}\label{thm:asympt_distribution}
Suppose that $(X,Y)$ is a random pair with continuous margins and a
copula $C$ that satisfies Assumption~\ref{assumption:copula}. Let
$(\gtrans_j,\htrans_j)$, $j \in \{1,\dots, J\}$ be pairs of
transformations that satisfy Assumptions~\ref{assA1.new}\footnote{We
  impose Assumption~\ref{assA1.new} since this is sufficient for the
  estimation of regular
  basis correlations in Section~\ref{sec:persp-appl}. However, the
  estimation theory could be extended to non-normalized transformations.} and \ref{assA3.new}, ${\boldsymbol{\rho}}=({\rho}_{\{\gtrans_1,\htrans_1\}}, \dots, {\rho}_{\{\gtrans_J,\htrans_J\}})$ with ${\rho}_{\{\gtrans_j,\htrans_j\}} = {\rho}_{\{\gtrans_j,\htrans_j\}}(X,Y)$ be the vector of the corresponding generalized Spearman correlations, and $\hat{\boldsymbol{\rho}}=(\hat{\rho}_{\{\gtrans_1,\htrans_1\}}, \dots, \hat{\rho}_{\{\gtrans_J,\htrans_J\}})$ be its estimator. 
Then as $n \to \infty$, $\sqrt{n}(\hat{\boldsymbol{\rho}} - \boldsymbol{\rho}) \rightsquigarrow \boldsymbol{Z}$, where $\rightsquigarrow$ denotes convergence in distribution and $ \boldsymbol{Z}=(Z_1,\dots, Z_J)$ is normally distributed with zero mean. Furthermore, 
\[
Z_j =  \int_0^1 \int_0^1 \{\mathbb{B}_C (u,v) - \dot C_1(u,v) \mathbb{B}_C(u,1) - \dot C_2(u,v) \mathbb{B}_C(1,v)\} \gtrans^\prime_j(u) \htrans^\prime_j(v) \rd u \rd v
\]
for $j \in \{1,\dots, J\}$, where
$\mathbb{B}_C$ is a tight centred Gaussian process such that for all $u_1,u_2,v_1,v_2 \in [0,1]$, $
\cov(\mathbb{B}_C(u_1,v_1), \mathbb{B}_C(u_2,v_2)) = C(\min(u_1, u_2),\min( v_1, v_2)) - C(u_1,v_1) C(u_2,v_2)
$.
%
\end{theorem}

\noindent The asymptotic behaviour of estimators of basis correlation matrices
is obtained as a corollary.

\begin{corollary}\label{cor:asympt_distribution}
Suppose that $\orthosystem$ is a regular correlation basis and that
the underlying copula satisfies
Assumption~\ref{assumption:copula}. Consider a finite set
$\mathcal{I}$ of pairs of indices $(j, k)$ and set $\hat{\boldsymbol{\rho}}_{\mathcal{I}}^\orthobase=(\hat{\rho}_{jk}, (j,k) \in \mathcal{I})$, $\boldsymbol{\rho}_{\mathcal{I}}^{\orthobase}=(\rho_{jk}^{\orthobase}, (j,k) \in \mathcal{I})$. 
Then as $n \to \infty$, $\sqrt{n}(\hat{\boldsymbol{\rho}}_{\mathcal{I}}^{\orthobase} - \boldsymbol{\rho}_{\mathcal{I}}^{\orthobase}) \rightsquigarrow \boldsymbol{Z}_{\mathcal{I}}$, where $\boldsymbol{Z}_{\mathcal{I}} = (Z_{jk},(j,k) \in \mathcal{I})$ with $Z_{jk} =  \int_0^1 \int_0^1 \{\mathbb{B}_C (u,v) - \dot C_1(u,v) \mathbb{B}_C(u,1) - \dot C_2(u,v) \mathbb{B}_C(1,v)\} \orthobase^\prime_{j}(u) \orthobase^\prime_{k}(v) \rd u \rd v$ is a normal random vector with zero mean.
\end{corollary}

  \subsection{Three applications}\label{sec:three-applications}
\begin{figure}[htb!]
  \centering
    \includegraphics[width=0.9\textwidth]{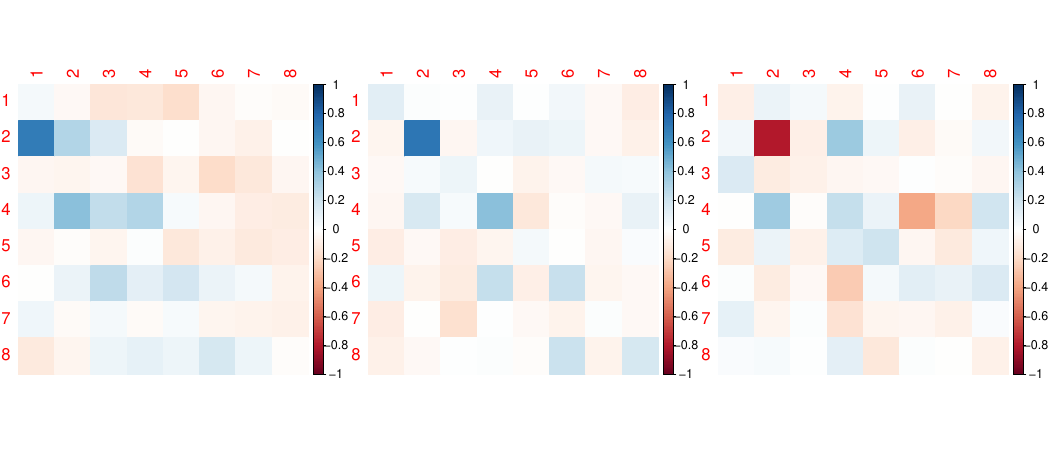}\vspace{-1cm}
    \caption{\label{fig:chessboards}Heat maps of the estimated
      Legendre basis correlation matrices $\hat P^{\blegendre}_8$ for the three datasets of 200 points
    in Figure~\ref{fig:motivate}.}
\end{figure}

\paragraph{Analysis of the features and symmetries of dependence structures.}
  An exploratory analysis of data exhibiting possible non-monotonic
  dependencies can be based on an estimated matrix of basis
  correlations. We apply the estimator~\eqref{eq:B(3)} to estimate
  $8\times 8$ matrices of Legendre correlations $\hat P_8^\blegendre$ for each
  of the three datasets and show these as heat maps in Figure~\ref{fig:chessboards} similar to those for
  t-copulas in
  Section~\ref{sec:basis-corr-spec}. The first heat map shows a
  particularly high value for the angularity measure
  $\hat\rho_{21}^\blegendre$ and the clear asymmetry of the image about
  the leading diagonal is
  evidence that the underlying copula 
  is non-exchangeable.
 The second and third heat maps are more symmetric about the leading
 diagonal and are characterized by high positive and negative cruciformity
 $\hat\rho_{22}^\blegendre$ respectively. In the latter two datasets
 there is a suggestion that the underlying dependence structure might
 be exchangeable and possibly also radially symmetric. To test the
 former we would want to test that differences $\rho_{jk}^\blegendre -
 \rho_{kj}^\blegendre$ in the underlying population measures are zero
 for a collection of pairs $(j,k)$. For
 the latter we would want to test that $\rho_{jk}^\blegendre =0$ for
 a collection of pairs $(j,k)$ with $j+k$ is odd. Formal inference
 could be based on the asymptotic result of
 Corollary~\ref{cor:asympt_distribution} 
 but full elaboration of details is left for future
   work.

\begin{figure}[t!]
  \centering
    \includegraphics[width=0.9\textwidth]{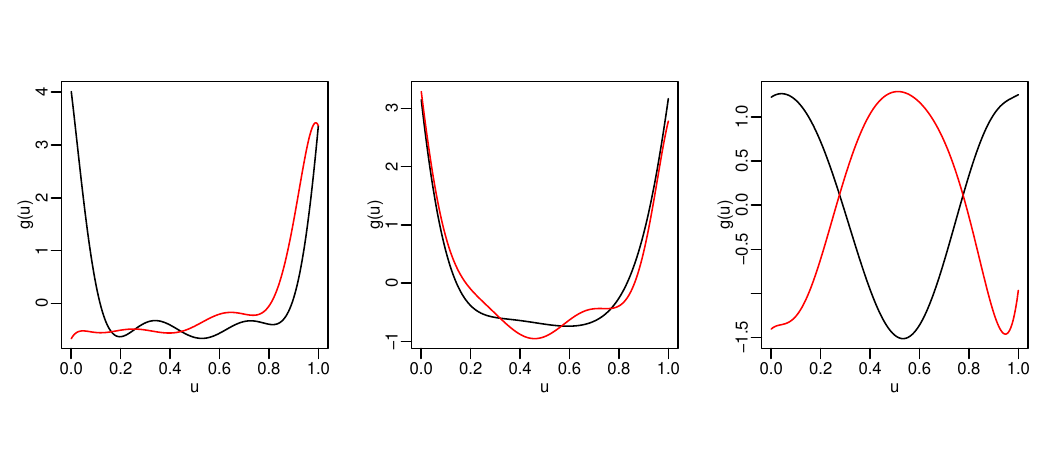}\vspace{-1cm}
    \caption{\label{fig:basisexpansions} Functions $\gtrans_N(u)
  = \sum_{j=1}^8 \alpha_{1j} \blegendre_j(u)$ (black) and $\htrans_N(u)
  = \sum_{j=1}^8 \alpha_{2j} \blegendre_j(u)$ (red)
  maximizing $\bm{\alpha}_1^\top \hat P^\blegendre_8
    \bm{\alpha}_2$ for the three datasets of 200 points
    in Figure~\ref{fig:motivate}.}
\end{figure}

\paragraph{Maximizing generalized Spearman correlation.} The second potential application uses the basis
 expansion technique from Section~\ref{sec:expansions}. Recall from~\eqref{eq:3} that our aim is to
  find unit vectors $\bm{\alpha}_1 =(\alpha_{11},\ldots,\alpha_{1N})^\top$ and $\bm{\alpha}_2 =(\alpha_{21},\ldots,\alpha_{2N})^\top$
such that, if we set
 $\gtrans_N(u)
  = \sum_{j=1}^N \alpha_{1j} \blegendre_j(u)$ and $\htrans_N(u)
  = \sum_{j=1}^N \alpha_{2j} \blegendre_j(u)$, the generalized
  Spearman correlation
  \begin{equation}\label{eq:6}
    \rho_{\{\gtrans_N,\htrans_N\}}(X,Y) = \bm{\alpha}_1^\top P^\blegendre_N
 \bm{\alpha}_2
\end{equation}
is maximized. To solve this problem we can implement a singular value
 decomposition of the basis correlation matrix $P^\blegendre_N$, that
 is a decomposition of the form $P^\blegendre_N = A_1 D_N A_2^\top$ for
 orthogonal matrices $A_1,A_2\in \R^{N \times N}$ (i.e.~matrices satisfying $A_i^\top A_i =
 A_iA_i^\top = I_N$ for $i\in \{1,2\}$) and a diagonal matrix
 $D_N = \text{diag}(d_1, \ldots,d_N)$ with diagonal elements arranged from largest to
 smallest. $A_1$ and $A_2$ are known as the matrices of left and right singular
 vectors and $D_N$ is the matrix of singular values. It is a
 well-known property of this decomposition that we can find the
 unit vectors 
 maximizing~\eqref{eq:6}  by setting $\bm{\alpha}_1$ and
 $\bm{\alpha}_2$
 equal to the first columns of $A_1$ and $A_2$
 respectively. Moreover, the maximum value of~\eqref{eq:6} is the first singular
 value $d_1$\footnote{More generally, the columns of $A_1$ and $A_2$ could be used to
   construct approximations to the canonical basis functions in the
   diagonal decomposition of~\cite{lancaster:1958} and the elements
 of $D_N$ would provide estimates of the canonical correlations; see Remark~\ref{rem:lancaster}.}. We apply the singular value decomposition
 to the estimated Legendre correlation matrices $\hat P_8^{\blegendre}$
 underlying Figure~\ref{fig:chessboards} and show
 the resulting functions $\gtrans_N$ and $\htrans_N$ in
 Figure~\ref{fig:basisexpansions}.
 As would be expected, $\gtrans_N$ is broadly u-shaped in
  all three pictures while $\htrans_N$ is roughly monotonic in the first,
  u-shaped in the second and inverse-u-shaped in the third.
  
\paragraph{Constructing parametric models for non-monotonic
  dependencies.}  For the final application we consider constructing parametric copulas to
model the dependence structures of the three datasets.
     For these datasets interpolating copula families $C_\theta$
  with densities of the form  $c_\theta(u,v) =
     c_\theta^*(\udp_\gtrans(u),\udp_\htrans(v))$, corresponding to independent bivariate
     stochastic inversion of base copula models $C^*_\theta$, seem appropriate 
     to capture the non-monotonic dependencies. This judgement is
     based on Proposition~\ref{prop:mult-stoch-invers-symmetry} and
     the symmetries evident in Figure~\ref{fig:motivate}.

    \begin{figure}[t]
  \centering
    \includegraphics[width=0.9\textwidth]{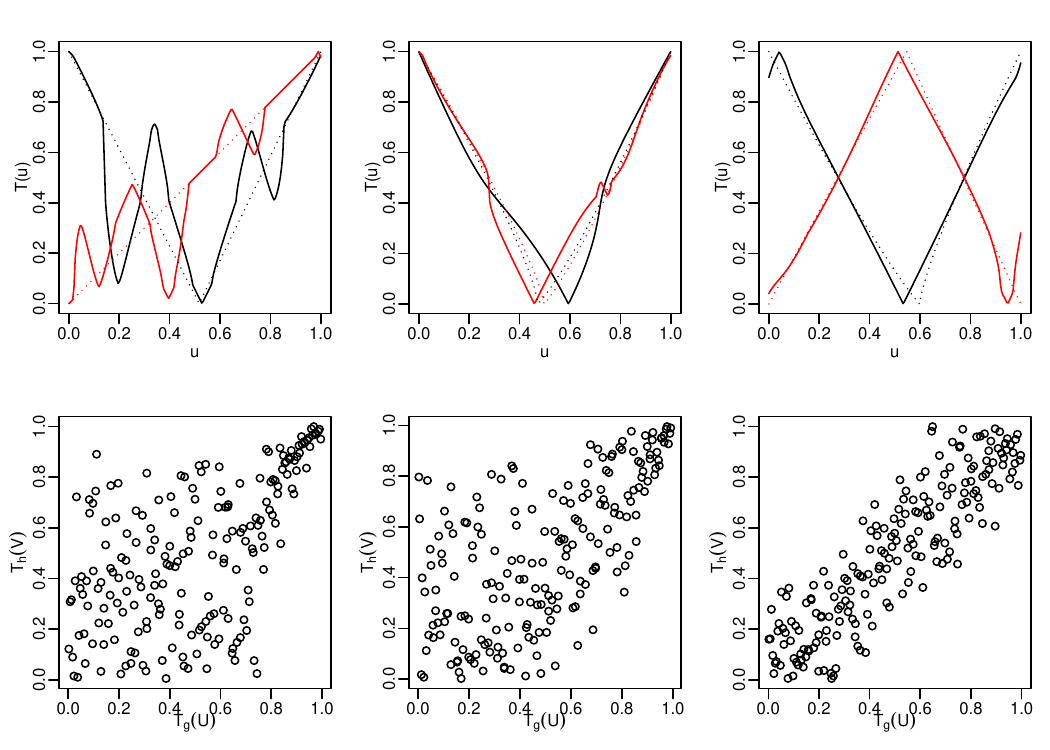}
    \caption{\label{fig:estimationdata} Top row: (solid lines): udp functions
    $\udp_\gtrans$ and $\udp_\htrans$ corresponding to $\gtrans=\gtrans_N$ and
    $\htrans=\htrans_N$ in Figure~\ref{fig:basisexpansions}. Top row (dotted
    lines): udp functions estimated in Approach 2. Bottom
    row: pseudo-copula data $\{(\udp_\gtrans(R_i/(n+1)),
     \udp_\htrans(S_i/(n+1)))^\top, i=1,\ldots,n\}$.}
\end{figure}

 We first need to select base copula families $C_\theta^*$. Recalling that 
 if $(U,V) \sim C_\theta$ then $(T_\gtrans(U), T_\htrans(V)) \sim
 C^*_\theta$, we evaluate  $\udp_\gtrans=\pbase_\gtrans \circ \gtrans$ and $\udp_\htrans =
  \pbase_\htrans \circ \htrans$  for the functions $\gtrans=\gtrans_N$ and
  $\htrans=\htrans_N$ in Figure~\ref{fig:basisexpansions}. We use numerical techniques
 to accurately approximate the distribution functions
  $\pbase_\gtrans$ and $\pbase_\htrans$ according to the general
  formula~\eqref{eq:S62} in
  Corollary~\ref{prop:Scalc-formulas}. In the top row of
     Figure~\ref{fig:estimationdata} the resulting udp functions
     $\udp_\gtrans$ and $\udp_\htrans$ are shown
    as solid black and red lines respectively.
  The data $\{(\udp_\gtrans(R_i/(n+1)),
     \udp_\htrans(S_i/(n+1)))^\top, i=1,\ldots,n\}$ shown in the bottom row of
  Figure~\ref{fig:estimationdata} can be viewed as a pseudo-sample from $C_\theta^*$. As expected,
     these data show monotonic dependence. They also give clues
     about the copula family $C_\theta^*$. For the first two datasets, a Gumbel 
     or rotated Clayton copula
     with upper tail dependence may be appropriate while, for the third
     dataset, a radially-symmetric Gauss or Frank copula may work
     well.

\begin{figure}[htb!]
  \centering
    \includegraphics[width=0.8\textwidth]{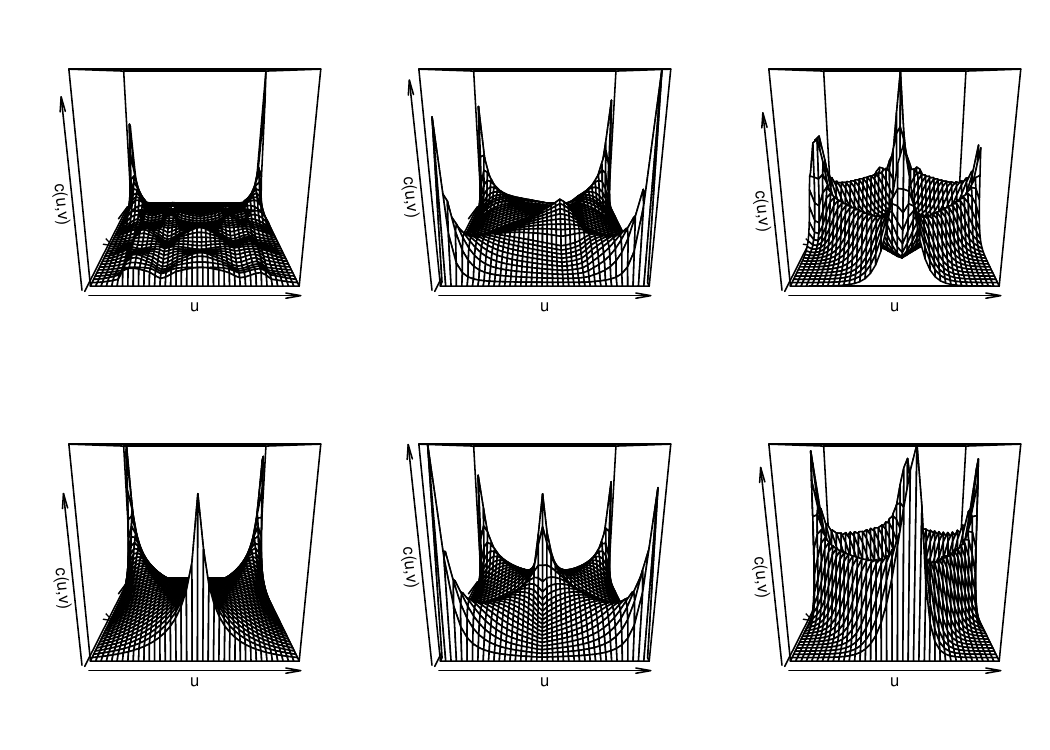}\vspace{-0.8cm}
    \caption{\label{fig:densityplots} Perspective plots of the
      densities of copulas fitted to the three datasets in
      Figure~\ref{fig:motivate} by Approach 1 (top row) and Approach 2
      (bottom row). In the top row the first two pictures are based on
      a rotated Clayton density $c^*_\theta$ with estimated parameters
      $\hat\theta = 1.69$ and 1.73 and the final picture is based on a Frank
      density with estimated parameter 10.9. In the bottom
      row the pictures are based on rotated Clayton, Gumbel and Frank
      copulas with estimated parameters 1.93, 1.93 and 11.7. The estimated parameters of the
      v-transforms are: $\delta= 0.52, \kappa=1.00$ (first picture);
      $\delta_1= 0.47, \kappa_1=0.67,\delta_2 = 0.50, \kappa_2=0.70$
      (second picture); and
      $\delta_1= 0.60, \kappa_1=1.36,\delta_2 = 0.55, \kappa_2=1.02$ (third picture).}
\end{figure}

   We implement two approaches to estimation of the
     copula $C_\theta$. In Approach
     1 we simply fit copulas with
     densities $c^*_\theta$ to the data in the bottom row of
     Figure~\ref{fig:estimationdata} using the method of maximum
     likelihood. For the first two datasets the rotated Clayton copula
     gives a better fit than the Gumbel copula. For the third dataset the Frank
     copula is favoured over the Gauss copula. The corresponding densities
     $c_\theta(u,v) =
     c_\theta^*(\udp_\gtrans(u),\udp_\htrans(v))$
are shown as
     perspective plots in the top row of
     Figure~\ref{fig:densityplots}.
     In Approach 2 we look for models with smoother
     densities by choosing smooth functions for $\gtrans$ and
     $\htrans$. Figure~\ref{fig:basisexpansions} suggests
that $\gtrans$ could be taken to be u-shaped in all cases and
$\htrans$ could be taken to be
monotonic for the first dataset, u-shaped
     for the second and inverse-u-shaped for the third. Based on the
     insight from Example~\ref{ex:ex8} that a flexible family of
     asymmetric u-shaped functions about a minimum value $0 < \delta <
     1$ corresponds to a family of v-transforms~\eqref{eq:v-transform}
     with fulcrum $\delta$ and generator $\Psi(x)
= 1 -(1-x)^\kappa$, $\kappa > 0$,  we introduce the notation
$\udp_{\delta,\kappa}$ for this family and make explicit use of it in our models. For the first dataset 
we fit a copula with density $c_\theta^*(\udp_{\delta,\kappa}(u),v)$. For
the second dataset we use the density
$c_\theta^*(\udp_{\delta_1,\kappa_1}(u),\udp_{\delta_2,\kappa_2}(v))$
and for the third dataset we use
$c_\theta^*(\udp_{\delta_1,\kappa_1}(u),1-\udp_{\delta_2,\kappa_2}(v))$. All
parameters are estimated in a single step using
maximum-likelihood. The densities $c_\theta^*$ selected are
the same as before for
datasets 1 and 3 (rotated Clayton and Frank) but a better fit is obtained if we change to a Gumbel
copula for dataset 2. Perspective plots of the fitted models are shown in the bottom
row of Figure~\ref{fig:densityplots} and the estimated udp
transformations are shown as dotted lines in the top row of
Figure~\ref{fig:estimationdata}.

\section{Code and reproducibility}

The \texttt{basiscor} R library available at \texttt{github.com/ajmcneil/basiscor} can be used to calculate population
and sample Legendre and cosine correlations. All of the examples and
illustrations in this paper can be reproduced using this library and
additional code found at \texttt{github.com/ajmcneil/papers}.


\clearpage
\setcounter{section}{0}
\setcounter{equation}{0}
\setcounter{figure}{0}
\setcounter{table}{0}
\setcounter{proposition}{0}
\setcounter{corollary}{0}
\setcounter{lemma}{0}
\setcounter{example}{0}
\setcounter{remark}{0}
\setcounter{assumpP}{0}
\renewcommand{\theequation}{S\arabic{equation}}
\renewcommand{\thesection}{S\arabic{section}}
\renewcommand{\thesubsection}{S\arabic{section}.\arabic{subsection}}
\renewcommand{\thetable}{S\arabic{table}}
\renewcommand{\theproposition}{S\arabic{proposition}}
\renewcommand{\thecorollary}{S\arabic{corollary}}
\renewcommand{\theexample}{S\arabic{example}}
\renewcommand{\thelemma}{S\arabic{lemma}}
\renewcommand{\theremark}{S\arabic{remark}}
\renewcommand{\thefigure}{S\arabic{figure}}
\renewcommand{\theassumpP}{P\arabic{assumpP}}

\section*{Appendix: Supplementary Material}
\addcontentsline{toc}{section}{Appendix: Supplementary Material}

\section{Technical foundations}

 Before giving all the proofs of results in the main paper in
 Section~\ref{sec:Sproofs}, we include
 some technical material on piecewise strictly monotonic
 transformations in Section~\ref{sec:Spiec-strictly-monot} and on
 uniform-distribution preserving (udp) transformations in
 Section~\ref{sec:Sudp-theory}. The proofs draw heavily on these
 technical foundations.
 
 \subsection{Piecewise strictly monotonic
  transformations}\label{sec:Spiec-strictly-monot}

  
We will consider an arbitrary piecewise monotonic function  $\psi:[0,a] \to
  \mathbb{R}$, $a \in (0,\infty)$ and a partition $a_0 = 0 < a_1 < \dots < a_{M-1} < a_M
= a$ for some $M \in \{1, 2, \dots\}$. We denote the restriction of $\psi$
to the set $A_m =(a_{m-1}, a_{m})$ by $\psi\vert_{A_m}$ and its range by $R_m$; we also set $A = \cup_{m=1}^M A_m$ and $R = \cup_{m=1}^M R_m$. We also impose one or more of the following assumptions on the functions
$\psi\vert_{A_m}$ as necessary.

\begin{assumpP}\label{as:Spiecewise-monotonic}
 For $m \in \{1,\dots, M\}$, $\psi\vert_{A_m}$ is
 continuous and strictly monotonic.
\end{assumpP}

\begin{assumpP}\label{as:Sturning-points}
For $m \in
\{1,\ldots,M-1\}$, $\psi\vert_{A_m}$ is increasing and
$\psi\vert_{A_{m+1}}$ is decreasing, or vice versa.
\end{assumpP}

\begin{assumpP}\label{as:Scon-diff-inverse}
For $m \in
\{1,\ldots,M\}$, $\psionAm^{-1}$ is
continuously differentiable on $R_m$.
\end{assumpP}

In what follows we find tractable expressions for the distribution
$F_{\psi(W)}$ and density $f_{\psi(W)}$ of the random variable
$\psi(W)$ where $W \sim
  \mathcal{U}(0,a)$.
The first useful observation concerns the continuity and strict monotonicity of $F_{\psi(W)}$.
\begin{proposition}\label{prop:Sproperties-Fpsi}
If  $\psi$ satisfies Assumption~\ref{as:Spiecewise-monotonic} and $W \sim
  \mathcal{U}(0,a)$ then
$F_{\psi(W)}$ is continuous. If, in addition, $\psi$ is continuous, $F_{\psi(W)}$ is strictly monotonic on the range of $\psi$.
\end{proposition}
\begin{proof}
Continuity of $F_{\psi(W)}$ holds because $\psi(W)$ is a continuous
random variable; for any fixed $x \in \mathbb{R}$, the number of roots
of the equation $\psi(w) =x$, $w \in A$, is at most $M$ and hence
finite, implying that $\Pr(\psi(W) = x) = 0$. Now suppose that $\psi$ is continuous. Assumption~\ref{as:Spiecewise-monotonic} guarantees that the range of $\psi$ is not a single point, and hence it must be a closed interval, say $[c,d]$ with $c < d$. For any $x_1, x_2$ such that $c \le x_1 < x_2 \le d$, the set $\{ w \in [0,a] : \psi(w) \in (x_1,x_2)\}$ is not empty by the intermediate value property of $\psi$, and open by the continuity of $\psi$. Thus there exists $w_0$ and $\delta > 0$ such that $(w_0-\delta, w_0 + \delta) \subset \{ w \in [0,a] : \psi(w) \in (x_1,x_2)\}$. Consequently, $F_{\psi(W)}(x_2) - F_{\psi(W)}(x_1) =  \Pr(\psi(W) \in (x_1, x_2]) \ge \Pr(W \in (w_0-\delta, w_0 + \delta) ) = 2\delta > 0$. 
\end{proof}


\begin{proposition}\label{prop:SCDF-transfo}
Suppose that $\psi$ satisfies Assumption~\ref{as:Spiecewise-monotonic}
and $W \sim \mathcal{U}(0,a)$. Then, for all $x \in \mathbb{R}$,
the distribution function $F_{\psi(W)}$ of $\psi(W)$ satisfies
\begin{equation}\label{eq:SCDF-transfo}
F_{\psi(W)}(x) = \frac{1}{a} \Bigl\{\sum_{m \in I_x} 
      \left(\psionAm^{-1}(x) - a_{m-1}\right) + \sum_{m \in D_x} \left(a_{m} -
      \psionAm^{-1}(x)\right) + \sum_{m \in O_x} \left(a_{m} - a_{m-1}\right)\Bigr\},
\end{equation}
where
 \begin{align*}
    I_x &= \{m \in \{1,\dots, M\} :  \psionAm\; \text{is increasing and}\, x \in R_m
          \} ,\\
     D_x &= \{m \in \{1,\dots, M\} :  \psionAm\; \text{is decreasing and}\,  x \in R_m
           \} ,\\
    O_x &= \{m \in \{1,\dots, M\} : x \not \in R_m, R_m \subset
          (-\infty,x] \}.
  \end{align*}
  \end{proposition}
  
  \begin{proof}
 Note that the sets $I_x$, $D_x$ and $O_x$ are disjoint and set $N_x = \{1,\dots, M\} \setminus (I_x \cup D_x \cup O_x)$. Because $W$ is a continuous random variable, 
\[
F_{\psi(W)}(x) = \P(\psi(W) \le x) = \P(\psi(W) \le x, W \in A) = \sum_{m=1}^M \P(\psi(W) \le x, W \in A_m).
\]
Fix an arbitrary $m \in \{1,\dots, M\}$ and observe that $R_m = \psionAm(A_m)$ is an open interval, say $(c_m, d_m)$. To calculate $\P(\psi(W) \le x, W \in A_m) = \P(\psionAm(W) \le x, W \in A_m)$, there are four possibilities to consider.  First, if $x \le c_m$, $ \P(\psionAm(W) \le x, W \in A_m)=0$. In this case, $m \in N_x $ and hence does not feature on the right-hand side of \eqref{eq:SCDF-transfo}. Second, if $x \ge d_m$, $m \in O_x$. At the same time, $\P(\psionAm(W) \le x, W \in A_m) = \P(W \in A_m) =(a_m - a_{m-1})/a$, corresponding to the summand of the third sum in \eqref{eq:SCDF-transfo}. Third, if $x \in (c_m, d_m)$ and $\psionAm$ is strictly increasing, $m \in I_x$ and $\P(\psionAm(W) \le x, W \in A_m) = \{\psionAm^{-1}(x) - a_{m-1}\}/a$. Finally, if $x \in (c_m, d_m)$ and $\psionAm$ is strictly decreasing, $m \in D_x$ and $\P(\psionAm(W) \le x, W \in A_m) = \{a_m - \psionAm^{-1}(x)\}/a$.
\end{proof}

\noindent A simplification of \eqref{eq:SCDF-transfo} occurs when $\psi$ is continuous and the points
$a_m$, $m \in \{1,\dots, M-1\}$ are all turning points in the sense of Assumption~\ref{as:Sturning-points}.
\begin{corollary}\label{cor:SCDF-transfo-turningpts}
  Suppose that $W \sim \mathcal{U}(0,a)$ and that $\psi$ is continuous and satisfies Assumptions~\ref{as:Spiecewise-monotonic} and~\ref{as:Sturning-points} on the same partition. Then, for all $x \in \mathbb{R}$,
\begin{equation}\label{eq:S60}
F_{\psi(W)}(x)  = \frac{1}{a} \Bigl\{\sum_{m \in I_x} 
      \psionAm^{-1}(x)  - \sum_{m \in D_x}   \psionAm^{-1}(x) + a \Ind{M \in D_x \cup O_x}\Bigr\}.
\end{equation}
\end{corollary}

\begin{proof}
Fix an arbitrary $x \in \mathbb{R}$. Recall that $a_0 = 0$ and observe that $a_M = a$ appears in the summations in \eqref{eq:SCDF-transfo} at most once with a plus sign. This occurs if either $M \in D_x$ or $M \in O_x$.  For any $m \in \{1,\dots, M-1\}$ such that $m \in D_x$ we must have either $m+1 \in I_x$ or $m+1 \in O_x$. Either way, $a_m$ appears in \eqref{eq:SCDF-transfo} exactly twice with opposite signs and drops out. For $m \in \{1,\dots, M-1\}$ such that $m \in I_x$, $a_m$ does not appear in 
 \eqref{eq:SCDF-transfo} at all. Finally, for $m \in \{1,\dots, M-1\}$ such that $m \in O_x$, we will have either $m+1 \in O_x$ or $m+1 \in I_x$ if $\psi\vert_{A_{m+1}}$ is increasing and $\lim_{t \uparrow a_{m+1}}\psi(t) > x$. Either way, $a_m$ appears in \eqref{eq:SCDF-transfo} exactly twice with opposite signs and drops out. 
\end{proof}

The next proposition shows that $\psi(W)$ has a density under
Assumptions~\ref{as:Spiecewise-monotonic}
and~\ref{as:Scon-diff-inverse}, where the latter imposes an additional smoothness
condition on $\psi$.

\begin{proposition}\label{prop:Sdensity-transfo}
Suppose that $\psi$ satisfies Assumptions~\ref{as:Spiecewise-monotonic} and~\ref{as:Scon-diff-inverse} with the same partition
and $W \sim \mathcal{U}(0,a)$. Then the density of $\psi(W)$ is given, for almost all $x \in \mathbb{R}$, by
\begin{equation}\label{eq:S61}
f_{\psi(W)}(x) = \sum_{m\in I_x \cup D_x} \frac{1}{a} \Bigl | \frac{1}{\psionAm^{\prime}\{\psionAm^{-1}(x)\}}\Bigr |.
\end{equation}
Furthermore, $F_{\psi(W)}$ is differentiable on $R \setminus N_\psi$ with $N_\psi =\{ \psi_{+}(a_m), m =0,\dots, M-1\} \cup \{\psi_{-}(a_m), m=1,\dots, M\}$, where for any $x \in [0,a]$, $\psi_{+}(x)$ and $\psi_{-}(x)$ denote the right-hand and left-had limits of $\psi$ at $x$, respectively. The derivative of $F_{\psi(W)}$ is given, for all $x \in R\setminus N_\psi$, by \eqref{eq:S61}.
\end{proposition}

\begin{proof}
For any $m\in \{1,\dots, M\}$, $W_m = W | W\in A_m$ is uniform on $A_m$. Write 
\[
F_{\psi(W)}(x) = \sum_{m=1}^M \P(\psi(W) \le x | W \in A_m) \P(W \in A_m) = \sum_{m=1}^M \frac{a_{m} - a_{m-1}}{a} \P(\psionAm(W_m) \le x).
\]
By the theorem for densities of monotone transformations, see, e.g., Theorem~2.1.5 in~\cite{bib:casella-berger-02}, $\psionAm(W_m)$ has a density given by
\[
\frac{1}{a_m - a_{m-1}} \Bigl | \frac{1}{\psionAm^{\prime}\{\psionAm^{-1}(x)\}}\Bigr |  \Ind{m\in I_x \cup D_x}.
\]
Interchanging the sum and the integral and simplifying leads to $F_{\psi(W)}(x) = \int_{-\infty}^x f_{\psi(W)}(t) \rd t$, as claimed.

Finally, take an arbitrary $x \in R\setminus N_\psi$. For such $x$, we have, for all $m \in \{1,\dots, M\}$, either $x \in R_m$ or $x \not \in \bar R_m$, where $\bar R_m$ denotes the closure of $R_m$. Consequently, one can find some small $\delta > 0$ such that for all $t \in (x-\delta, x+\delta)$, $t \in R_m$ if and only if $x \in R_m$, which is if and only if $m \in I_x \cup D_x$. Hence $I_t = I_x$, $D_t = D_x$, and $O_t = O_x$ for all $t \in (x-\delta, x+\delta)$. The differentiability then follows from \eqref{eq:SCDF-transfo} given that $\psionAm^{-1}$ is differentiable at $x$ for each $m \in I_x \cup D_x$.
\end{proof}

Finally, we give a corollary that rewrites~\eqref{eq:S60} and~\eqref{eq:S61} in 
alternative forms and emphasizes that calculating the functions that describe the distribution of $\psi(W)$ is
primarily an exercise in root finding.

\begin{corollary}\label{prop:Scalc-formulas}
  Let $W \sim \mathcal{U}(0,a)$ and let $\{r_1(x),\dots,
r_{n(x)}(x)\} = \{w \in A : \psi(w) =x\}$ denote the roots  $w \in A$ of
$\psi(w) = x$, $s_i(x) = 1$ if $\psi$ is increasing at
$r_i(x)$ and $s_i(x) = -1$ if $\psi$ is decreasing at
$r_i(x)$ for $i \in \{1,\dots, n(x)\}$. Then the following holds.
\begin{enumerate}
\item[(i)]  For a continuous $\psi$ satisfying Assumptions~\ref{as:Spiecewise-monotonic} and~\ref{as:Sturning-points} with the same partition,
  \begin{align}
F_{\psi(W)}(x)  &= \frac{1}{a} \Bigl\{ \sum_{i=1}^{n(x)} s_i(x)
                  r_i(x) +a \Ind{M \in D_x \cup
                  O_x}\Bigr\}. \label{eq:S62}
  \end{align} 
  \item[(ii)] For $\psi$ satisfying Assumptions~\ref{as:Spiecewise-monotonic}
  and~\ref{as:Scon-diff-inverse} with the same partition, 
  \begin{align}
    f_{\psi(W)}(x) &= \Ind{x \in R}\sum_{i=1}^{n(x)} \frac{1}{a} \Bigl |
    \frac{1}{\psi^\prime(r_i(x))} \Bigr | \label{eq:S63}
  \end{align}
for almost all $x \in \mathbb{R}$. Furthermore, $F_{\psi(W)}$ is differentiable with $F^\prime_{\psi(W)}(x) = f_{\psi(W)}(x)$ for any $x \in R \setminus N_\psi$, where $N_\psi=  \{ \psi_{+}(a_{m-1}), \psi_{-}(a_m) , m=1,\dots, M\}$.
\end{enumerate}
\end{corollary}
\begin{proof}
By noting that $| I_x \cup D_x| = n(x) \leq M$ and that the values
$\psionAm^{-1}(x)$, $m \in I_x \cup D_x$ are precisely the roots $w \in A$ of
$\psi(w) = x$ these formulas follow easily from~\eqref{eq:S60} and~\eqref{eq:S61}. The differentiability of $F_{\psi(W)}$ has been established in Proposition~\ref{prop:Sdensity-transfo}.
 \end{proof}

\subsection{Udp
  transformations}\label{sec:Sudp-theory}


While there is an existing
literature on
udp functions, many of the main
results have been derived in a
non-probabilistic context; see, for
example,~\citet{bib:porubsky-salat-strauch-88}. We provide the results
we need in a probabilistic form.
We recall from Definition~\ref{def:udp} in the main paper that $T:[0,1] \to [0,1]$
is uniform distribution preserving (udp) if $T(U) \sim
\mathcal{U}(0,1)$ for $U \sim \mathcal{U}(0,1)$.

Clearly, if $\udp$ is udp, there can be no interval $(a,b) \subset [0,1]$ such that $\udp(x) = c$ for all $x \in (a,b)$. If this were a case, we would have $\P(\udp(U) = c) \ge P(U \in (a,b)) = b-a \neq 0$, which contradicts the fact that $\udp(U) \sim \mathcal{U}(0,1)$ is a continuous random variable. In fact, more is true.

\begin{lemma}\label{lem:SUDP}
Suppose that a udp transformation is continuously differentiable on $(a,b) \subset [0,1]$. Then $\udp^\prime(x) \neq 0$ for all $x \in (a,b)$ and $\udp$ is strictly monotone on $(a,b)$.
\end{lemma}

\begin{proof}
This statement follows by contradiction. Suppose there exists $x \in (a,b)$ such that $\udp^\prime(x) = 0$. Pick some small $0 < \varepsilon < 1$. By continuity of $\udp^\prime$ at $x$ there exists some $\delta > 0$ such that $ a < x-\delta < x+\delta < b$ and that for all $t \in (x-\delta, x+\delta)$, $|\udp^\prime(t) | < \epsilon$.  Because $\udp$ is continuous on $[x-\delta, x+\delta]$, it attains its minimum and maximum there at some $x_{\min}$ and $x_{\max}$, respectively. The set $\{ y : y=\udp(t), t \in [x-\delta, x+\delta]\} = [\udp(x_{\min}), \udp(x_{\max})]$ and by the mean-value theorem, $\udp(x_{\max})-\udp(x_{\min}) = \udp^\prime(c) (x_{\max}-x_{\min})$ for some $c$ between $x_{\min}$ and $x_{\max}$. Hence, 
\[
\udp(x_{\max})-\udp(x_{\min}) = |\udp(x_{\max})-\udp(x_{\min})| = |\udp^\prime(c) | |x_{\max}-x_{\min}| < 2\epsilon \delta.
\]
Let $U \sim \mathcal{U}(0,1)$. Because $\udp$ is uniformity preserving, $\P(\udp(U) \in [\udp(x_{\min}), \udp(x_{\max})]) = \udp(x_{\max})-\udp(x_{\min}) < 2\epsilon \delta$. At the same time, $\P(\udp(U) \in [\udp(x_{\min}), \udp(x_{\max})]) \ge P(U \in [x-\delta, x+\delta]) = 2\delta$. This gives $2 \delta < 2\epsilon \delta$ which is a contradiction. The fact that $\udp$ is strictly monotone on $(a,b)$ follows easily from the intermediate value property of $\udp^\prime$.
\end{proof}

We recall from Definition~\ref{def:udp-regular} that a udp
transformation is said to be regular if there exists a finite partition $a_0 = 0 < a_1 < \dots < a_{L-1} < a_L = 1$ for $L \in \{1, 2, \dots\}$ such that $\udp$ is continuously differentiable on $A_\ell =(a_{\ell-1}, a_{\ell})$ for all $\ell \in \{1,\dots, L\}$. For any $\ell \in \{1,\dots, L\}$, $\udponAl$ denotes the restriction of $\udp$ to $A_\ell$ and $\udponAl(A_\ell)$ the range of $\udponAl$. Finally, $A= \cup_{\ell=1}^L A_\ell$ and $\udp(A) = \cup_{\ell=1}^L \udponAl(A_\ell)$. In the next proposition we establish the properties of the $T$-partition, consisting of the ordered distinct points $b_0 < \dots < b_K$ of the set $\{T_{-}(a_{\ell}), T_{+}(a_{\ell-1}), \ell \in \{1,\dots, L\}\}$. The latter set is of course well-defined because $\udponAl$ is continuous by assumption and strictly monotone by Lemma~\ref{lem:SUDP}. Because $\udp$ is uniformity preserving, we must have that $\{T_{-}(a_{\ell}), T_{+}(a_{\ell-1}), \ell \in \{1,\dots, L\}\} \subset [0,1]$.

\begin{proposition}\label{prop:ST-partition}
Suppose that $\udp$ is a regular uniformity preserving transformation
as in Definition~\ref{def:udp-regular} and  $b_0 < \dots < b_K$ the corresponding $T$-partition. Then the following holds.
\begin{itemize}
\item[(i)] $b_0 = 0$ and $b_K = 1$.
\item[(ii)] For any $k \in \{1,\dots, K\}$, $(b_{k-1}, b_k) \subseteq \udp(A_\ell)$ for at least one $\ell\in \{1,\dots, L\}$.
\item[(iii)] The sets $\udp(A)$ and $B  = \cup_{k=1}^K (b_{k-1},b_k)$ both have Lebesgue measure one.
\end{itemize}
\end{proposition}

\begin{proof}
To show part (i), it suffices to consider the claim $b_0=0$ because the proof that $b_K=1$ is analogous. By definition, $b_0 = \min\{T_{-}(a_{\ell}), T_{+}(a_{\ell-1}), \ell \in \{1,\dots, L\}\}$. We have already explained above that $b_0 \in [0,1]$, so we merely need to show that $b_0 > 0$ leads to a contradiction. Indeed, first observe that the set $D=\{ u \in A: \udp(u) \in (0,b_0)\}$ is then necessarily empty since any $u^* \in D$ must lie in $A_\ell$ for some $\ell \in \{1,\dots, L\}$ so we have that $\udp_{+}(a_{\ell-1}) < \udp(u^*) < b_0$ if $\udponAl$ is increasing or that $\udp_{-}(a_{\ell}) < \udp(u^*) < b_0$ if $\udponAl$ is decreasing. Either way, $b_0$ could not have been the smallest element of $\{T_{-}(a_{\ell}), T_{+}(a_{\ell-1}), \ell \in \{1,\dots, L\}\}$. The just established fact that $D$ is empty leads to a contradiction because on the one hand $\P(\udp(U) \in (0, b_0)) = b_0$ because $\udp$ is uniformity preserving and on the other $\P(\udp(U) \in (0, b_0)) = \P(U \in D) = 0$.

Turning to part (ii), fix an arbitrary $k \in \{1,\dots, K\}$ and observe that for any $\ell\in \{1,\dots, L\}$, the intersection between $(b_{k-1}, b_k)$ and $\udp(A_\ell)$ is either empty or $(b_{k-1}, b_k)$ itself. To see this, recall that $\udp(A_\ell) = (\min(\udp_{+}(a_{\ell-1}), \udp_{-}(a_\ell)), \max(\udp_{-}(a_{\ell}), \udp_{+}(a_{\ell-1})))$ because $\udponAl$ is continuous and strictly monotone. However, $b_{k-1} < \min(\udp_{+}(a_{\ell-1}), \udp_{-}(a_\ell)) < b_k$ or $b_{k-1} < \max(\udp_{-}(a_{\ell}), \udp_{+}(a_{\ell-1})) < b_k$ cannot occur because it would contradict the way the $\udp$-partition has been constructed. Now suppose that $(b_{k-1}, b_k) \cap \udp(A_\ell) = \emptyset$ for all $\ell \in \{1,\dots, L\}$. This  is impossible: on the one hand, $ \P(\udp(U) \in (b_{k-1}, b_k)) = b_k-b_{k-1} >0$ because $\udp$ is uniformity preserving. On the other hand,
\begin{align*}
\P(\udp(U) \in (b_{k-1}, b_k))& = \P(\udp(U) \in (b_{k-1}, b_k), U \in A)= \sum_{\ell=1}^L \P(\udp(U) \in (b_{k-1}, b_k), U \in A_\ell) \\
& \le \sum_{\ell=1}^L \P(\udp(U) \in (b_{k-1}, b_k) \cap \udp(A_\ell)) =0.
\end{align*}
 This is obviously a contradiction, so there must exist at least one $\ell$ such that $(b_{k-1}, b_k) \subseteq \udp(A_\ell)$.
 
Part (iii) is now immediate. $B$ has Lebesgue measure $1$ by part (i), and $B \subseteq \udp(A)$ by part (ii), so that $\udp(A)\subseteq (0,1)$ has Lebesgue measure $1$ as well.
\end{proof}

In the next proposition, we gather several useful properties of regular udp transformations that allow us to define
and investigate their stochastic inverses in Section~\ref{sec:bounds} of the main
paper. 

\begin{proposition}\label{prop:Supd-properties}
Suppose that $\udp$ is a regular uniformity preserving transformation
as in Definition~\ref{def:udp-regular} and write $A = \cup_{\ell=1}^L A_\ell$. Then the following holds.
\begin{itemize}
\item[(i)] $\udponAl$ is strictly monotonic and $\udponAl^{-1}$ is continuously differentiable, that is, $\udp$ satisfies Assumptions ~\ref{as:Spiecewise-monotonic} and~\ref{as:Scon-diff-inverse} for the same partition as in Definition~\ref{def:udp-regular}.
\item[(ii)] For any $x \in \udp(A)$, the pre-image $\udp^{-1}(x) = \{u : \udp(u) = x, u \in A\}$ of the point $x$ intersected with $A$ is finite, i.e., 
$\udp^{-1}(x) = \{r_1(x), \ldots, r_{n(x)}(x)\}$. 
\item[(iii)] For any $a \in (0,1]$, define $F_{a}$ for all $x \in
  \mathbb{R}$ by $F_a(x) = \P(\udp(U) \le x , U \le a)$ where $U \sim
  \mathcal{U}(0,1)$. Then for almost all $x \in \{
  \udp(u) : u < a, u \in A\}$, the derivative of $F_a$ is
\[
F_{a}^\prime(x) = \sum_{i = 1}^{n(x)} \left |\frac{1}{\udp^\prime(r_i(x))} \right | \Ind{r_i(x) < a},
\]
where $n(x)$ and $r_1(x), \dots, r_{n(x)}(x)$ are as in (ii).
\item[(iv)]  For any $x \in B$ and $r_1(x), \dots, r_{n(x)}(x)$ as in (ii),
\[
\sum_{i = 1}^{n(x)} \left |\frac{1}{\udp^\prime(r_i(x))} \right | = 1.
\]
\end{itemize}
\end{proposition}

\begin{proof}
Part (i) follows from Lemma~\ref{lem:SUDP}, which implies that, for
each $\ell \in \{1,\dots, L\}$, $\udponAl$ is strictly monotone with a
derivative that is non-zero on $A_\ell$. Because $\udponAl$ is continuously differentiable by assumption, $\udponAl^{-1}$ is also continuously differentiable.


Part (ii) follows at once because $\udp$ is regular.

Part (iii) follows from part (ii) of Corollary~\ref{prop:Scalc-formulas}; the null set of points in $ \{
  \udp(u) : u < a, u \in A\}$ for which the statement doesn't hold is thus at most finite. Indeed, set $\udp\vert_a$ to be the restriction of $\udp$ to
 $[0,a]$. By (i), $\udp\vert_a$ satisfies Assumptions ~\ref{as:Spiecewise-monotonic} and~\ref{as:Scon-diff-inverse} with the same partition of size 
 size $L_a \le L$. Furthermore, $U$ conditionally on $U \in [0,a]$ is uniform on $[0,a]$, we have that for $W \sim \mathcal{U}(0,a)$, $F_{\udp\vert_a (W)}(x)=\P(\udp(U) \le x | U \le a) = \P(\udp\vert_a (W) \le x)$.

Finally, to show part (iv), note first that by the definition of $B$ and part (i) of Proposition~\ref{prop:ST-partition}, $(0,1) \setminus \{T_{-}(a_{\ell}), T_{+}(a_{\ell-1}), \ell \in \{1,\dots, L\}\} = B$. Because $B \subseteq T(A) \subseteq (0,1)$ by Proposition~\ref{prop:ST-partition}, we conclude that $B = T(A) \setminus \{T_{-}(a_{\ell}), T_{+}(a_{\ell-1}), \ell \in \{1,\dots, L\}\}$. Consequently, part (ii) of Corollary~\ref{prop:Scalc-formulas} shows that for all $x \in B$, the derivative $F_{\udp(U)}^\prime(x)$ is given by $\sum_{i = 1}^{n(x)} 1/|\udp^\prime(r_i(x))|$. At the same time, $F_{\udp(U)}(x) = x$ for all
 $x \in [0,1]$, since $\udp$ is uniform-distribution preserving. Hence
 $F_{\udp(U)}^\prime(x) = 1$ for all $x \in (0,1)$. 
\end{proof}

\section{Proofs}\label{sec:Sproofs}

This section contains the proofs organized by section of the
paper. 

\subsection{Proofs from Section~\ref{sec:2}} \label{sec:SS2}

\begin{proof}[Proof of Proposition \ref{prop:HardyKrause}.]
Observe that the function given, for all $u,v \in [0,1]$ by $\gtrans(u)\htrans(v)$ has bounded variation in the sense of Hardy--Krause \citep{Clarkson/Adams:1933}. This follows easily from the fact that the derivatives $\gtrans^\prime$ and $\htrans^\prime$ are bounded and continuous on $(0,1)$. Moreover, any copula  $C$ is continuous and grounded, i.e., $C(0,v) = C(u,0) = 0$ for any $u,v \in [0,1]$. By integration by parts for Lebesgue--Stieltjes integrals \citep{Young:1917,Ansari:2024}, we can rewrite $\rho_{\{\gtrans,\htrans\}}$ as
\begin{align*}
 \int_0^1 \gtrans(u) \htrans(v) \rd C(u,v) & = \gtrans(0) \htrans(0)C(1,1) + \int_0^1 (1-u) \rd \gtrans(u)\htrans(0) + \int_0^1 (1-v) \rd \gtrans(0)\htrans(v) \\
 & \phantom{=} +  \int_0^1 \int_0^1\{ C(1,1) - C(u,1) - C(1,v) + C(u,v)\} \rd \gtrans(u) \rd \htrans(v).
\end{align*}
Since $C(1,1) = 1$, the first term on the right-hand side is simply $\gtrans(0) \htrans(0)$. As for the second and third terms, note that
\begin{equation}\label{eq:Sibpunivariate1}
\int_0^1 \rd \gtrans(u) = \gtrans(1) - \gtrans(0), \quad \int_0^1 \rd \htrans(v) = \htrans(1) - \htrans(0), 
\end{equation}
while a renewed use of the integration by parts formula for Lebesgue--Stieltjes integrals \citep[Theorem 6.2.2]{Carter/vanBrunt:2000} gives
\begin{equation}\label{eq:Sibpunivariate2}
\int_0^1 u \rd \gtrans(u) = \gtrans(1) - \int_0^1 \gtrans(u) \rd u =  \gtrans(1), \quad \int_0^1 u \rd \htrans(u) = \htrans(1) - \int_0^1 \htrans(v) \rd v = \htrans(1)
\end{equation}
Gathering terms leads to
\[
\int_0^1 \int_0^1 (1-u) \rd \gtrans(u)\htrans(0) = \int_0^1 (1-v) \rd \gtrans(0)\htrans(v) = - \gtrans(0) \htrans(0).
\]
Finally, $C(1,1) - C(u,1) - C(1,v) + C(u,v)  =  1- u - v + C(u,v)$, while \eqref{eq:Sibpunivariate1} and \eqref{eq:Sibpunivariate2} give that
\begin{multline*}
\int_0^1 \int_0^1\{ 1- u - v + C(u,v)\} \rd \gtrans(u) \rd \htrans(v)  = \{\gtrans(1) - \gtrans(0)\}\{\htrans(1) - \htrans(0)\}  \\ - \gtrans(1) \{\htrans(1) - \htrans(0)\} - \htrans(1)\{\gtrans(1) - \gtrans(0)\} + \int_0^1 \int_0^1 C(u,v) \rd \gtrans(u) \rd \htrans(v)
\end{multline*}
The expression on the right-hand side simplifies to
\[
\int_0^1 \int_0^1 C(u,v) \rd \gtrans(u) \rd \htrans(v) + \gtrans(0)\htrans(0) - \gtrans(1) \htrans(1).
\]
Gathering therms yields that $\rho_{\{\gtrans,\htrans\}}(X,Y)$ equals
\[
\int_0^1 \int_0^1 C(u,v) \rd \gtrans(u) \rd \htrans(v)  - \gtrans(1) \htrans(1) =\int_0^1 \int_0^1 C(u,v) \gtrans^\prime(u) \htrans^\prime(v) \rd u \rd v - \gtrans(1) \htrans(1),
\]
as was to be shown. Alternatively, this result follows from Theorem~1 in \cite{bib:cuadras-15}; see also \cite{bib:cuadras-02}.
\end{proof}

\begin{proof}[Proof of Lemma \ref{lem:1}.]
Equation~\eqref{eq:27} is derived by first writing
\begin{displaymath}
  c(u,v) =\sum_{j=0}^\infty \sum_{k=0}^\infty a_{jk} \orthobase_j(u) \orthobase_k(v)
\end{displaymath}
and then using the orthonormality of the functions $\orthobase_j$ to conclude that 
\begin{align*}
  a_{jk} & = \int_0^1 \int_0^1 \orthobase_j(u) \orthobase_k(v) c(u,v) \rd
  u \rd v \\
  &= \E\left(\orthobase_j(U)\orthobase_k(V)\right) =
  \begin{cases}
    1& \text{if $j=k=0$,}\\
    0& \text{ if one of $j$ or $k$ is 0,}\\
   \rho_{jk}^B(U,V)& \text{otherwise.}
    \end{cases}
  \end{align*}
  This concludes the argument.
\end{proof}

\begin{remark}\label{rem:Ssupp:expansionC}
Note that the expansion of the copula density in Lemma~\ref{lem:1} leads to an expansion of the copula itself, in which the basis correlations appear as the
coefficients of products of the integrals of the basis functions
$\orthobase_j$:
  \begin{equation*}
    \label{eq:S26}
    C(u,v) = uv + \sum_{j=1}^\infty \sum_{k=1}^\infty
    \rho_{jk}^B(U,V) \int_0^u \orthobase_j(x) \rd x \int_0^v \orthobase_k(y)
    \rd y.
  \end{equation*}
\end{remark}

\begin{proof}[Proof of Lemma \ref{lem:2}.]
Because $\mathcal{B}$ is a complete orthonormal system in $\mathcal{L}^2([0,1])$, $\gtrans$ and $\htrans$ can be expanded as
\[
g(u) = \sum_{j=0}^\infty \alpha_j(\gtrans)\orthobase_j(u), \quad h(v) = \sum_{k=0}^\infty \alpha_k(\htrans)\orthobase_k(v)
\]
By Assumption~\ref{assA1.new}, $\alpha_0(\gtrans) = \alpha_0(\htrans)
= 0$, so that the above summations effectively start at $j,k=1$. Moreover, $\sum_{j=1}^\infty
  \alpha_j(\gtrans)^2 = \sum_{j=1}^\infty
  \alpha_j(\htrans)^2 = \int_0^1 g^2(u) \rd u = \int_0^1 h^2(v) \rd v = 1$ by Parseval's Identity and Assumption~\ref{assA1.new}. Because the map $(g,h) \mapsto \int_0^1 \int_0^1 g(u) h(v) \rd C(u,v)$ defined on $\mathcal{L}^2([0,1]) \times \mathcal{L}^2([0,1])$ is continuous, we obtain that, as soon as $\gtrans$ and $\htrans$ satisfy the normalization Assumption~\ref{assA1.new}, 
\begin{align*}
\rho_{\{\gtrans,\htrans\}} (X,Y)& = \int_0^1 \int_0^1 \gtrans(u) \htrans(v) \rd C(u,v) \\
&= \int_0^1 \int_0^1 \lim_{n\to \infty} \sum_{j=1}^n \alpha_j(\gtrans)\orthobase_j(u) \sum_{k=1}^n \alpha_k(\htrans)\orthobase_k(v) \rd C(u,v) \\
& = \lim_{n\to \infty} \sum_{j=1}^n \sum_{k=1}^n \alpha_j(\gtrans) \alpha_k(\htrans)  \rho^B_{jk}(X, Y) =   \sum_{j=1}^\infty \sum_{k=1}^\infty \alpha_j(\gtrans) \alpha_k(\htrans)  \rho^B_{jk}(X, Y),
\end{align*}
as claimed. 
\end{proof}

 
 \begin{proof}[Proof of Proposition \ref{prop:legendre-perfect-dependence}.]
  Using the fact that for comonotonic  random variables the
  probability-transformed variables satisfy $(F_X(X),F_Y(Y))^\top \eqdis (U,U)^\top$ for a uniform random
  variable $U$ and~\eqref{eq:orthog} implies that
  \begin{displaymath}
    \rho^B_{jk}(X,
Y) = \E \left ( \orthobase_j(U) \orthobase_k(U) \right) = \int_0^1 \orthobase_j(u) \orthobase_k(u) \rd u = \delta_{jk}. 
\end{displaymath}
 Using the fact that for countermonotonic  random variables the
  probability-transformed variables satisfy $(F_X(X),F_Y(Y)) ^\top \eqdis (1-U,U)^\top$ for a uniform random
  variable $U$ together with property (iii) in Definition~\ref{def:naturalbasis}
  and the previous result implies that
  \begin{displaymath}
    \rho^B_{jk}(X,
    Y) = \E \left ( \orthobase_j(1-U) \orthobase_k(U) \right) = (-1)^j \E \left ( \orthobase_j(U) \orthobase_k(U) \right)
  = (-1)^j \delta_{jk}. 
  \end{displaymath}
\end{proof}

    \begin{proof}[Proof of Proposition \ref{prop:symmetry}.]
We prove only (i) since the other assertions follow easily from
similar arguments to (i). Property (iii) in Definition~\ref{def:naturalbasis} implies
\begin{align*}
  \rho^B_{jk}(X,Y) = \rho\left( \orthobase_j(U), \orthobase_k(V) \right) =
  \rho\left( \orthobase_j(U), \orthobase_k(1-V) \right) & = (-1)^k \rho\left(
    \orthobase_j(U), \orthobase_k(V) \right) \\ &= (-1)^k  \rho^B_{jk}(X,Y)
\end{align*}
and hence $\rho^B_{jk}$ must be zero if $k$ is odd.
\end{proof}

\subsection{Proofs from Section~\ref{sec:udp}}

\begin{proof}[Proof of Proposition \ref{prop:nonmonotonic-udp}.]
Under Assumption~\ref{assA2.new}, $\gtrans$ and $\htrans$ satisfy 
  Assumption~\ref{as:Spiecewise-monotonic} in
  Section~\ref{sec:Spiec-strictly-monot} and hence, by Proposition~\ref{prop:Sproperties-Fpsi}, the distribution functions $F_{\gtrans}$ and
$F_{\htrans}$ are continuous. The remainder of part (i) claiming that
$\udp_\gtrans$ and $\udp_\htrans$ are udp thus holds by the property of the
probability-integral transform. The same proposition also implies part (ii). Finally, part (iii) follows by the quantile transformation. In fact, because $\gtrans(U)$ and $\htrans(V)$ are continuous random variables by part (i), the random pair $(F^{-1}_g (T_g(U)), F^{-1}_h(T_h(V)))^\top =  (F^{-1}_\gtrans (F_\gtrans(\gtrans(U)), F^{-1}_\htrans(F_\htrans(\htrans(V)))^\top $ is equal to $(\gtrans(U), \htrans(V))^\top$ almost surely. 
\end{proof}

\begin{proof}[Proof of Proposition \ref{prop:udp-vtransform}.]
We first observe that $F_{\htrans}(x) = \htrans_r^{-1}(x) -
\htrans_\ell^{-1}(x)$ for $x \in \htrans([0,1])$. It follows that
\begin{displaymath}
  T_\htrans (u) =
  \begin{cases}
    \htrans_r^{-1} \circ \htrans_\ell(u) - u, & u \leq \delta,\\
    u - \htrans_\ell^{-1} \circ \htrans_r(u),& u > \delta,
    \end{cases}
  \end{displaymath}
  and this coincides with~\eqref{eq:v-transform} on setting $\Psi(x) =
(1- \htrans_r^{-1} \circ \htrans_\ell(\delta x))/(1-\delta)$. Moreover, $\Psi$ fulfills the conditions of a distribution
function on $[0,1]$. The
final assertion follows since symmetry about $\delta =
0.5$ implies $\htrans_\ell(x/2) = \htrans_r(1-x/2)$ for $x \in [0,1]$
and hence $\Psi(x) = x$ and~\eqref{eq:v-transform} is the
symmetric v-transform~$\udp_\vee$.
\end{proof}

\begin{proof}[Proof of Proposition \ref{prop:udp-orthosystem-regular}.]
Clearly, it suffices to show that $T_\gtrans$ is a regular udp transformation.
Assumptions~\ref{assA2.new} and~\ref{assA3.new} imply that Assumptions~\ref{as:Spiecewise-monotonic},~\ref{as:Sturning-points}
and~\ref{as:Scon-diff-inverse} in Section~\ref{sec:Sudp-theory} all hold for $\gtrans$ on the same
finite partition
$0 =a _0 < a_1 < \dots < a_{M-1} < a_M=1$, where
$a_1,\dots, a_{M-1}$ are the turning points of $\gtrans$. 
Since $F_\gtrans$ is continuous and
strictly increasing on the range of $\gtrans$ by
Proposition~\ref{prop:nonmonotonic-udp} and since $\gtrans$
is continuous, it follows that
$T_\gtrans = \pbase_{\gtrans} \circ \gtrans$ is 
continuous and piecewise monotonic with the same $M-1$
turning points $a_1 < \dots <
a_{M-1}$ as $\gtrans$. Moreover, Assumption~\ref{as:Scon-diff-inverse} implies that
$\gtrans\vert_{A_m}$, the restriction of $\gtrans$ to $A_m
= (a_{m-1}, a_m)$, has a continuously differentiable inverse for $m \in
\{1,\dots, M\}$ and we can use
Proposition~\ref{prop:Sdensity-transfo} to infer that that
$F_{\gtrans}$  has a continuous derivative on $R =
(\cup_{m=1}^M R_m) \setminus \{ g(a_m) : m = 0,\dots, M\}$ where $R_m = \gtrans\vert_{A_m}(A_m)$. Thus
$T_\gtrans$ is continuously differentiable on $(0,1)$ except
possibly at the points in the finite set $S = \{ u \in (0,1) :
\gtrans(u) = \gtrans(a_m), m \in \{0,\dots, M\}\}$. Thus we
can find a partition with$L = |S|+1$ for which $T_\gtrans$
satisfies the requirements of Definition~\ref{def:udp-regular}.
\end{proof}

\begin{proof}[Proof of Lemma \ref{lemma:tv}.]
$\orthobase_1$ and $\orthobase_2$ are regular and therefore continuous. By the natural-basis property (ii), $\orthobase_1$ has no turning point and is therefore strictly monotonic. By property (i) it is increasing near 1, hence increasing everywhere. Thus $\orthobase_1$ is a continuous, strictly increasing function
with unique inverse $\orthobase_1^{-1}$ and so we obtain
$\udp_1^B(u) = \pbase_{\orthobase_1(U)}(\orthobase_1(u)) =
\P( \orthobase_1(U) \leq \orthobase_1(u)) = \P(U \leq
\orthobase_1^{-1}(\orthobase_1(u))) = u$. By properties (ii) and (iii) of a natural basis function, $\orthobase_2$ is even about a single turning point at 0.5 with $\orthobase_2(0) = \orthobase_2(1)$. By property (i) it is decreasing on $[0,0.5]$ and increasing on $[0.5,1]$. Hence Proposition~\ref{prop:udp-vtransform} applies with $\delta =0.5$ and symmetric $h$, giving $T_2^\orthobase = T_\vee$.
 \end{proof}

\begin{proof}[Proof of Theorem \ref{prop:nonmonotonic-bounds}.]
From the classical work of~\citet{bib:hoeffding-40}
 and~\cite{bib:frechet-57} (see
also~\citet{bib:mcneil-frey-embrechts-15}, Theorem 7.28, page 241) we
know that, for random variables $W$ and $Z$ with fixed marginal
distributions and unspecified joint distribution, the attainable correlations $\rho\left(W,Z\right)$ form a closed interval
$[\rho_{\min},\rho_{\max}]$ with $\rho_{\min} < 0 < \rho_{\max}$. The
upper bound
$\rho_{\max}$ corresponds to the case where $W$ and $Z$ are comonotonic variables and
the lower bound $\rho_{\min}$ corresponds to the case of countermonotonic variables. 

Considering Proposition~\ref{prop:nonmonotonic-udp}(iii), we apply these ideas to the random
variables $W = F_{\gtrans}^{-1}(T_\gtrans(U))$ and $Z =
F_{\htrans}^{-1}(T_\htrans(V))$ where $U =
                    F_X(X)$ and $V = F_Y(Y)$. Comonotonicity of
                    $(W,Z)^\top$ means that their dependence structure
                    is the Fr\'{e}chet upper bound copula so that
                    $\udp_\gtrans(U) \asequal   \udp_\htrans(V)$ and
                    hence the upper bound in~\eqref{eq:24a} is obtained
                    by evaluating the expectation
                    in~\eqref{eq:53} with the random variables
                    $T_\gtrans(U)$ and $T_\htrans(V)$ replaced by a
                    single common uniform random variable, say $\tilde
                    U$. Countermonotonicity of
                    $(W,Z)^\top$ means that their dependence structure
                    is the Fr\'{e}chet lower bound copula so that
                    $\udp_\gtrans(U) \asequal   1- \udp_\htrans(V)$ and
                    hence the lower bound in~\eqref{eq:24a} is obtained
                    by evaluating the expectation expression
                    in~\eqref{eq:53} with $T_\gtrans(U) = \tilde U$ and
                    $T_\htrans(V) = 1-\tilde U$ for $\tilde U$ standard uniform. The
                    values 1 when $F_{\gtrans} = F_{\htrans}$
                    and -1 when $F_{\htrans}^{-1}(1-u) = -
F_{\gtrans}^{-1}(u)$ follow because Assumption~\ref{assA1.new}
implies $ \int_0^1 (F_{\gtrans}^{-1}(u))^2 \rd u =1$.
\end{proof}

\subsection{Proofs from Section~\ref{sec:bounds}}

\begin{proof}[Proof of Theorem~\ref{theorem:stochinverse}.]
Because $V$ and $Z$ are independent and $G_v^{-1}(Z)$ is a random variable with distribution function $G_v$, 
\[
\P(\udp^\leftarrow(V,Z) < u) = \E( \P(\udp^\leftarrow(V,Z) < u | V)) =  \int_0^1 \P(G_v^{-1}(Z) < u) \rd v =  \int_0^1 G_v(u-) \rd v
\]
for all $u \in (0,1)$. That this integral is indeed equal to $u$ can be seen as follows. Suppose that $U \sim \mathcal{U}(0,1)$ and set $W = \udp(U)$. Because $\udp$ is  a udp transformation, $W \sim \mathcal{U}(0,1)$, so the joint distribution of $(U,W)$ is a copula, say $C$. Because $\udp$ is regular, part (iii) of Proposition~\ref{prop:Supd-properties} shows that $C(u,w) = F_u(w)$ and that for all $u \in (0,1)$ and almost all $w$,
\[
\frac{\partial C(u,w)}{\partial w}  = \sum_{i = 1}^{n(w)} \left |\frac{1}{\udp^\prime(r_i(w))} \right | \Ind{r_i(w) < u} = G_w(u-).
\]
Because the margins of $C$ are uniform, we get that
\[
u = \int_0^1 \frac{\partial C(u,w)}{\partial w} \rd w =  \int_0^1 G_w(u-) \rd w,
\]
and hence $\udp^\leftarrow(V,Z)$ is uniform, as claimed.
\end{proof}

\begin{proof}[Proof of Corollary~\ref{cor:stochinversemult}.]
  Uniformity of the margins comes from Theorem~\ref{theorem:stochinverse} applied componentwise. Moreover, for each $j$,
  $\udp_{j}(\udp_{j}^\leftarrow(V_j,Z_j)) = V_j$, almost surely, since $V_j$ lies in the $T$-partition set of $\udp_j$, almost surely. Hence all events hold simultaneously, almost surely.
\end{proof}

 \begin{proof}[Proof of Corollary~\ref{sec:mult-stoch-invers-2}.]
  By Theorem~\ref{theorem:stochinverse},  $U \sim
  \mathcal{U}(0,1)$, $V \sim \mathcal{U}(0,1)$ and $\tilde V \sim
  \mathcal{U}(0,1)$ and hence the distributions
  of $(U,V)^\top$ and $(U, \tilde V)^\top$ are copulas. By Corollary~\ref{cor:stochinversemult},
  $(\udp_\gtrans(U),\udp_\htrans(V))^\top \asequal (U^*,U^*)^\top$ so that
  $(\udp_\gtrans(U),\udp_\htrans(V))^\top$ is a comonotonic pair
  giving the maximum generalized Spearman correlation. Similarly
  $(\udp_\gtrans(U),\udp_\htrans(\tilde V))^\top
  \asequal (U^*,1- U^*)^\top $  so that
  $(\udp_\gtrans(U),\udp_\htrans(\tilde V))^\top$ is a countermonotonic pair
  giving the minimum generalized Spearman correlation.
  \end{proof}

\begin{proof}[Proof of Proposition~\ref{prop:mult-stoch-invers-symmetry}.]
  We show only (i) since the other parts follow easily.
If $\udp_{1}$ is symmetric around 0.5, its stochastic inverse
(see Definition~\ref{def:si}) has the property that
$\udp_{1}^\leftarrow(x, 1-Z) = 1 - \udp_{1}^\leftarrow(x,Z)$ and hence
$\udp_{1}^\leftarrow(x, Z) \eqdis 1 - \udp_{1}^\leftarrow(x,Z)$. It follows by independence of $Z_1$ and $Z_2$ that
\begin{align*}
  \P(U \leq u, V \leq v \mid U^* = x) 
&=  \P(\udp_{1}^\leftarrow(U^*, Z_1) \leq u\mid U^* = x) 
                                        \P(  \udp_{2}^\leftarrow(V^*, Z_2)   \leq v \mid U^* = x) \\
&=  \P(1-\udp_{1}^\leftarrow(U^*, Z_1) \leq u\mid U^* = x) 
                                         \P(V \leq v \mid U^* = x) \\
                                        &=  \P(1-U \leq u, V \leq v \mid U^* = x)
\end{align*}
from which we can conclude that $(U,V) \eqdis (1-U,V)$, which is v-symmetry.
\end{proof}

 \subsection{Proofs from Section~\ref{sec:estim-gener-spearm}}

Before proving Theorem~\ref{thm:asympt_distribution}, we will need to make a few preliminary observations. To this end, let $\gtrans$ and $\htrans$ be some arbitrary transformations that satisfy Assumptions~\ref{assA1.new} and \ref{assA3.new}. 

Assumption~\ref{assA3.new} guarantees that 
\begin{equation}\label{eq:Sbounds-in-proof}
\begin{split}
\sup_{x \in [0,1]} | \gtrans(x)| & =  a_\gtrans < \infty,  \quad \sup_{x \in (0,1)} | \gtrans^\prime(x)|  =  b_\gtrans < \infty, \\
\sup_{x \in [0,1]} | \htrans(x)|  & = a_\htrans < \infty,    \quad \sup_{x \in (0,1)} | \htrans^\prime(x)|  = b_\htrans < \infty.
\end{split}
\end{equation}
Next, introduce the auxiliary statistic
\[
\check \rho_{\{\gtrans,\htrans\}} = \frac{1}{n} \sum_{i=1}^n \gtrans \left( \frac{R_i}{n+1}\right) \htrans\left( \frac{S_i}{n+1}\right)
\]
and let $e_{\gtrans, n}$ and $v_{\gtrans, n}$ denote the sample mean and variance of $\gtrans(R_i/(n+1))$, $i \in \{1,\dots, n\}$, respectively, and similarly for $e_{\htrans, n}$ and $v_{\htrans, n}$. With these notations, we can write
\begin{equation}\label{eq:SB(3)alternative}
\hat{\rho}_{\{\gtrans,\htrans\}} = \frac{\check{\rho}_{\{\gtrans,\htrans\}} - e_{\gtrans,n} e_{\htrans, n}}{\sqrt{v_{\gtrans, n} v_{\htrans, n}}}.
\end{equation}
With probability one, we have that 
\[
e_{\gtrans, n} = \frac{1}{n} \sum_{i=1}^n \gtrans\left(\frac{i}{n+1}\right) = \frac{n+1}{n(n+1)} \sum_{i=1}^{n+1} \gtrans\left(\frac{i}{n+1}\right) - \frac{1}{n} \gtrans(1).
\]
Obviously, $ \gtrans(1) (\sqrt{n}/n) \to 0$ as $n \to \infty$. Moreover, $(1/(n+1))
\sum_{i=1}^{n+1}\gtrans(i/(n+1))$ is a Riemann sum, which,
because $\gtrans$ is Riemann integrable, converges to
$\int_0^1 \gtrans(u) \rd u$ as $n \to \infty$. By Assumption~\ref{assA1.new}, this integral is $0$. Because
$\gtrans$ is of bounded variation, we even have that
$(1/(n+1)) \sum_{i=1}^{n+1}\gtrans(i/(n+1)) = O(1/n)$. This
result, attributed to P\'olya and Szeg\H{o}, can be found, e.g., in \citep[Theorem~1 (b)]{Chui:1971}. From this we obtain that $e_{\gtrans, n} = O(1/n)$ almost surely. Similarly, with probability one,
\[
v_{\gtrans, n} =  \frac{1}{n} \sum_{i=1}^n \gtrans^2\left(\frac{i}{n+1}\right) -  e_{\gtrans, n}^2.
\]
Because $\gtrans^2$ is absolutely continuous and Assumption~\ref{assA1.new} holds, Theorem~1 (b) of \cite{Chui:1971} implies that $(1/(n+1)) \sum_{i=1}^{n+1}\gtrans^2(i/(n+1)) - 1 = O(1/n)$, which leads to $v_{\gtrans, n} - 1 = O(1/n)$. Analogously, we conclude that $e_{\htrans, n} = O(1/n)$ as well as $v_{\htrans, n} - 1 = O(1/n)$ almost surely.  Using \eqref{eq:SB(3)alternative} we obtain
\begin{align*}
\sqrt{n}(\hat{\rho}_{\{\gtrans,\htrans\}} - \check{\rho}_{\{\gtrans,\htrans\}}) & = \frac{\sqrt{n}(\check{\rho}_{\{\gtrans,\htrans\}} ( 1- \sqrt{v_{\gtrans, n} v_{\htrans, n}}) - e_{\gtrans, n} e_{\htrans, n})}{\sqrt{v_{\gtrans, n} v_{\htrans, n}}} \\
& = \check{\rho}_{\{\gtrans,\htrans\}} \frac{\sqrt{n}( 1- \sqrt{v_{\gtrans, n} v_{\htrans, n}})}{\sqrt{v_{\gtrans, n} v_{\htrans, n}}} - \frac{\sqrt{n} (e_{\gtrans, n} e_{\htrans, n})}{\sqrt{v_{\gtrans, n} v_{\htrans, n}}}
\end{align*}
From \eqref{eq:Sbounds-in-proof}, $| \check{\rho}_{\{\gtrans,\htrans\}}| \le a_\gtrans a_\htrans$, while $\sqrt{n}(e_{\gtrans, n} e_{\htrans, n})/\sqrt{v_{\gtrans, n} v_{\htrans, n}} \to 0$ almost surely, as well as 
\[
\frac{\sqrt{n}( 1- \sqrt{v_{\gtrans, n} v_{\htrans, n}})}{\sqrt{v_{\gtrans, n} v_{\htrans, n}}} = \frac{\sqrt{n}( 1- v_{\gtrans, n} v_{\htrans, n})}{\sqrt{v_{\gtrans, n} v_{\htrans, n}}(1+\sqrt{v_{\gtrans, n}v_{\htrans, n}})} = \frac{\sqrt{n}( 1- v_{\gtrans, n} + v_{\gtrans, n}(1-v_{\htrans, n}))}{\sqrt{v_{\gtrans, n} v_{\htrans, n}}(1+\sqrt{v_{\gtrans, n}v_{\htrans, n}})} \to 0
\]
almost surely. Hence, 
\begin{equation}\label{eq:Sequiv_conv_check}
\sqrt{n}(\hat{\rho}_{\{\gtrans,\htrans\}} - \check{\rho}_{\{\gtrans,\htrans\}}) \to 0 \quad \text{almost surely},
\end{equation}
 which means that we can study the asymptotic behaviour of $\check{\rho}_{\{\gtrans,\htrans\}}$ instead of $\hat{\rho}_{\{\gtrans,\htrans\}}$ without loss of generality. Next, recall the empirical copula given, for all $u,v \in [0,1]$, by
\[
C_n(u,v) = \frac{1}{n} \sum_{i=1}^n \Ind{R_i/n \le u, S_i/n \le v}.
\]
and its bilinear interpolation $C_n^\maltese$, which makes it a proper copula \cite{Deheuvels-Raoult:1979,Genest/Neslehova/Remillard:2014, Genest/Neslehova/Remillard:2017}. In a sample with no ties, it is easiest to define it through its density $c_n^\maltese$, which takes the value $n$ on any rectangle of the form $((R_i-1)/n, R_i/n] \times ((S_i -1)/n, S_i/n]$ for $i \in \{1,\dots, n\}$, and zero elsewhere. Now introduce another, final, auxiliary statistic
\begin{align}\label{eq:SB(5)}
\tilde{\rho}_{\{\gtrans,\htrans\}}& =   \int_0^1 \int_0^1 \gtrans(u) \htrans(v) dC^\maltese_n(u,v) \\
& = n \sum_{i=1}^n \left( I_\gtrans\left(\frac{R_i}{n}\right) -
        I_\gtrans\left( \frac{R_i -1}{n}\right)\right) \left( I_\htrans\left(\frac{S_i}{n}\right) -
        I_\htrans\left( \frac{S_i -1}{n}\right)\right) \notag
\end{align}
where for any $x \in [0,1]$, $I_\gtrans(x) = \int_0^x \gtrans(u) \rd
u$ and similarly $I_\htrans(x) = \int_0^x \htrans(u) \rd
u$. For any integer $r \in \{1,\dots, n\}$ and $\ell \in \{j,k\}$, the Mean Value Theorem gives that $n\{I_\gtrans(r/n) - I_\gtrans((r-1)/n)\} = \gtrans(t)$, for some $t \in ((r-1)/n, r/n)$. A renewed application of the Mean Value Theorem thus implies that $|n\{I_\gtrans(r/n) - I_\gtrans((r-1)/n)\} - \gtrans(r/n)| \le b_\gtrans/n$.   Now for any integer $r \in \{1,\dots, n\}$, $r/n - r/(n+1) = r/n(n+1) \le 1/(n+1)$. Hence, yet another application of the Mean Value Theorem and \eqref{eq:Sbounds-in-proof} gives that 
\[
\left |n \left \{I_\gtrans\left(\frac{r}{n}\right) - I_\gtrans\left(\frac{r-1}{n}\right)\right\} - \gtrans\left(\frac{r}{n+1}\right)\right| \le b_\gtrans \left( \frac{1}{n} + \frac{1}{n+1}\right) = b_\gtrans \frac{2n+1}{n(n+1)}.
\]
These observations allow us to calculate that
\begin{align*}
|\tilde{\rho}_{\{\gtrans,\htrans\}}- \check{\rho}_{\{\gtrans,\htrans\}}| & \le  \frac{1}{n} \sum_{i=1}^n  \left| n^2\left( I_\gtrans\left(\frac{R_i}{n}\right) -
        I_\gtrans\left( \frac{R_i -1}{n}\right)\right) \left( I_\htrans\left(\frac{S_i}{n}\right) -
        I_\htrans\left( \frac{S_i -1}{n}\right)\right) \right.\\
        & \phantom{=} \qquad \left. - \gtrans\left( \frac{R_i}{n+1}\right) \htrans\left( \frac{S_i}{n+1}\right)\right|\\
& \le  \frac{1}{n} \sum_{i=1}^n \left\{ a_\htrans \left| n\left( I_\gtrans\left(\frac{R_i}{n}\right) -
        I_\gtrans\left( \frac{R_i -1}{n}\right)\right) - \gtrans\left( \frac{R_i}{n+1}\right)\right| \right. \\
        & \phantom{=} \qquad \left. + a_\gtrans\left| n\left( I_\htrans\left(\frac{S_i}{n}\right) -
        I_\htrans\left( \frac{S_i -1}{n}\right)\right) - \htrans\left( \frac{S_i}{n+1}\right)\right| \right\}\\
& \le \frac{(a_\htrans b_\gtrans + a_\gtrans b_\htrans)(2n+1)}{n(n+1)}.
\end{align*}
Clearly, the latter inequality implies that as $n\to \infty$,
\begin{equation}\label{eq:Sequiv_conv_tilde}
\sqrt{n}(\check{\rho}_{\{\gtrans,\htrans\}} - \tilde{\rho}_{\{\gtrans,\htrans\}}) \to 0 \quad \text{almost surely}.
\end{equation}

\begin{proof}[Proof of Theorem~\ref{thm:asympt_distribution}.]

In view of \eqref{eq:Sequiv_conv_check} and \eqref{eq:Sequiv_conv_tilde}, we can work with $\tilde{\boldsymbol{\rho}}=(\tilde \rho_{\{\gtrans_1,\htrans_1\}},\dots, \tilde \rho_{\{\gtrans_J,\htrans_J\}})$ instead of $\hat{\boldsymbol{\rho}}$. 
Because $C_n^\maltese$ is itself a copula, Proposition~\ref{prop:HardyKrause} allows us to write
\[
\sqrt{n}(\tilde{\boldsymbol{\rho}} - \boldsymbol{\rho}) = \left( \int_0^1 \int_0^1 \mathbb{C}_n^\maltese(u,v) \gtrans^\prime_{j}(u) \htrans^\prime_{j}(v) \rd u \rd v, j \in \{1,\dots, J\}\right).
\]
Under Assumption~\ref{assumption:copula}, we also have that $\mathbb{C}_n^\maltese=\sqrt{n}(C_n^\maltese - C)$ converges weakly to $\mathbb{C}$, because $\|C_n- C_n^\maltese\| \le d/n$ almost surely \citep[Remark~2]{Genest/Neslehova/Remillard:2017}.  The right-hand side of the displayed expression above is a continuous functional of  $\mathbb{C}_n^\maltese$, and hence converges weakly to $\boldsymbol{Z}$ by the Continuous Mapping Theorem. Because the latter functional is also linear and $\mathbb{C}$ is tight, the limit $\boldsymbol{Z}$ is multivariate Normal~\citep[Lemma~3.9.8]{Vaart/Wellner:1996}. Interchanging expectation and integration, we can easily see that $\boldsymbol{Z}$ has  zero mean, given that $\mathbb{C}$ is a zero-mean Gaussian process.
\end{proof}



 

\section{Additional material}\label{sec:Sadditional-material}

\subsection{Calculating $F_j^\orthobase$ and $f_j^\orthobase$ for
  cosine and Legendre bases}\label{sec:ScalcFjfj}

The calculation is particularly simple for the cosine basis and
follows by application of Proposition~\ref{prop:Sdensity-transfo} to
the basis functions upon setting $\psi = \ocos_j$ and replacing $W \sim
\mathcal{U}(0,a)$ with $U \sim\mathcal{U}(0,1)$. 

\begin{example}\label{ex:Scosine-pbase}
The cosine basis function $\ocos_j(u) = (-1)^j \sqrt{2}\cos(j \pi u)$
satisfies Assumptions~\ref{as:Spiecewise-monotonic} and ~\ref{as:Scon-diff-inverse} with a partition in
which $a=1$, $M=j$,
 $A_m = ((m-1)/j,m/j)$ for $m=1,\ldots,j$ and $R_m =\ocos_j\vert_{A_m}(A_m) =
 (-\sqrt{2},\sqrt{2})$. For $x \in R_m$ we have $I_x \cup D_x =
 \{1,\ldots,j\}$. Moreover,
 $\ocos_j\vert_{A_m}^{-1}(x) = \arccos(x/\sqrt{2})/(j\pi)
+ (m-1)/j$ or $\ocos_j\vert_{A_m}^{-1}(x) =  m/j - \arccos(x/\sqrt{2})/(j\pi)$, depending on the
parity of $m$ and $j$. Either way, the absolute value of the
derivative of the inverse is $1/(j \pi\sqrt{2-x^2})$, which is
independent of $m$.
Summing up over $m$, we obtain 
$\dbase^{\ocos}_j(x)  = 1/(\pi\sqrt{2-x^2})$, which is independent
of $j$. The corresponding distribution function is
$\pbase^{\ocos}_j(x)  = 1-\pi^{-1}\arccos(x/\sqrt{2})$. The insight 
that the distribution functions $\pbase^{\ocos}_j$ are identical
for all $j$ can also be obtained directly by observing that
$\cos(j\pi U ) \eqdis \cos(\pi U)$ for all $j \in
 \mathbb{N}$ and $U \sim \mathcal{U}(0,1)$.
\end{example}

\noindent In order to derive equivalent expressions in the Legendre
case, we use Corollary~\ref{prop:Scalc-formulas}, noting that
any basis function $\orthobase_j$ of a natural correlation basis 
fulfills all of
Assumptions~\ref{as:Spiecewise-monotonic},~\ref{as:Sturning-points}
and~\ref{as:Scon-diff-inverse} for the same partition in which
$M=j$

\begin{example}\label{ex:Slegendre-pbase}
We use~\eqref{eq:S62} and~\eqref{eq:S63}  to calculate the distribution
functions $\pbase^{\blegendre}_j$ and densities $\dbase^{\blegendre}_j$ for the Legendre
polynomial basis for $j = 2,\ldots,6$, as shown in
Figure~\ref{fig:SFplotsLegendre}; note that $\pbase^{\blegendre}_1(x) =
0.5(x/\sqrt{3} +1)$ which has constant density.
In the application of the distribution function formula in~\eqref{eq:S62}, we have $M \not
\in D_x \cup O_x$ for $x \in R$, since the basis functions $\blegendre_j(u)$
increase to their maximum values at $u=1$.
  \begin{figure}[htb]
  \centering
  \includegraphics[width=16cm,height=8cm]{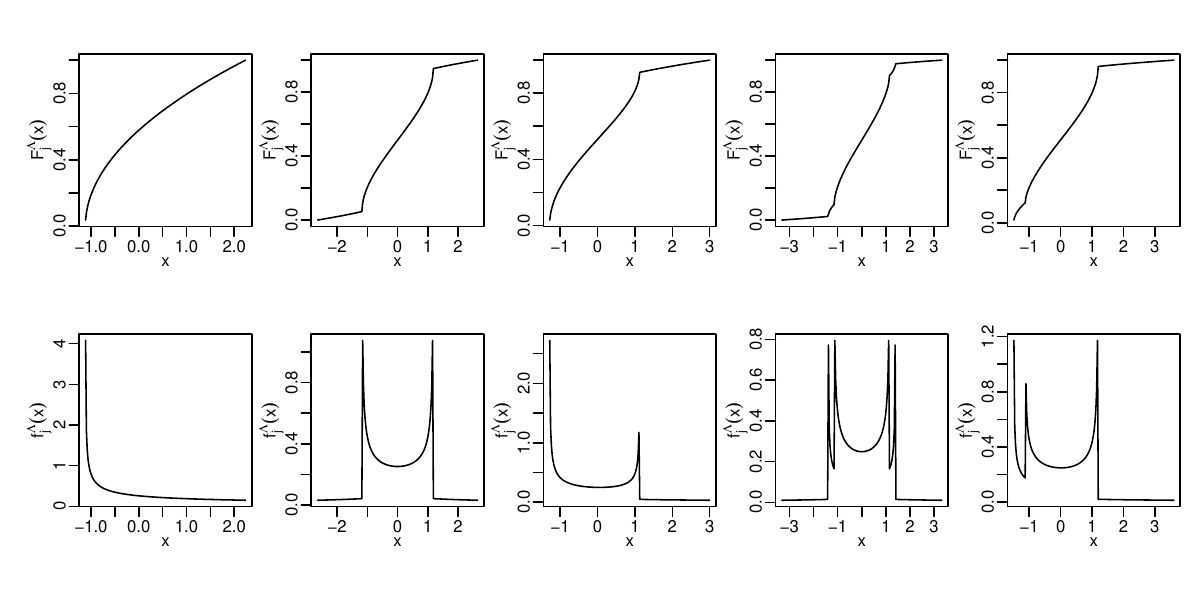}
   \caption{\label{fig:SFplotsLegendre} Plots of $\pbase_j^\blegendre$
     (upper row) and $\dbase^\blegendre_{j}$ (lower row) for the Legendre basis
     and $j = 2,\ldots,6$.}
 \end{figure}
\end{example}

\subsection{A further insight into the support sets
  $\mathsf{U}_{\{g,h\}}$ and
  $\mathsf{L}_{\{g,h\}}$}\label{sec:Ssupplement-support-sets}

\begin{proposition}\label{prop:Ssupport-sets}
Under   Assumptions~\ref{assA2.new} and~\ref{assA3.new}, let
   $0 =a_0 < a_{\gtrans, 1} < \cdots < a_{\gtrans, M_\gtrans} = 1$ be the partition of the
   unit interval defined by the ordered turning points of $\gtrans$ so that
   $M_\gtrans \in \N$ is one larger than the number of turning
   points. Let $0 =a_0 < a_{\htrans, 1} < \cdots < a_{\htrans,
     M_\htrans} = 1$ be the analogous partition defined by the ordered turning points of $\htrans$.
   For any rectangle $(a_{\gtrans,m_1-1}, a_{\gtrans,m_1}) \times (a_{\htrans,m_2-1},
  a_{\htrans,m_2})$ with $m_1 \in \{1,\dots, M_\gtrans\}$ and $m_2 \in \{1,\dots, M_\htrans\}$,
   if the intersection of the rectangle with the support set of the maximum or minimum copula is
   non-empty, the support set forms a continuous and strict monotonic
   curve on the rectangle.
 \end{proposition}

 \begin{proof}
Set
  $A_{\gtrans,m} = (a_{\gtrans,m-1}, a_{\gtrans,m})$ for $m \in \{1,\dots, M_\gtrans\}$, let
  $\udp_\gtrans\vert_{A_{\gtrans,m}}$ be the restriction of
    $\udp_\gtrans = F_\gtrans \circ \gtrans$ to $A_{\gtrans,m}$ and
    let $R_{\gtrans,m} =
    \udp_\gtrans\vert_{A_{\gtrans,m}}(A_{\gtrans,m})$. Use 
  analogous notation for the corresponding objects induced by the
  partition for $\htrans$. From Proposition~\ref{prop:nonmonotonic-udp}, $\udp_\gtrans$ and $\udp_\htrans$ are continuous and, as argued in the proof of Proposition~\ref{prop:udp-orthosystem-regular}, $\udp_\gtrans\vert_{A_{\gtrans,m}}$ and $\udp_\htrans\vert_{A_{\htrans,m}}$ are strictly monotonic.

$\mathsf{U}_{\{g,h\}}$ intersects the
  rectangular set only if $S_{m_1 m_2} = R_{\gtrans,m_1} \cap
 R_{\htrans,m_2}\neq \emptyset$ where we note that $S_{m_1,m_2}$ is an
 interval in $(0,1)$. In this case, since the functions $\udp_\gtrans\vert_{A_{\gtrans,m_1}}$
  and $\udp_\htrans\vert_{A_{\htrans,m_2}}$ are continuous and
  strictly monotonic, the
  support set in this rectangle is the curve $\{(\udp_\gtrans\vert_{A_{\gtrans,m_1}}^{-1}(s),
  \udp_\htrans\vert_{A_{\htrans,m_2}}^{-1}(s)) : s \in S_{m_1 m_2}\}$
and this can be reparameterized as $\{(u,
  \udp_\htrans\vert_{A_{\htrans,m_2}}^{-1} \circ
  \udp_\gtrans\vert_{A_{\gtrans,m_1}}(u)) : \udp_\gtrans\vert_{A_{\gtrans,m_1}}(u)
  \in S_{m_1 m_2}\}$ which describes a continuous and strictly
  monotonic curve on the sub-interval of $A_{\gtrans,m_1}$ that is
  mapped to $S_{m_1,m_2}$ by $T_\gtrans$.

$\mathsf{L}_{\{g,h\}}$ intersects the
  rectangular set only if $S_{m_1 m_2} = R_{\gtrans,m_1} \cap
 \overline{R}_{\htrans,m_2}\neq \emptyset$ where
 $\overline{R}_{\htrans,m_2} = 
 \overline{\udp}_{\htrans}\vert_{A_{\htrans,m_2}}(A_{\htrans,m_2})$
  and $\overline{\udp}_{\htrans}\vert_{A_{\htrans,m_2}}(u) = 1-
  \udp_{\htrans}\vert_{A_{\htrans,m_2}}(u)$ for $u \in A_{\htrans,m_2}$. In this case, the
  support set in this rectangle is the curve $\{(\udp_\gtrans\vert_{A_{\gtrans,m_1}}^{-1}(s),
  \overline{\udp}_\htrans\vert_{A_{\htrans,m_2}}^{-1}(s)) : s \in S_{m_1 m_2}\}$
and this can be reparameterized as $\{(u,
  \overline{\udp}_\htrans\vert_{A_{\htrans,m_2}}^{-1} \circ
  \udp_\gtrans\vert_{A_{\gtrans,m_1}}(u)) : \udp_\gtrans\vert_{A_{\gtrans,m_1}}(u)
  \in S_{m_1 m_2}\}$ which again describes a continuous and strictly
  monotonic curve on the sub-interval of $A_{\gtrans,m_1}$ that is
  mapped to $S_{m_1,m_2}$ by $T_\gtrans$.
\end{proof}
 
\subsection{Two new copulas maximizing $\rho_{44}^\blegendre$}\label{sec:Srho44}

This section contains two examples of copulas maximizing
$\rho_{44}^\blegendre$. 
The first example has an explicit form, is jointly symmetric, and has support $\mathsf{U}^\blegendre_{44}$.
The second example is not jointly symmetric and is supported on a subset of 
$\mathsf{U}^\blegendre_{44}$ describing a
 ``prohibition sign''. Neither copula corresponds to the case of independent stochastic inversion.

\begin{example}\label{ex:Sconstr-extr-copul}
  We first observe that the mapping $(u,v) \mapsto (T_\vee(u),T_\vee(v))$ maps the support set $\mathsf{U}^\blegendre_{44}$ onto the set $\mathsf{Q}$ consisting of the line
line $v=u$ and the quarter circle with radius $R=2r$ where $r =
\sqrt{3/14}$ is the radius of the circular part of $\mathsf{U}^\blegendre_{44}$.
 We construct a random vector $(U^*, V^*)^\top$ distributed according to a copula $C^*$
 supported on $\mathsf{Q}$. The copula we construct is not the unique copula on $\mathsf{Q}$, but it has a tractable form.
If $Z_1$ and $Z_2$ are independent uniform variables, independent of  $(U^*, V^*)^\top$, then the random vector
  $(U,V)^\top = (\udp_\vee^\leftarrow(U^*, Z_1), \udp_\vee^\leftarrow(V^*, Z_2))^\top$ is supported on $\mathsf{U}^\blegendre_{44}$ and is jointly symmetric by Proposition~\ref{prop:mult-stoch-invers-symmetry}. Moreover, it is easy to verify that it has distribution function 
  \begin{displaymath}
  C(u,v) =
    \frac{2u+2v -1 + (-1)^{\indicator{(u-0.5)(v-0.5) < 0 }} C^*(\vtrans(u), \vtrans(v))}{4}.
  \end{displaymath}
 Although $C$ attains the bound $\overline{\rho}^\blegendre_{44}=1$, we will show that it cannot be the distribution function of  $((\udp_4^\blegendre)^\leftarrow(W, Z_1),
 (\udp_4^\blegendre)^\leftarrow(W, Z_2))^\top$ for iid uniform random variables $W, Z_1$ and $Z_2$, so it does not correspond to the copula obtained by independent bivariate stochastic
 inversion.

We take $U^* \sim \mathcal{U}(0,1)$, define $p_R(x) = \tfrac{x}{R}\indicator{x \leq R}$ and set
    \begin{equation}\label{eq:circle_construct}
      V^* =
      \begin{cases} U^* & \text{with probability $1-p_R(U^*)$,}
        \\
           \sqrt{R^2 -U^{*2}} & \text{with probability
             $p_R(U^*)$.}
         \end{cases}
       \end{equation}
It is obvious that this construction distributes points on $\mathsf{Q}$  but we need to verify that $V^*$ is uniformly distributed. We do this by showing that the df
$C^*$ of $(U^*,V^*)^\top$ is
  \begin{equation}
    \label{eq:S15}
    C^*(u,v) =
 \min(u,v) +  \indicator{(u,v) \in
        [0,R]^2}  \frac{\max(u^2 + v^2 - R^2,0) - \min(u,v)^2}{2R} 
    \end{equation}
which must be a copula since
 $C^*(0,v) = C^*(u,0) = 0$ and $C^*(u,1) =u$
and $C^*(1,v) =v$. 

For a point $(u,v) \in [0,R]^2$ we have that
\begin{align*}
  \P\left(U^* \leq u, V^* \leq v\right) &=
                                          \int_0^u \P\left(V^* \leq v
                                          \mid U^* = x\right) \rd x \\
                                        &=
                                          \int_0^u\left( \frac{x}{R}
  \indicator{\sqrt{R^2 - x^2}\leq v} +
                                          \left(1-\frac{x}{R} \right)
                                          \indicator{x \leq v}\right) \rd x\\
  &= \int_0^u \frac{x}{R}
  \indicator{x \geq \sqrt{R^2 - v^2}} \rd x + \int_0^u
                                          \left(1-\frac{x}{R} \right)
                                          \indicator{x \leq v} \rd x \\
  &= \int_{ \sqrt{R^2 - v^2}}^{\max(u,\sqrt{R^2 - v^2})}\frac{x}{R}
    \rd x + \int_0^{\min(u,v)}  \left(1-\frac{x}{R} \right) \rd x 
\end{align*}
which is equal to $C^*(u,v)$ in~\eqref{eq:S15} when evaluated. For a
point $(u,v) \in [0,R] \times (R,1]$ we simply have $\P(V^* \leq v
\mid U^* = x) = 1$ for $x \in [0,u]$, so that $\P(U^*
\leq u, V^* \leq v) = u = \min(u,v)$. Thus we have verified that
$\P(U^* \leq u, V^* \leq v) = C^*(u,v)$ when $u \in [0,R]$. It
remains to consider the case $u \in (R,1]$. We can use the previous case
to establish that
\begin{align*}
  \P(U^* \leq u, V^* \leq v) &=   \P(U^* \leq R, V^* \leq v)
                               + \int_{R}^u \P\left( V^* \leq v \mid
                               U^* =x\right) \rd x \\ &= \min(R,v) +
                               \int_{R}^u \indicator{x \leq v} \rd x
                               = \min(u,v) = C^*(u,v).
    \end{align*}

To see that the copula we have constructed cannot correspond to the case of independent stochastic inversion, note first that the probability that $C$ places a point on the circle
rather than one of the diagonals is
\begin{align*}
 \P(V^* \neq U^*) =  \int_0^R p_R(x) \rd x = R/2 = \sqrt{3/14}.
\end{align*}
We show that for the construction $((\udp_4^\blegendre)^\leftarrow(W, Z_1),
   (\udp_4^\blegendre)^\leftarrow(W, Z_2))^\top$ the corresponding probability must be strictly less than $R/2$.
   
In order for the latter to place a point on the circle we require $W=x \in (0,R)$, so that the cardinality of the pre-image $(\udp_4^\blegendre)^{-1}(\{x\})$ is 4. We also require that one of the two stochastic inverses  $(\udp_4^\blegendre)^\leftarrow(x, Z_1)$ and
$(\udp_4^\blegendre)^\leftarrow(x, Z_2)$ corresponds to an outer arm of $\udp_4^\blegendre$ while the other corresponds to an inner arm; any other configuration would place a point on one of the diagonals. Let us write the probabilities of the four different ordered values of the stochastic inverse at $x$ as $p_{1}(x),\ldots,p_4(x)$ and observe that, by the symmetry of $\udp_4^\blegendre$ (see Figure~\ref{fig:Tplots}), we have $p_{1}(x) = p_{4}(x)$ and $p_{2}(x) =
p_{3}(x)$. Writing the probability of an outer arm as $p_o(x) = p_1(x) + p_4(x)$ we then have that
\begin{displaymath}
  \P(\text{circle} \mid W = x) = 2p_o(x)(1-p_o(x)) \leq 0.5
\end{displaymath}
and it follows that $\P(\text{circle}) \leq R/2$ with equality only in the case that $p_o(x) = 0.5$ for almost all $x \in (0,R)$. However this would only be true if $p_1(x) = p_2(x) = p_3(x) = p_4(x) = 1/4$ for almost all $x \in (0,R)$ and this contradicts the properties of $\udp_4^\blegendre$, since the gradients of the outer arms differ from the gradients of the inner arms.
The copula corresponding to independent stochastic inversion can be derived by defining a different function $p_R(x)$ in~\eqref{eq:circle_construct}, but the resulting expression for $C^*$ is more complex.
\end{example}

The copula $C$ in the previous example must correspond to a bivariate stochastic inverse construction 
$ (U,V)^\top = ((\udp_4^\blegendre)^\leftarrow(W, Z_1),
  (\udp_4^\blegendre)^\leftarrow(W, Z_2))^\top$ where the copula of the randomizer vector $(Z_1,Z_2)^\top$ depends on $W$. In the next example, we show how dependent randomizers can be used to construct another maximizing copula.

\begin{example}[Prohibition sign copula]\label{ex:Sprohibition}
 Let the realized value $W =
 x$ be a value in the $T$-partition of $\udp_4^\blegendre$.
The cardinality $n(x)$ of the
  pre-image set $(\udp_4^\blegendre)^{-1}(\{x\})$ is either 2 or 4. Moreover, for values $x$ where $n(x) =2$
  the udp function $\udp_4^\blegendre$ is linear with gradient 2. For
  each stochastic inverse we take ${\udp_4^\blegendre}^\leftarrow(x, Z_i) =
  G_x^{-1}(Z_i)$ where $G_x$ is the distribution function of a
  discrete random variable taking values $r_1(x) = (1-x)/2$ and $r_2(x) = (1+x)/2$ each
  with probability $0.5$. In this case we take $Z_1$ and $Z_2$ to be
  comonotonic uniform random variables.

  Now consider the case where $n(x) = 4$ and, as in the previous example, let $p_{1}(x),\ldots,p_4(x)$
  denote the four probabilities associated with the four values
  $r_{1}(x),\ldots,r_4(x)$ in ${\udp_4^\blegendre}^{-1}(\{x\})$, where
  the latter are ordered by size. As we have noted in Example~\ref{ex:Sconstr-extr-copul}, these
probabilities take 2 different values in
general and we write
$p_{1}(x) = p_{4}(x) = p_A(x)$ and $p_{2}(x) =
p_{3}(x) = p_B(x)$. It is also a feature of this case that we
must have $p_B(x) \geq p_A(x)$, since the absolute gradients of the outer
arms of $\udp_4^\blegendre$ in Figure~\ref{fig:Tplots} are greater
than the absolute gradients of the inner arms. In this case we
choose the copula $C_x$ of $Z_1$ and $Z_2$ to be one that gives
the joint probabilities
\begin{equation}\label{eq:S46}
 p_{jk}(x) = \P\left(G_x^{-1}(Z_1) =
   r_{j}(x),G_x^{-1}Z_2) = r_{k}(x)
 \right),\quad j= 1,\ldots,4, \; k = 1,\ldots,4,
\end{equation}
in Table~\ref{tab:Sprohibition}. Note how this set of joint
probabilities gives the correct
marginal probability distribution $G_x$ for $ (\udp_4^\blegendre)^\leftarrow(x, Z_i) = G_x^{-1}(Z_i)$ for $i=1,2$. Note also that we have assigned
zero probability to cases where  $(\udp_4^\blegendre)^\leftarrow(x,
Z_1) =  {\udp_4^\blegendre}^\leftarrow(x, Z_2)$.
 Figure~\ref{fig:Sprohibition} shows the subset of $\mathsf{U}^\blegendre_{44}$ on which the resulting
 ``prohibition sign copula'' is concentrated.

\begin{table}[h]
  \centering
\begin{tabular}{r|rrrr|r}
 $p_{jk}(x)$ & $k=1$ & $2$ & $3$ & $4$ & $p_j(x)$
                                                          \\ \hline
  $j=1$ & 0 & $\frac{p_A(x)}{2}$ & $\frac{p_A(x)}{2}$ & 0
                                                                 &
                                                                   $p_A(x)$ \\
   $2$ & $\frac{p_A(x)}{2}$ & 0 & $p_B(x) - p_A(x)$ &
                                                              $\frac{p_A(x)}{2}$
                                                                 & $p_B(x)$\\
    $3$ & $\frac{p_A(x)}{2}$ & $p_B(x) - p_A(x)$ & 0 &
                                                               $\frac{p_A(x)}{2}$
                                                                 & $p_B(x)$ \\
   $4$ & 0 & $\frac{p_A(x)}{2}$ & $\frac{p_A(x)}{2}$ & 0
                                                                 &
                                                                   $p_A(x)$\\
  \hline
 $p_k(x)$ & $p_A(x)$ & $p_B(x)$ & $p_B(x)$ & $p_A(x)$ & 1
\end{tabular}
\caption{\label{tab:Sprohibition} Contingency table showing joint
  probabilities $p_{jk}(x)$, as defined in~\eqref{eq:S46}, that are used to specify the copula $C_x$ of
  $(Z_1,Z_2)^\top$ for $n(x) =4$ in the construction of a
bivariate random vector $(U,V)^\top$ that maximizes $\rho_{44}^\blegendre$ by dependent
stochastic inversion; see Example~\ref{ex:Sprohibition}.}
\end{table}
 \begin{figure}[t]
  \centering
   \includegraphics[width=3in,height=3in]{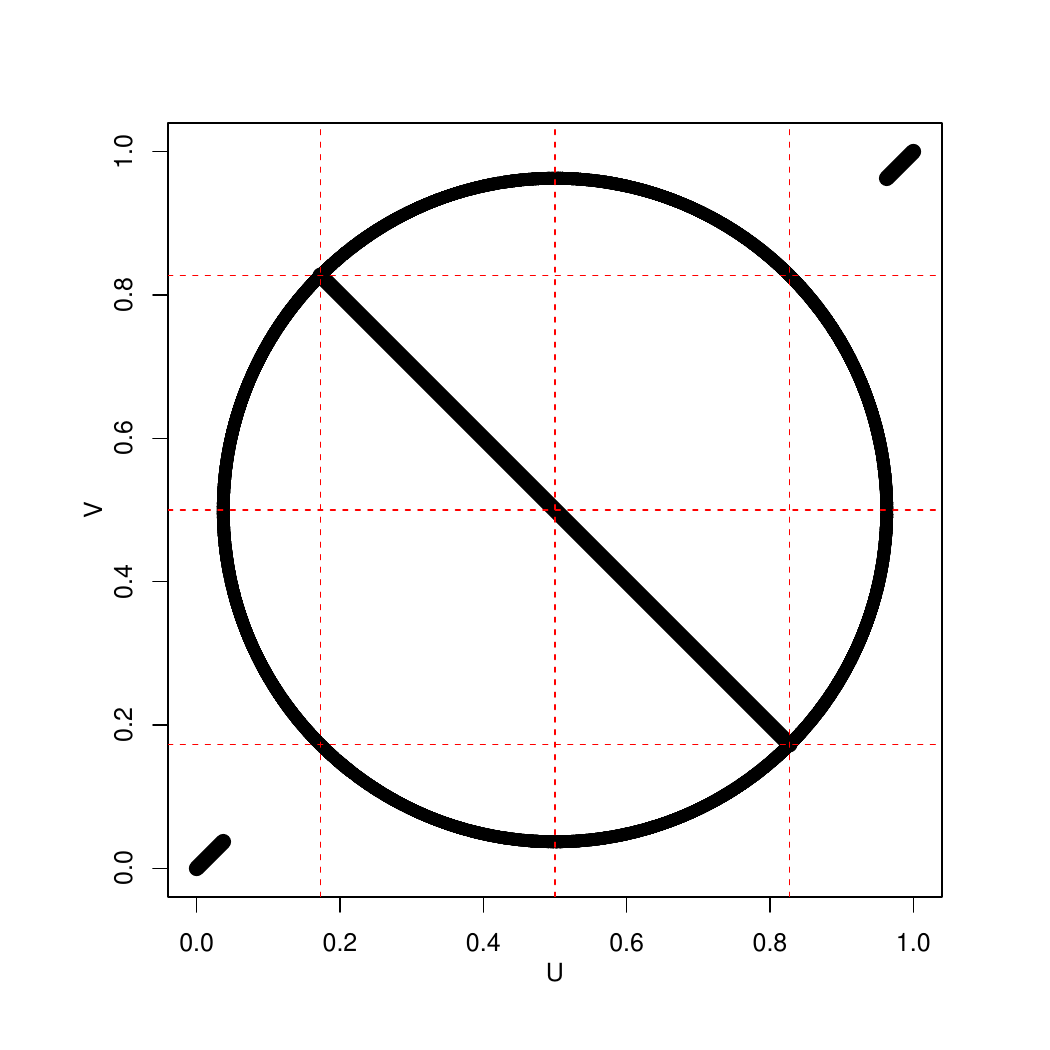}
   \caption{\label{fig:Sprohibition} The
subset of $\mathsf{U}^\blegendre_{44}$ on which the 
 ``prohibition sign copula'' of Example~\ref{ex:Sprohibition} is concentrated.}
\end{figure}

\end{example}

\end{document}